\documentclass[conference]{IEEEtran}
\IEEEoverridecommandlockouts
\usepackage{microtype}                 
\PassOptionsToPackage{warn}{textcomp}  
\usepackage{textcomp}                  
\usepackage{mathptmx}                  
\usepackage{times}                     
\usepackage{cite}                      
\usepackage{tabu}                      
\usepackage{booktabs}                  
\usepackage{adjustbox}
\usepackage{tabularx} 
\usepackage{array}
\usepackage{caption}
\usepackage{subcaption}
\usepackage{gensymb}
\usepackage{amsmath}
\usepackage{xcolor}
\usepackage{makecell}
\usepackage{algorithm}
\usepackage{algpseudocode}
\usepackage{multirow}
\usepackage{enumitem}
\usepackage{float}
\usepackage{hyperref}
\newcolumntype{P}[1]{>{\centering\arraybackslash}p{#1}}
\newcolumntype{M}[1]{>{\centering\arraybackslash}m{#1}}
\newcolumntype{C}{>{\centering\arraybackslash}X}
\def\BibTeX{{\rm B\kern-.05em{\sc i\kern-.025em b}\kern-.08em
    T\kern-.1667em\lower.7ex\hbox{E}\kern-.125emX}}
\begin{document}

\title{Scalable Volume Visualization for Big Scientific Data Modeled by Functional Approximation\\
}


\author{\IEEEauthorblockN{Jianxin Sun\textsuperscript{1}, David Lenz\textsuperscript{3}, Hongfeng Yu\textsuperscript{1,2}, Tom Peterka\textsuperscript{3}}
\IEEEauthorblockA{\textit{\textsuperscript{1}School of Computing, University of Nebraska-Lincoln, Lincoln, NE, USA} \\
\textit{\textsuperscript{2}Holland Computing Center, University of Nebraska-Lincoln, Lincoln, NE, USA}
\\
\textit{\textsuperscript{3}Argonne National Laboratory, Lemont, IL, USA}
}
}

\maketitle

\begin{abstract}
Considering the challenges posed by the space and time complexities in handling extensive scientific volumetric data, various data representations have been developed for the analysis of large-scale scientific data. Multivariate functional approximation (MFA) is an innovative data model designed to tackle substantial challenges in scientific data analysis. It computes values and derivatives with high-order accuracy throughout the spatial domain, mitigating artifacts associated with zero- or first-order interpolation. However, the slow query time through MFA makes it less suitable for interactively visualizing a large MFA model. In this work, we develop the first scalable interactive volume visualization pipeline, MFA-DVV, for the MFA model encoded from large-scale datasets. Our method achieves low input latency through distributed architecture, and its performance can be further enhanced by utilizing a compressed MFA model while still maintaining a high-quality rendering result for scientific datasets. We conduct comprehensive experiments to show that MFA-DVV can decrease the input latency and achieve superior visualization results for big scientific data compared with existing approaches.
\end{abstract}


\section{Introduction}
\label{sec:Introduction}
Advancement in instrumentation and technology has enabled scientists in diverse fields, such as medical imaging, meteorology, materials science, and physical simulations, to generate or obtain 3D volumetric datasets at an increasingly rapid rate, which necessitate scalable volume visualization solutions to facilitate scientists to explore and gain discoveries from their datasets~\cite{sarker2021data, el20193d}. However, creating volumetric visualizations for large-scale scientific datasets presents distinctive performance and quality challenges that require comprehensive solutions to address them holistically. 

First, managing substantial scientific datasets on hardware systems with constrained memory resources and I/O bandwidths presents significant space and time complexities. These complexities can easily incur performance bottlenecks and further complicate the task of delivering a smooth user experience for interactively visualizing large-scale datasets. Researchers have developed various techniques to address this performance challenge. Out-of-core techniques\cite{8695851} partition the dataset into manageable chunks that can be rendered independently, ensuring memory usage remains constant. 
Multi-resolution volume rendering methods\cite{6876002} decrease the volume of data sent to the rendering pipeline based on zoom levels, rendering data at higher resolutions only when required. 
Data compression techniques~\cite{son2014data} aim to reduce the size of datasets generated in scientific research while preserving the essential information for processing and visualization.
Data streaming strategies\cite{beyer2013exploring} facilitate an incremental rendering of a dataset as its availability progresses by employing push and pull models. However, these methods often compromise rendering quality (data compression and multi-resolution methods) or input latency (out-of-core and streaming methods).

Second, accurately retrieving values and gradients from arbitrary positions within a 3D volumetric space is challenging but necessary for volume visualization algorithms to generate high-quality images. 
Popular interpolation methods for volume visualization include low-order filters, such as nearest neighbor search (NNS) and trilinear interpolation, and high-order filters, such as tricubic and Catmull-Rom. However, these filters can relatively easily generate artifacts in the interpolated values and gradients. Functional data analysis\cite{ramsay2002applied, ferraty2006nonparametric} uses functional approximation to represent the original discrete dataset for querying off-grid value and gradient with higher accuracy through the computing of geometric bases. An important consideration in functional data analysis involves selecting the appropriate family of basis functions. Common choices include Fourier\cite{boashash2015time}, wavelet\cite{jansen2005second}, and geometric splines\cite{de1978practical} bases. 
More recently, Austin et al.\cite{austin2016parallel} introduced the Tucker decomposition as a low-rank alternative. Majdisova and Skala also suggested the use of radial basis functions for particle data\cite{majdisova2017radial}. 
Multivariate functional approximation (MFA)~\cite{peterka_ldav18} is a new function approximation model with advantages of high-order evaluation of both value and derivative anywhere in the spatial domain, compact representation for large-scale volumetric data, and uniform representation of both structured and unstructured data. The initial discrete data require an initial preprocessing step involving encoding or prefiltering\cite{8130306, 875199} to obtain the sequence of coefficients of the interpolating functions.
However, the primary challenge when employing functional approximation lies in its computational complexity, making it less suitable for real-time applications with stringent demands for information query latency. This latency becomes more severe when handling large-scale datasets.

In this work, we aim to tackle both quality and performance challenges, and propose a scalable volume visualization solution for big scientific datasets. First, we select functional approximation, specifically MFA, in our pipeline for its evaluation accuracy and better rendering quality. Second, we utilize distributed computing to accelerate both data fetching latency and rendering latency, so that the overall input latency of visualizing a large-scale MFA model can satisfy the requirements of interactive visualization. We name our proposed pipeline as MFA-based distributed volume visualization (MFA-DVV). To the best of our knowledge, this is the first work utilizing MFA as a data representation with distributed computing to speed up the rendering process for responsive user exploration of large-scale scientific datasets. The main contributions of this work include:
\begin{itemize}[noitemsep]
  \item A novel interactive volume visualization framework (MFA-DVV) using the MFA model with distributed computational architecture to visualize big scientific datasets with high quality and low input latency. 
  \item Detailed performance analysis of critical components of the proposed pipeline.
  \item Comprehensive experiments studying the effect of using the compressed MFA model to optimize the performance.
\end{itemize}
The subsequent sections of this paper are organized as follows: we first provide an overview of related work in Section~\ref{sec:Related} and the background of MFA in Section~\ref{sec:Backgroud}. Section~\ref{sec:Methods} outlines the architecture and implementation of the proposed MFA-DVV. Experimental results and evaluation are presented in Section~\ref{sec:results}. We draw our conclusions in Section~\ref{sec:Conclusions}.
\section{Related Work}
\label{sec:Related}


\subsection{Large-scale Volume Visualization}

Researchers have developed various algorithms to visualize large-scale scientific data, encompassing tasks such as isosurface computation, streamline computation, and I/O-efficient volume rendering~\cite{https://doi.org/10.1111/cgf.12605, https://doi.org/10.1111/cgf.13671}.
%
Through the consideration of the distance between the camera view and individual data segments within the current view, multi-resolution techniques\cite{6876002, https://doi.org/10.1111/cgf.12102} intelligently load data segments at different levels of detail. This approach reduces the volume of data loaded for rendering while preserving a comparable level of rendering quality.
Efficiently accessing raw data in real-time visualization tasks such as progressive slicing and particle traces can be facilitated by employing strategies like an optimized disk data layout~\cite{10.1145/582034.582036}, or by utilizing pre-computed lookup tables~\cite{10.5555/502125.502134}.
Cox and Ellsworth~\cite{10.5555/266989.267068} introduce a framework for out-of-core scientific visualization systems. This framework involves modifying the I/O subsystem to implement application-controlled demand paging. The approach capitalizes on the observation that many crucial visualization tasks only require access to a small portion of extensive datasets at a given time. Similarly, Ueng et al.~\cite{646239}, and Leutenegger and Ma~\cite{leutenegger1999fast} adopt a comparable strategy to load data on demand for visualization tasks by reorganizing the physical data on disk and employing structures like octrees or R-tree partitions to handle both structured and unstructured data. 


\subsection{Data Compression}

The substantial size of large-scale scientific dataset results in prolonged data retrieval times due to constraints in bandwidths of various I/O types. Lossy compression\cite{JAYASANKAR2021119} can be used to mitigate issues arising from I/O intensive workloads during data storage and transfer. TTHRESH\cite{8663447} (using Tucker decomposition\cite{10.1007/s00371-015-1130-y}), TAMRESH\cite{https://doi.org/10.1111/cgf.12102} (using global tensor approximation\cite{doi:10.1137/07070111X} factor matrices), ZFP\cite{6876024} (a floating-point compressor using custom transform matrices), SZ\cite{7516069} (using best fit curve-fitting compression) and SQ\cite{10.1007/978-3-642-32820-6_83} (preserving connected and coherent regions) are popular choices. Although some compression algorithms, like ZFP, support random-access decompression\cite{1250385}, none have the ability to query at arbitrary locations away from the sample locations of the original discrete dataset, meaning they are subject to interpolation artifacts. Multivariate functional approximation (MFA) can model large-scale raw volumetric data into an MFA model with a compact size. Various volume visualization techniques can directly render results from the MFA model without referencing the original data, enabling large-scale volume visualization under limited memory resources.

\subsection{Acceleration Techniques}

Visualizing large datasets can be accelerated by leveraging environments and platforms like multi-core CPUs, GPUs, and high-performance computing (HPC) clusters~\cite{zhang2005survey}.
Piringer et al.~\cite{5290719} present a generic multithreaded visualization architecture that helps avoid pitfalls related to multithreading with visual feedback. Piringer et al.~\cite{Liu2015OnVL} develop POIViz using radial representation to compute a 2D layout of the multidimensional dataset in parallel on CPU and GPU to improve the visualization of large datasets on a single computer. GPU-based large-scale volume visualization \cite{https://doi.org/10.1111/cgf.12605} combines the parallel processing power with out-of-core methods and data streaming to improve interactivity. In this work, we exploit distributed computing power to develop a scalable volume visualization solution.

\section{Multivariate Functional Approximation}
\label{sec:Backgroud}
\subsection{Background}

\begin{figure}[t]
 \centering 
 \includegraphics[width=0.59\linewidth]{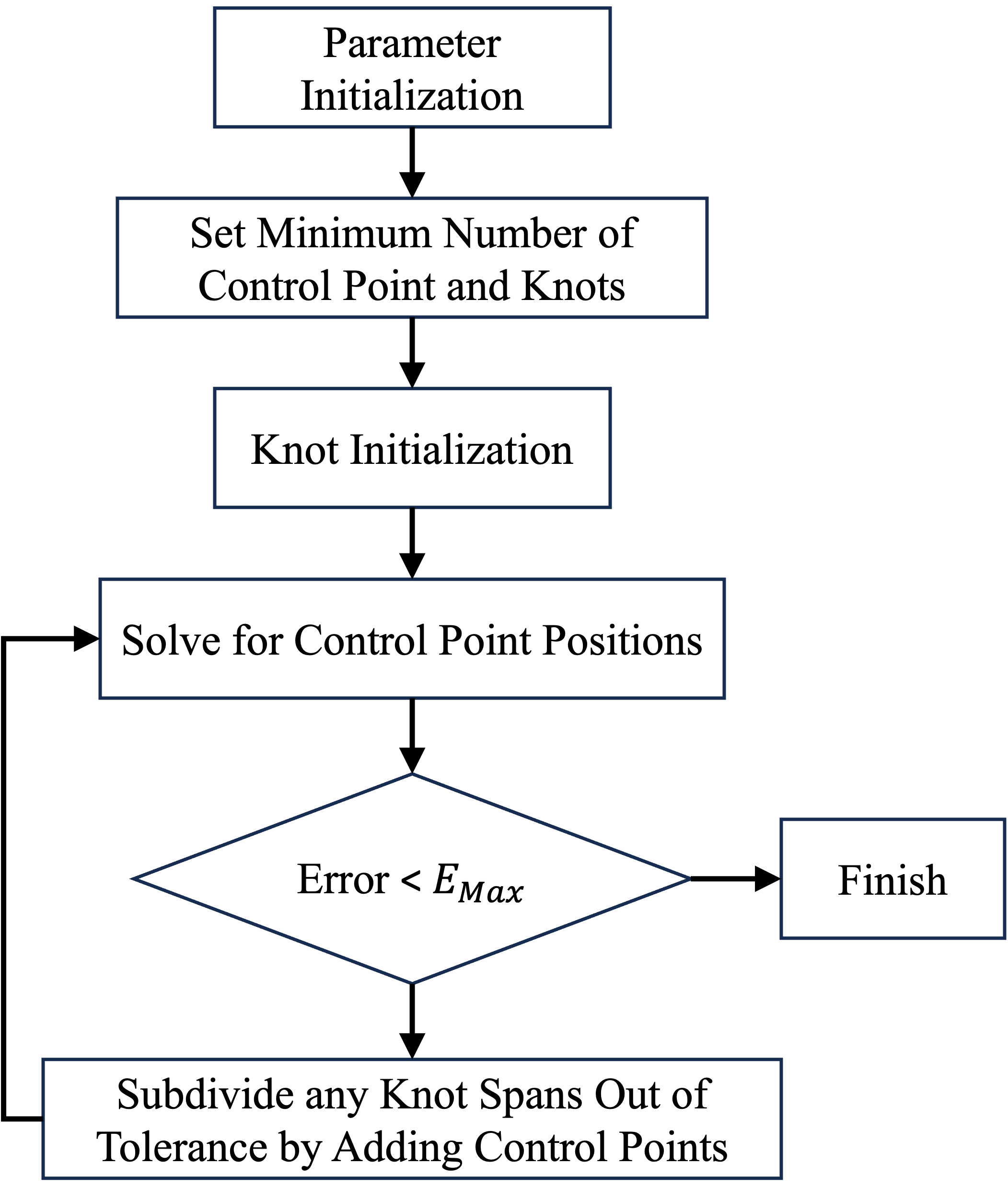}
 \caption{Key steps to construct an MFA model from raw input data. $E_{Max}$ is the maximum allowable relative error.}
 \label{fig:mfaEncoding}
\end{figure}

Multivariate Functional Approximation (MFA)~\cite{peterka_ldav18
} represents discrete high-dimensional scientific data using a functional basis representation derived from the tensor product of B-spline functions. The geometric characteristics of the field and the values of the scientific data are captured and efficiently represented through a collection of control points and knots. \autoref{fig:mfaEncoding} illustrates the key steps to construct an MFA model from input data. Initially, the system computes parameterization and establishes an initial knot distribution with the minimum required number of control points. Additional control points and knots are dynamically introduced until all evaluated points within each knot span meet the specified maximum allowable relative error compared to the original points. During this iterative process, knot spans exceeding the tolerance threshold are subdivided, and the functional approximation is recalculated. From the input data to its MFA model, various levels of compression can be achieved by adjusting the number of control points to generate a compact MFA model.

\begin{figure*}[!ht]
 \centering 
 \includegraphics[width=1.0\linewidth]{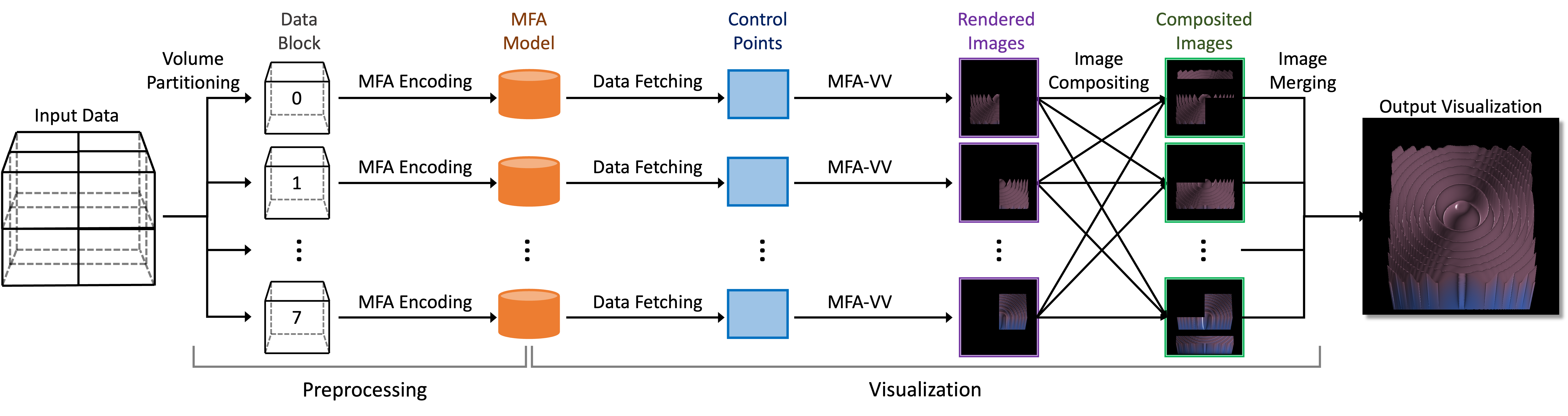}
 \caption{Architecture of our MFA-DVV. In this example, the input data is partitioned into eight octants with identical size.}
 \label{fig:overview}
\end{figure*}

\subsection{Advantages of MFA}

In contrast to other local filters that aim to address numerous small local optimization problems, MFA represents the resolution of a solitary global optimization spanning the entire domain. As a result, MFA gives more accurate approximation than other high-order local filters like tricubic and Catmull-Rom~\cite{sun2023mfa}. The detailed advantages of MFA are: 
\begin{itemize}[noitemsep]
  \item While local filters either interpolate or precisely match input points, MFA approximates them, which permits the smoothing of high-frequency discretization artifacts while retaining information about underlying low-frequency patterns.
  \item Local filters tend to reduce continuity between successive filter applications, whereas MFA, functioning as a global model, maintains high-order continuity consistently across the entire domain.
  \item Local filters determine their size and position by adapting to the distribution of input points, effectively behaving like a sliding window over these points. Conversely, MFA distributes its piecewise polynomials based on predetermined knot positions, which remain independent of the input point distribution. These knot locations can accurately reflect the data's complexity rather than relying on the input point distribution. 
  \item Local filters may rely on finite approximations for gradient calculations, resulting in approximate high-order derivatives, especially at cell boundaries. On the other hand, MFA provides analytical high-order derivatives, ensuring precision in its gradient computations.
  \item Although the model is an approximation, its derived values, including gradients, are analytically accurate. Additionally, it offers analytical availability for higher-order derivatives, extending up to the polynomial degree.
\end{itemize}


\subsection{Opportunities for MFA}\label{mfaDisadvantage}
To achieve effective volume visualization, existing techniques necessitate searching or retrieving values and gradients from arbitrary locations within the volume. As a result, the latency at which this information is retrieved determines the overall performance of a volume visualization system. This retrieval latency is particularly pronounced when handling large datasets. While MFA excels at modeling large-scale scientific data in situ and provides more accurate evaluations than many local filters, its query time becomes the primary performance bottleneck for visualizing the MFA model. Although MFA's query time outperforms popular high-order local filters such as tricubic and Catmull-Rom, its query performance falls short when compared to straightforward local filters like nearest neighbor and trilinear interpolation. We aim to improve the performance of volume visualization of the MFA model encoded from large-scale datasets. We build a distributed volume visualization architecture for rendering the MFA model with improved overall input latency. Our work also utilizes the compressed MFA model to achieve a better performance and rendering quality than the general distributed volume rendering using the raw discrete data through trilinear interpolation~\cite{5219060}.
\section{Methods}
\label{sec:Methods}

\subsection{Architecture}

\autoref{fig:overview} shows the architecture of our developed MFA-based distributed volume visualization pipeline, MFA-DVV, for handling large-scale scientific datasets using multiple processors of a computing cluster or supercomputer. There are six key steps to generate the final visualization result.


\subsubsection{Volume Partitioning}

The input large-scale volume is first partitioned into blocks that will be distributed among processors. This step is a recursive binary partitioning until the target number of partitions is achieved. The partition follows the round-robin order among the volume's x, y, and z directions. If $n$ recursions of partitioning are executed, then the total number of partitioned blocks is $2^n$. The partitioned data block will be the input of the MFA encoder to generate the MFA model. 

\subsubsection{MFA Encoding}

Each partitioned block is encoded using the MFA encoder with a configuration where the key encoding parameters are the number of control points and the polynomial degree. Control points serve as coefficients that scale piecewise-continuous basis functions within the hypervolume of B-splines. Typically, the quantity of control points is equal to or fewer than the number of the input sample points. Meanwhile, the polynomial degree represents the degree of the basis function employed for modeling. The number of control points is the key parameter to determine the actual size of the encoded MFA model, which will be further evaluated in Section~\ref{compress}. Encoding is done following the previous partition step, and partitioned blocks are processed together for encoding on corresponding nodes in parallel.

\subsubsection{Data Fetching}

When the visualization starts, MFA models are first fetched to the memory of each target node for rendering. Each processing node only retrieves the MFA model file encoded from the corresponding partitioned block. The fetching time, determined by the I/O bandwidth, is one of the key contributors to the overall input latency. 

\subsubsection{MFA-VV}

An MFA-based volume visualization (MFA-VV) pipeline generates the required visualizations, like volume rendering and isosurface rendering, by considering the user configurations. Various volume visualization methods can utilize the query interfaces to directly query value and gradients from the MFA model to construct the visual representation. In this paper, the MFA-VV is ray-casting-based volume visualization utilizing the over operation between the neighboring samples on the ray segment. The rendering time is the main contributor to the final input latency. The complexity of rendering time is $O(n)$ where $n$ is the number of queries for value and gradient within the volume space. Since such queries of a functional approximation are expensive, as discussed in Section~\ref{mfaDisadvantage}, we parallelize the rendering pipeline on $m$ computing cores such that each node only needs to perform $n/m$ queries to finish its rendering of the local block. This approach significantly accelerates the rendering time on multiprocessor systems, thereby reducing the overall input latency of visualizing large-scale MFA models. The output of MFA-VV from each node is a partially-rendered image that must be aggregated with other images for the final resulting image.

\subsubsection{Image Compositing}

Partially-rendered images of partitioned blocks are combined together to construct the final image. Our image compositing utilizes the associative property of the over operation. As long as the distances of the two images to the viewpoint are comparable, the correct composited image can be obtained through the over operation in the pixel level of the two images. Image compositing requires communications between nodes for exchanging pixels, and the compositing time is determined by the resolution of the final image and how many nodes are involved in the compositing process. We utilize binary-swap to schedule the communication between nodes efficiently~\cite{5219060}. 
After the compositing, only a subset of the total pixels of each composited image are correct. Section~\ref{section:scalability} will investigate the compositing time with respect to the number of nodes employed for this task.

\subsubsection{Image Merging}

Correct pixels from each node are sent to the master node for concatenation into the final volume visualization image. Section~\ref{section:scalability} will also reveal how the image merging time scales when the number of nodes increases.

\subsection{Implementation}

During the course of the MFA-DVV pipeline, Steps 1 and 2 (i.e., volume partitioning and MFA encoding) are preprocesses done before the visualization. The visualization pipeline is from Step 3 (i.e., data fetching) to 6 (i.e., image merging) and executes in response to every data-dependent (e.g., changing transfer functions) or view-dependent (e.g., change view parameters) operation by the user. The parallel library we used to execute MFA encoding, data fetching, MFA-VV, and image compositing is DIY ("Do-It-Yourself" Parallel Analysis)~\cite{morozov_ldav16}, which is a block-parallel library for implementing scalable distributed- and shared-memory parallel algorithms that can run both in- and out-of-core. DIY supports data distribution in parallel through its write and read I/O functions, which are utilized for an MFA model writing to storage after encoding and data fetching from storage to system memory of each node. The binary-swap image compositing algorithm can be readily integrated by leveraging DIY's merge-reduce mechanism for managing communication. In addition, our MFA-DVV provides a user interface so that users can freely adjust data-dependent (e.g., setting for transfer functions and shading parameters) and view-dependent (e.g., exploring view parameters) operations. The setting of all operations will be passed to MFA-DVV for generating the corresponding visualization as desired.
\section{Results and Evaluation} 
\label{sec:results}

We design and conduct comprehensive experiments to evaluate the scalability and performance of the MFA-DVV. Additionally, we explore the impact of essential MFA encoding parameters on these factors.

\subsection{Datasets}
\subsubsection{Synthetic Datasets}

We use the Marschner-Lobb synthetic function to gain an accurate ground-truth reference for volume rendering. Marschner-Lobb is a function initially introduced for evaluating 3D resampling filters when applied to the conventional Cartesian cubic lattice\cite{346331}. We create corresponding discrete datasets derived from the Marschner-Lobb function to serve as the input for MFA-DVV and other discrete volume rendering algorithms. Accurately reconstructing a Marschner-Lobb signal from discrete samples is challenging due to its complex amplitude distribution across various frequencies. This complexity makes it a valuable benchmark for evaluating reconstruction quality. The Marschner-Lobb function $F_{ML}$ used in this paper is defined as:
\begin{equation}
F_{ML}(x, y, z)=\frac{1-\sin(\frac{\pi z}{2}) + \alpha(1+\rho_{r}(\sqrt{x^2+y^2}))}{2(1+\alpha)}
\end{equation}
where
\begin{equation*}
\rho_{r}(r) = \cos(2\pi f_{M} \cos(\frac{\pi r}{2}))
\end{equation*}
We use $f_{M} = 6$ and $\alpha = 0.25$ to generate discrete Marschner-Lobb datasets with different resolutions. The spatial boundaries on each dimension are $[0, 7]$. 

\subsubsection{Regular Real Datasets}

We further evaluate the rendering quality of MFA-DVV using real discrete datasets. These datasets are rendered using MFA models encoded at various compression levels. Specifically, we showcase the rendering quality on small datasets, such as Fuel and Nucleon, with sizes of $41^3$ and $64^3$, respectively. Fuel represents a simulation of fuel injection into a combustion chamber, while Nucleon simulates the two-body distribution probability of a nucleon in an atomic nucleus. Additionally, we examine larger datasets, Aneurysm and Bonsai, both with sizes of $256^3$. Aneurysm represents a rotational angiography scan of a head with an aneurysm, while Bonsai is a CT scan of a bonsai tree. The spatial boundaries for all the real datasets in each dimension range from $0$ to $255$.

\subsubsection{Large-scale Real Dataset}

The large-scale scientific dataset used to test the capability of MFA-DVV is the Richtmyer-Meshkov Instability (RMI) simulation~\cite{doi:10.1146/annurev.fluid.34.090101.162238}. The Richtmyer-Meshkov instability arises when a shock wave interacts with an interface separating two different fluids. The original time-varying RMI dataset is over 2TB. The input volume we use is the 160th time step with a spatial resolution of $1024 \times 1024 \times 1024$.

\subsection{Experiment Setup}

We run experiments using MFA-DVV on the Swan cluster at the Holland Computing Center (HCC) of the University of Nebraska-Lincoln. Swan is a massively parallel processing system with 9408 computing cores. There are 168 nodes in total, and each node has an Intel Xeon Gold 6348 CPU (56 cores) with 256GB RAM. Swan provides high-performance, low-latency communication for MPI jobs. We run our job in parallel using different processor numbers ranging from 1 to 1024. Due to the nature of the binary partition of the input data, the number of processors used is $2^n$, where $n$ is the times of partitions executed. Our test uses $n$ from 1 to 10.

\begin{table*}[t]
  \caption{Setting of volume partition and number of MFA control points with respect to the partition patterns.}
  \label{tab:partition}
  \scriptsize%
	\centering%
  \begin{adjustbox}{width=1.0\textwidth}
      \begin{tabu}{ c | c c  c c  c c  c c  c c c}
          \toprule
          Partition & $1\times1\times1$ & $2\times1\times1$ & $2\times2\times1$ & $2\times2\times2$ & $4\times2\times2$ & $4\times4\times2$ & $4\times4\times4$ & $8\times4\times4$ & $8\times8\times4$ & $8\times8\times8$ & $16\times8\times8$\\
          \midrule
          Block Size & $256\times256\times256$  & $128\times256\times256$ & $128\times128\times256$  & $128\times128\times128$ & $64\times128\times128$  & $64\times64\times128$ & $64\times64\times64$  & $32\times64\times64$ & $32\times32\times64$ & $32\times32\times32$ & $16\times32\times32$ \\
          \midrule
          Ctrl Points & $256\times256\times256$  & $128\times256\times256$ & $128\times128\times256$  & $128\times128\times128$ & $64\times128\times128$  & $64\times64\times128$ & $64\times64\times64$  & $32\times64\times64$ & $32\times32\times64$ & $32\times32\times32$ & $16\times32\times32$ \\
          \midrule
          Cores & 1 & 2 & 4 & 8 & 16  & 32  & 64  & 128 & 256 & 512 & 1024 \\
          \bottomrule
      \end{tabu}
 \end{adjustbox}
\end{table*}

\begin{figure}
    \centering
    \begin{subfigure}[b]{0.49\linewidth}
        \centering
        \includegraphics[trim=0 0 0 0,clip,width=\linewidth]{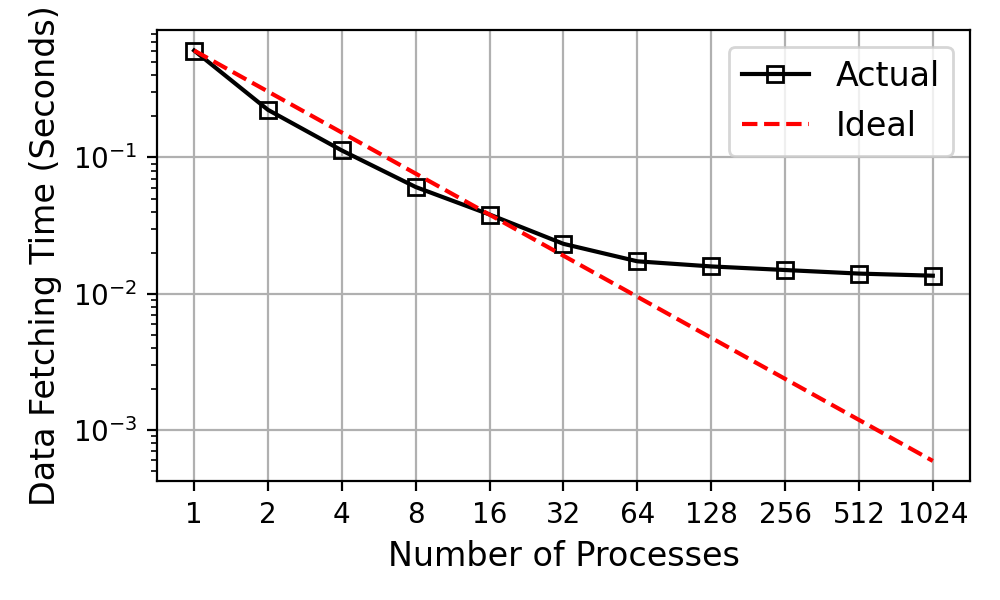}
        \caption{Data fetching time}
        \label{fig:fetchingTime}
    \end{subfigure}
    \hfill
    \begin{subfigure}[b]{0.49\linewidth}   
        \centering 
        \includegraphics[trim=0 0 0 0,clip,width=\linewidth]{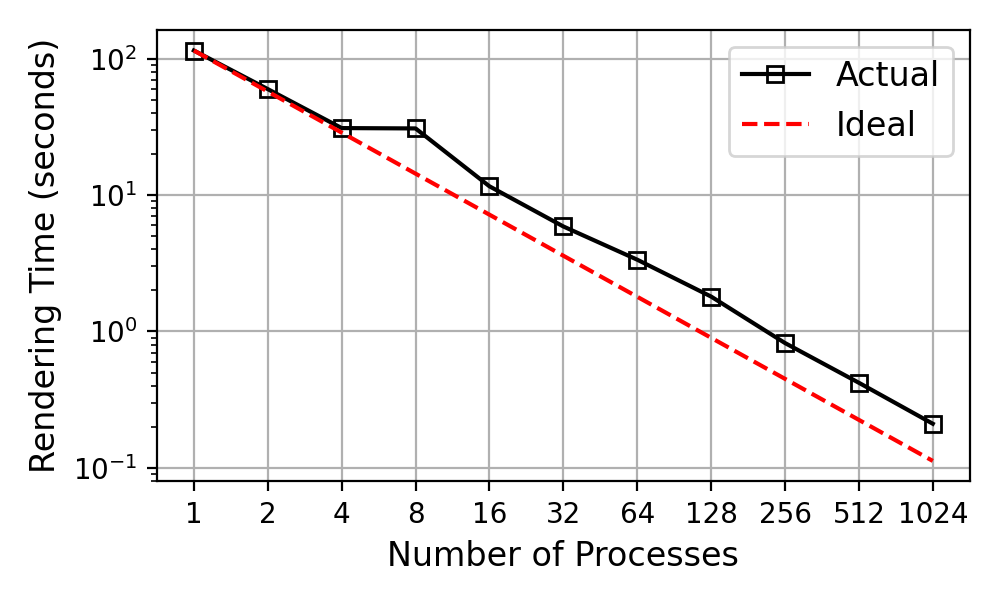}
        \caption{Rendering time}
        \label{fig:renderingTime}
    \end{subfigure}
    \vskip\baselineskip
    \begin{subfigure}[b]{0.49\linewidth}   
        \centering 
        \includegraphics[trim=0 0 0 0,clip,width=\linewidth]{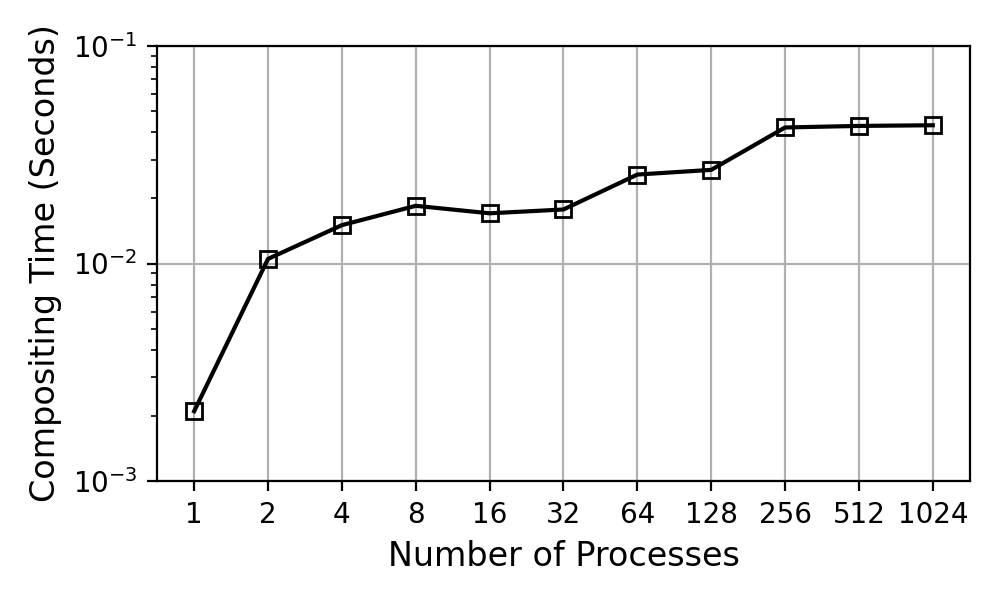}
        \caption{Image compositing time}
        \label{fig:compositingTime}
    \end{subfigure}
    \hfill
    \begin{subfigure}[b]{0.49\linewidth}   
        \centering 
        \includegraphics[trim=0 0 0 0,clip,width=\linewidth]{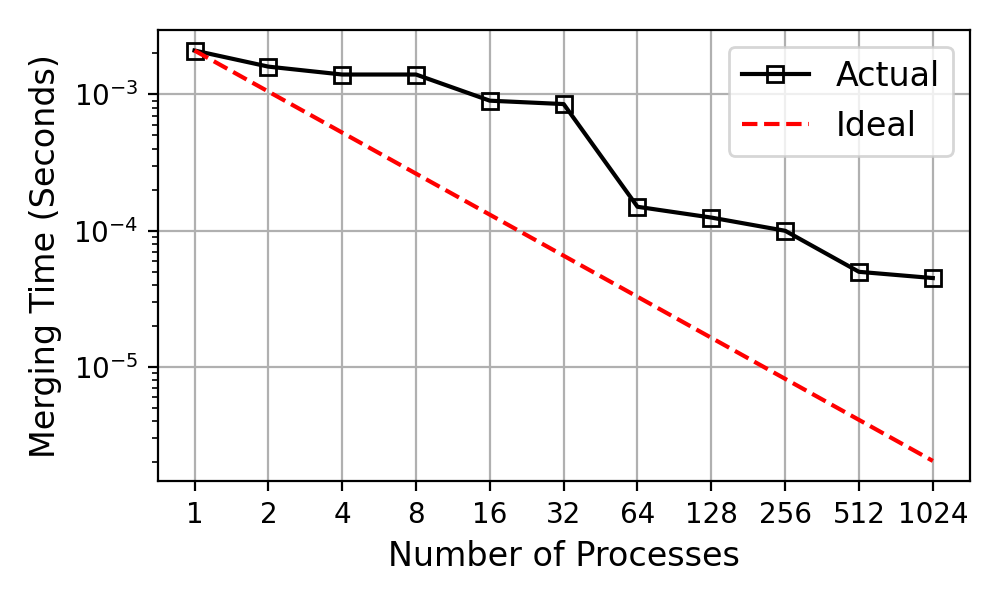}
        \caption{Image merging time}
        \label{fig:mergingTime}
    \end{subfigure}
    \vskip\baselineskip
    \begin{subfigure}[b]{0.49\linewidth}   
        \centering 
        \includegraphics[trim=0 0 0 0,clip,width=\linewidth]{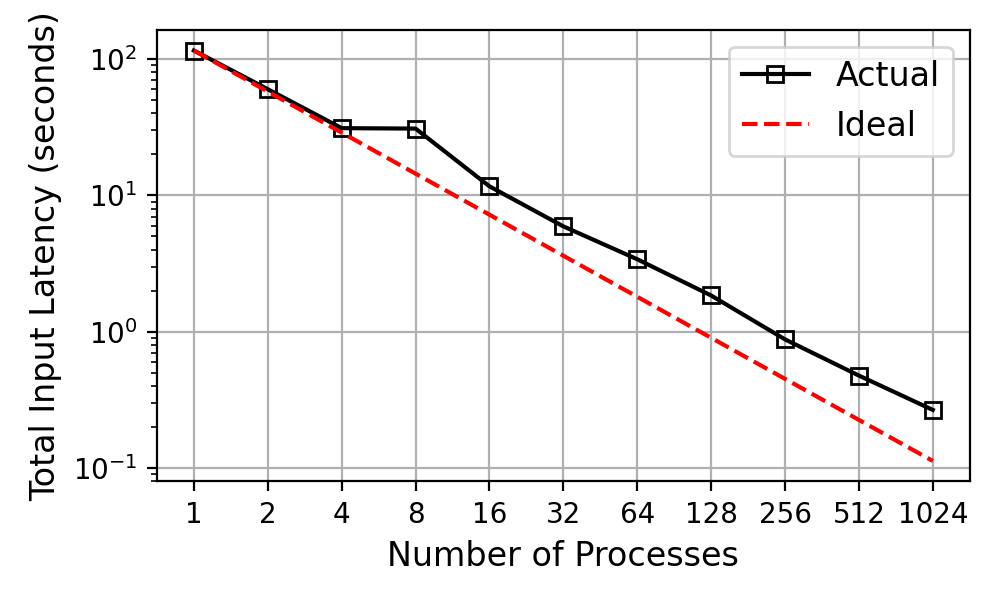}
        \caption{Total input latency}
        \label{fig:totalTime}
    \end{subfigure}
    \caption{Time breakdown of MFA-DVV rendering pipeline and total input latency.}
    \label{fig:scalabilityTest}
\end{figure}


\subsection{Scalability of MFA-DVV}
\label{section:scalability}

We perform the scalability test of MFA-DVV using the Marschner-Lobb dataset. The testing volume is a scalar field with a spatial resolution of $256\times256\times256$ with the float32 data type. We first preprocess the input discrete data with data partitioning and MFA encoding. The input data is encoded by MFA also using 256x256x256 control points, which is the same as the size of the input discrete sample of the dataset. The polynomial degree of the MFA encoder is set to 2 for non-linear approximation. \autoref{tab:partition} shows the partitioned block size, the number of control points for MFA encoding, and the parallel core counts with respect to the partition patterns. For high-quality visualization, a fine sample step is used for the ray-casting rendering step of MFA-DVV. The resolution of the final rendering image is $768\times768$. The total input latency of MFA-DVV on a large-scale dataset depends on the visualization time in its pipeline, including data fetching latency, MFA-VV rendering time, image compositing time, and image merging time. We summarize the time measurement for each of the four components and total input latency in \autoref{fig:scalabilityTest}. From the results, we can see that, for a general size volume, the main contributor to the input latency of MFA-DVV is the rendering time, as the times used in other states are comparably much smaller. The scalability results when handling large-scale data will be investigated in the next section.

The data fetching time depends on the size of the MFA model encoded for each partitioned block. Since the number of control points used for encoding the MFA model is the same as the size of the samples of input partitioned blocks, the MFA model has the same size as the input block. As the size of the blocks goes down together with the MFA model when using more processes, data prefetching time also decreases. Because the dataset used here is relatively small in size, and the partitioned blocks can become very small (e.g., as small as 64KB) as more processes are used, it reaches the saturation points of the I/O interfaces, as shown in \autoref{fig:fetchingTime}. For a large-scale dataset, the data fetching time still scales well because the partitioned data block will not reach such a small limit in file size. 

The rendering time scales with the number of processes as shown in \autoref{fig:renderingTime} for its computational cost are distributed evenly across all parallel processing nodes. 

\autoref{fig:compositingTime} shows the image compositing time. More communications are involved when more processes are used, as more images are needed for compositing. However, the compositing time increases marginally as the number of process increases, and each compositing operation between two images in each round only perform one over operation compared to many over operations executed in the MFA-VV step. As a result, the compositing time is normally not comparable with the rendering time.

The final image is merged in the master node by concatenating pixels sent from all the working nodes. As shown in \autoref{fig:mergingTime}, the image merging time also scales with the number of processes. More processes mean a smaller number of pixels are transmitted to the master node for image merging. Compared to the image compositing step that requires multiple rounds of communication among all working nodes, image merging only requires one communication between the working nodes to the master node. Thus, the image merging time is normally the fastest step of the pipeline. 

By summing up all the time used in each step, \autoref{fig:totalTime} shows that the total input latency of MFA-DVV scales well with respect to the number of processes, where the decreasing rate of the measured total input latency is almost the same as the ideal rate.

\subsection{MFA-DVV using Compressed MFA Model}
\label{compress}

\subsubsection{Quality Evaluation using Compressed Data}

\begin{figure}[t]
  \begin{subfigure}[b]{0.49\columnwidth}
    \includegraphics[trim=0 0 40 0,clip,width=\linewidth]{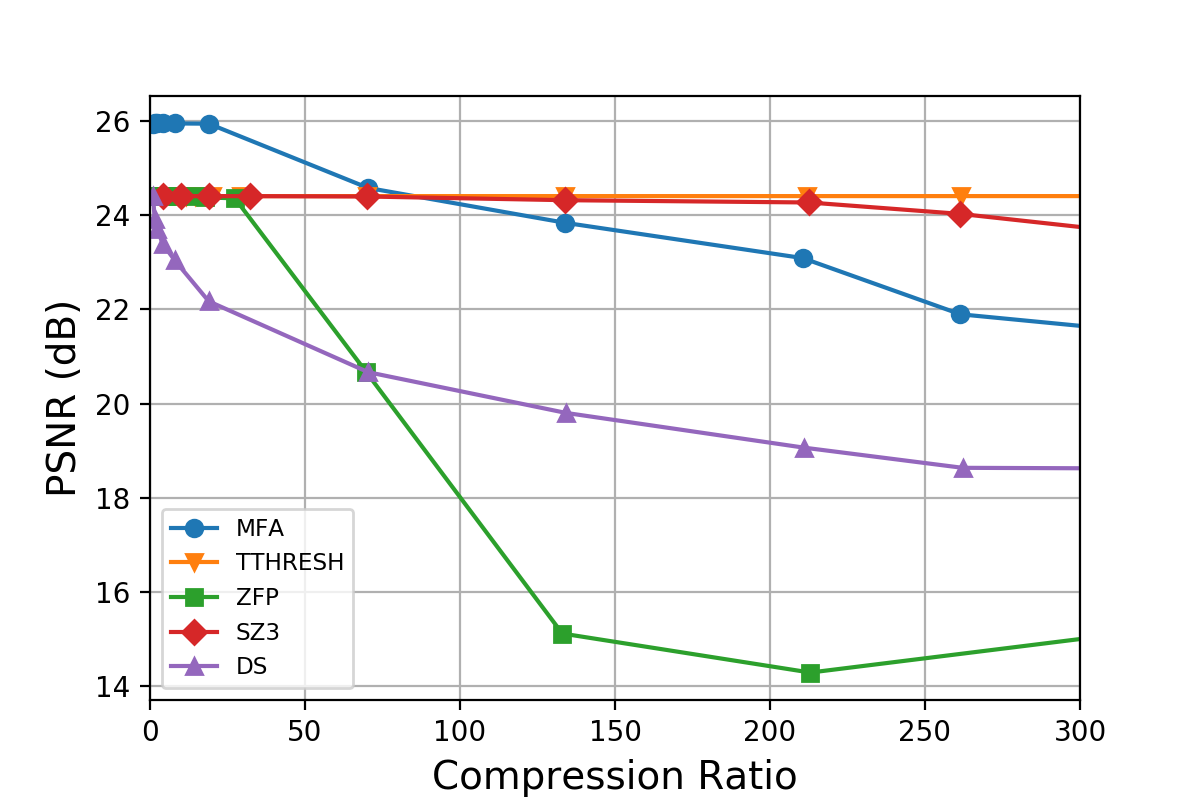}
    \caption{Image space quality}
    \label{fig:ratioVsPSNR}
  \end{subfigure}%
  \hspace*{\fill}   
  \begin{subfigure}[b]{0.49\columnwidth}
    \includegraphics[trim=0 0 40 0,clip,width=\linewidth]{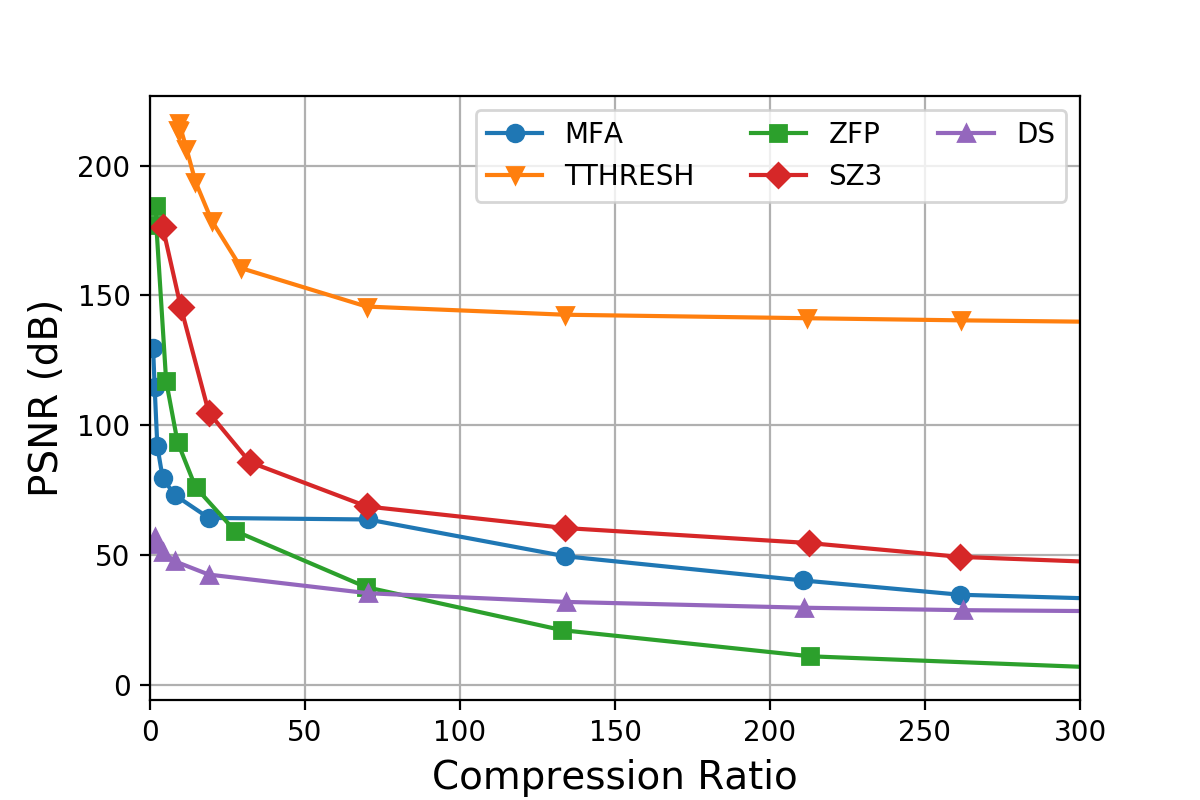}
    \caption{Volume space quality}
    \label{fig:ratioVsPSNR_volume}
  \end{subfigure}%
\caption{Compression evaluation using MFA and other compression algorithms.}
\label{fig:ratioVsQuality}
\end{figure}

\begin{figure}[t]
  \tiny
  \centering
\begin{tabular}{M{0.0\textwidth}M{0.091\textwidth}@{}M{0.091\textwidth}@{}M{0.091\textwidth}@{}M{0.091\textwidth}@{}M{0.091\textwidth}}
 &DS&SZ3&ZFP&TTHRESH&MFA-DVV\\
\rotatebox[origin=l]{90}{\thead{CR $\approx 20$}}&
\includegraphics[width=\linewidth]{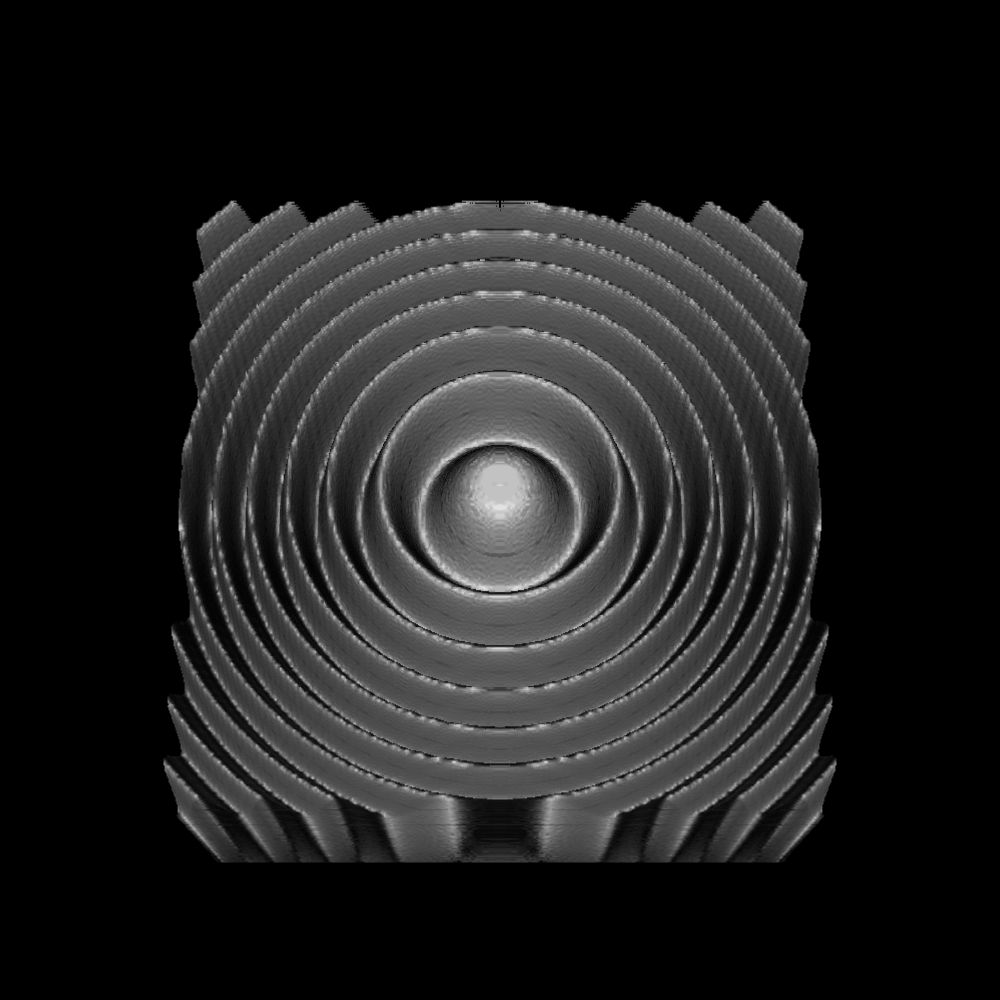}&
\includegraphics[width=\linewidth]{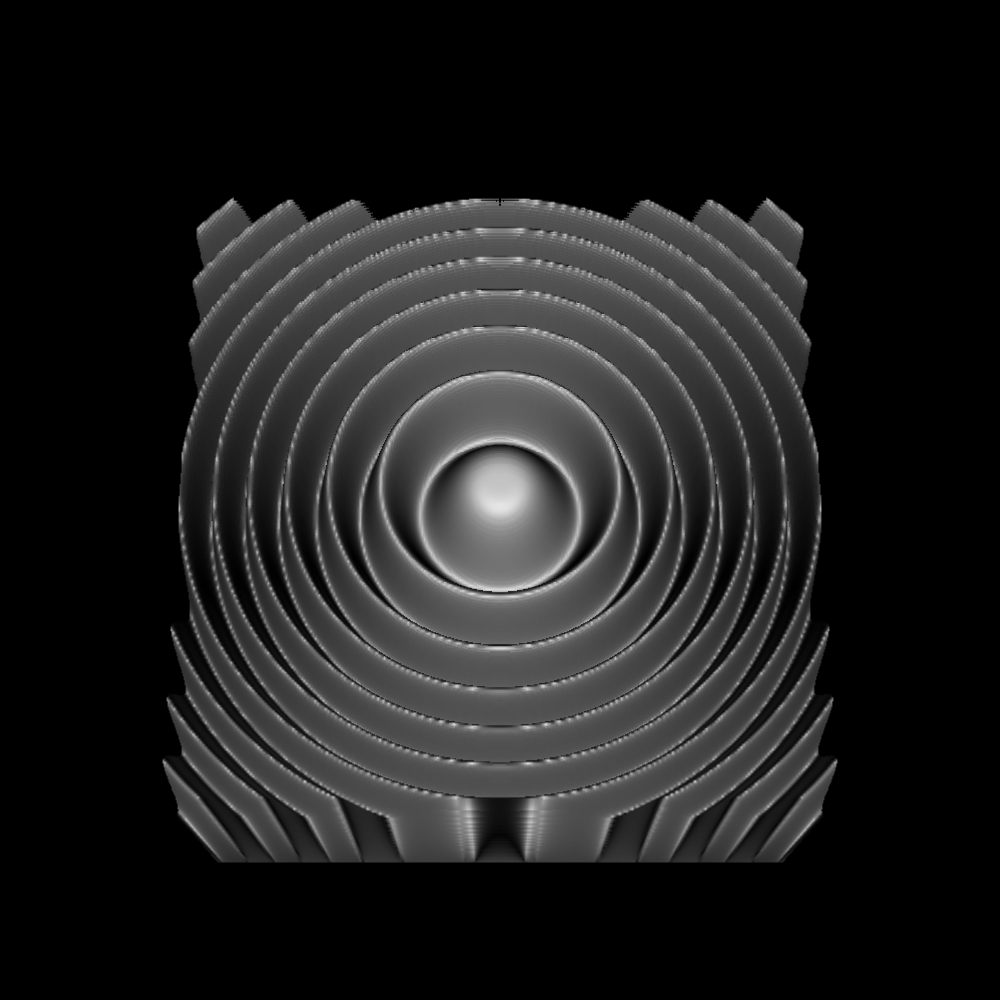}&
\includegraphics[width=\linewidth]{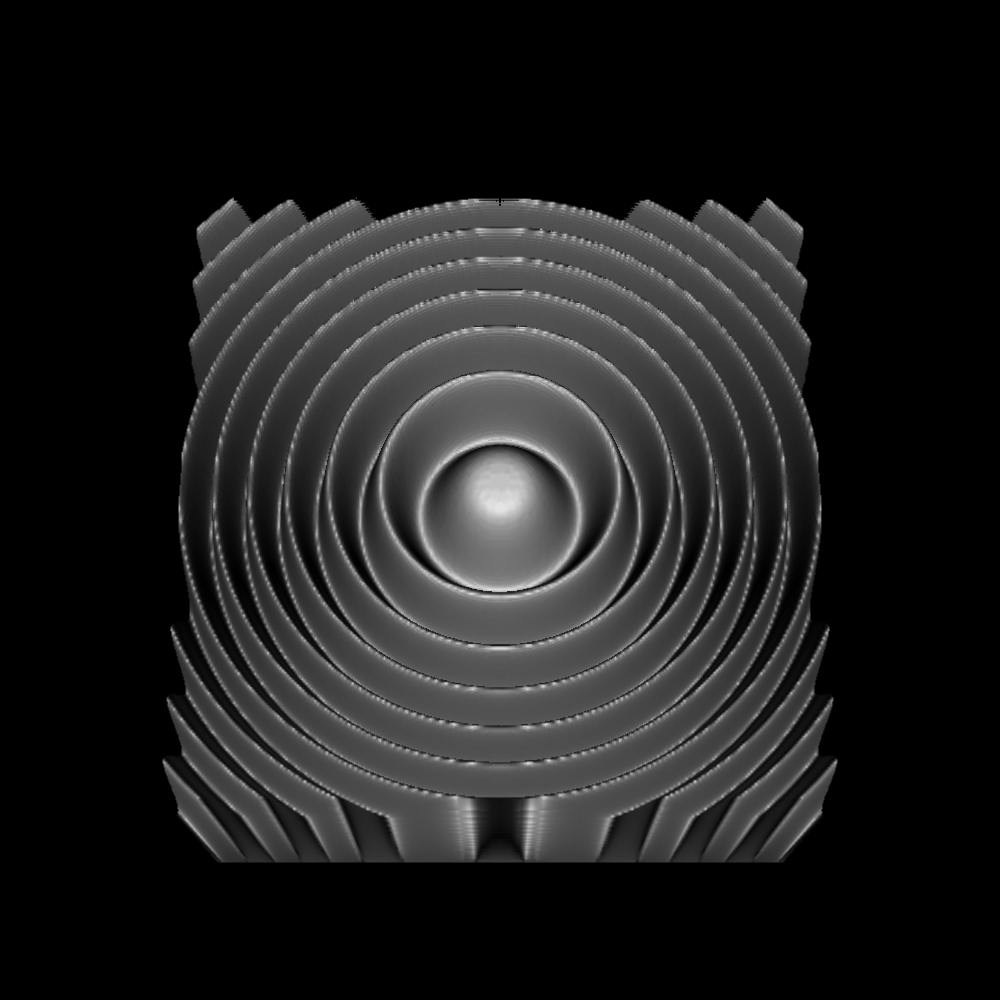}&
\includegraphics[width=\linewidth]{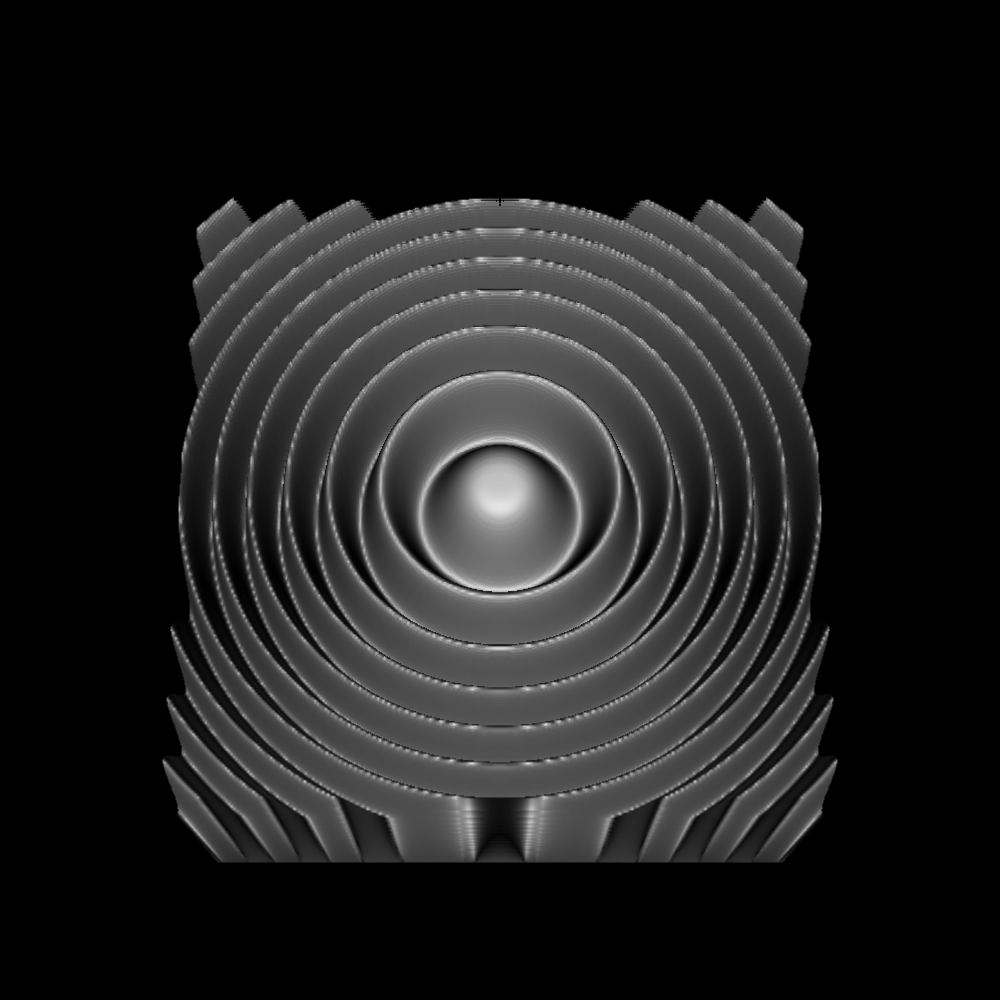}&
\includegraphics[width=\linewidth]{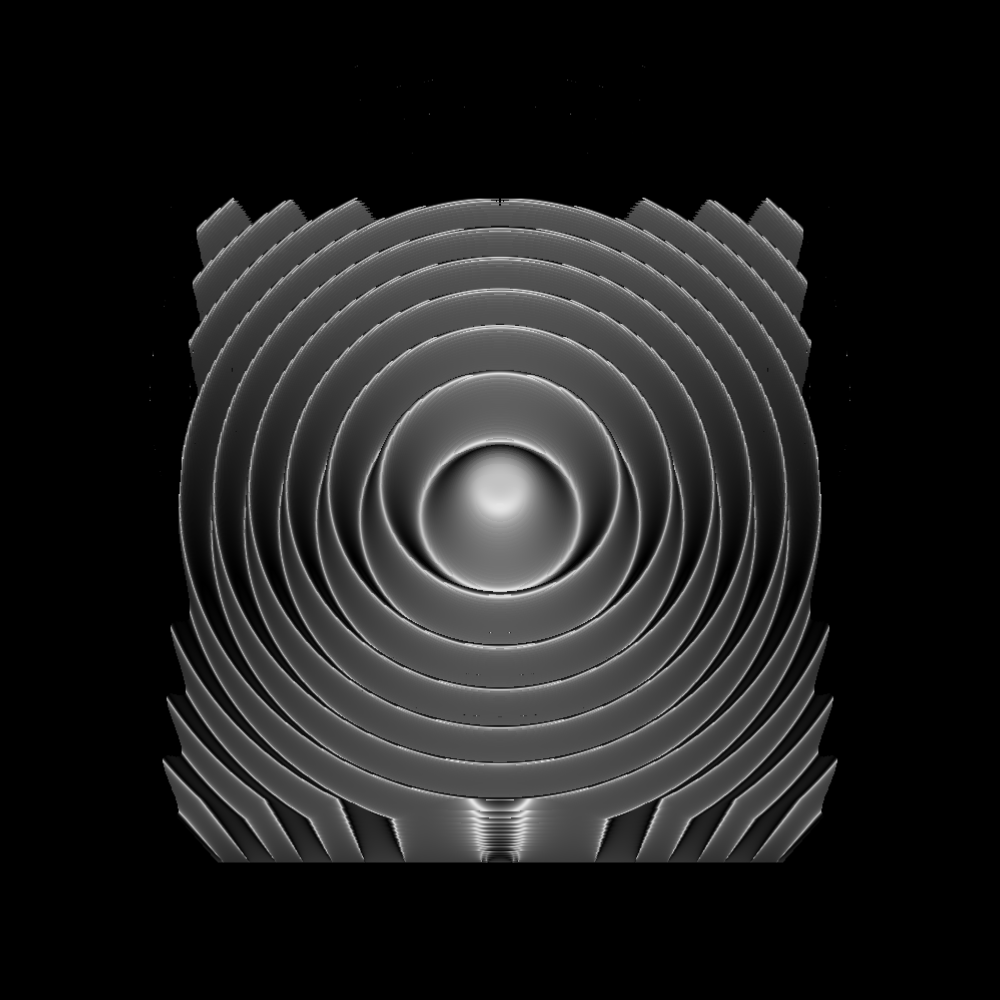}\\
 &PSNR: 22.171&PSNR: 24.406&PSNR: 24.379&PSNR: 24.406&PSNR: \textbf{25.944}\\
\rotatebox[origin=l]{90}{\thead{CR $\approx 69$}}&
\includegraphics[width=\linewidth]{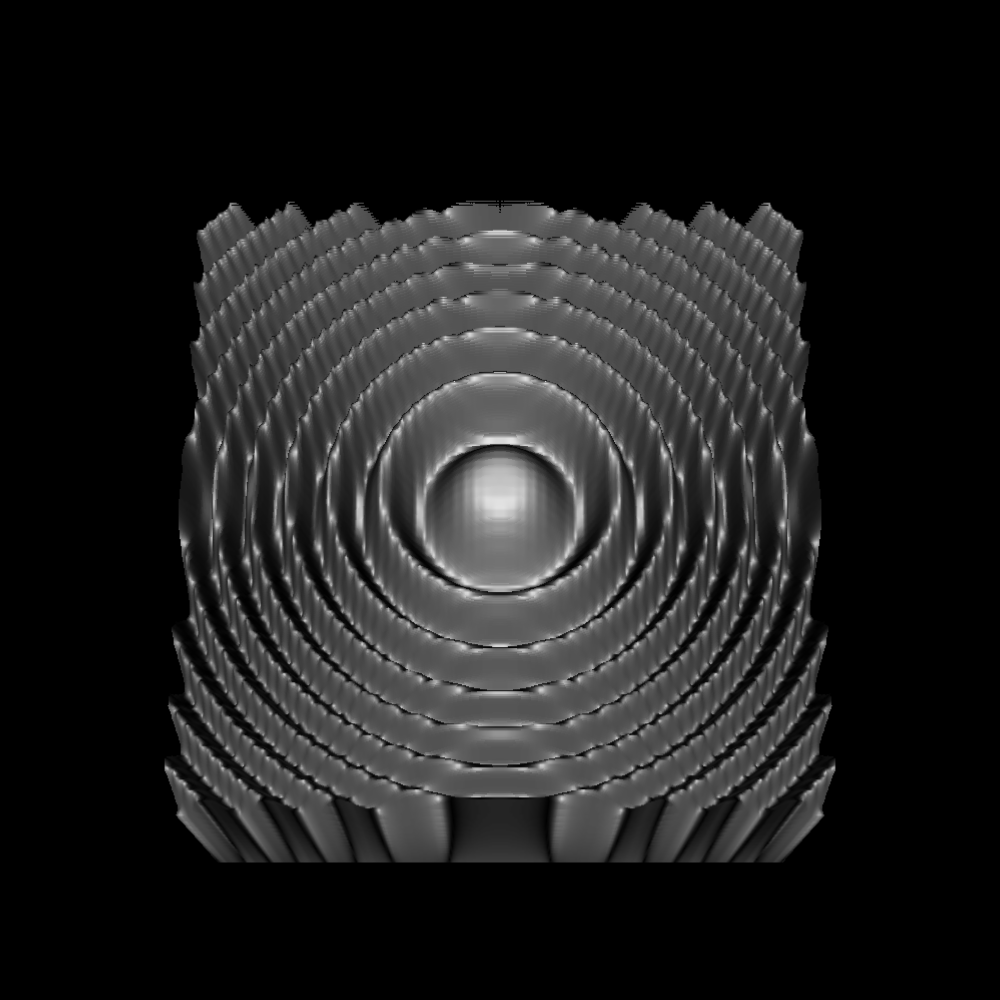}&
\includegraphics[width=\linewidth]{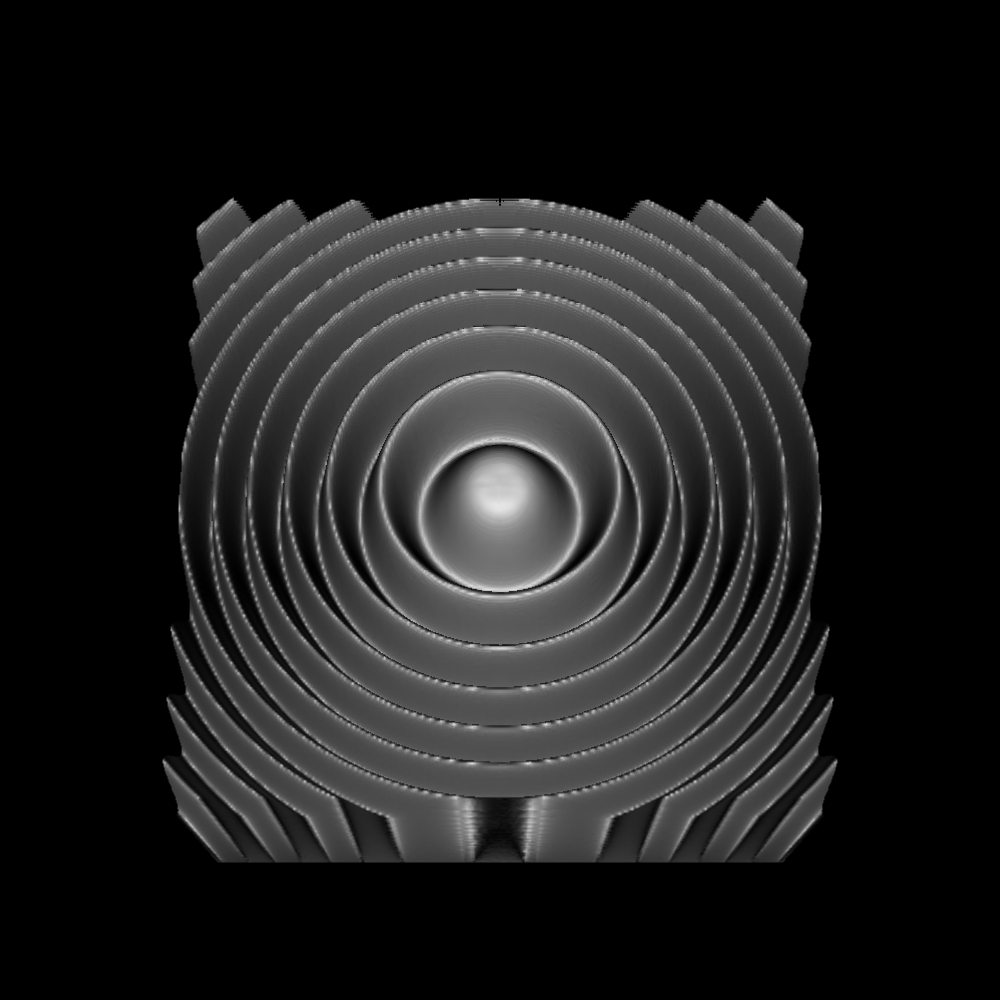}&
\includegraphics[width=\linewidth]{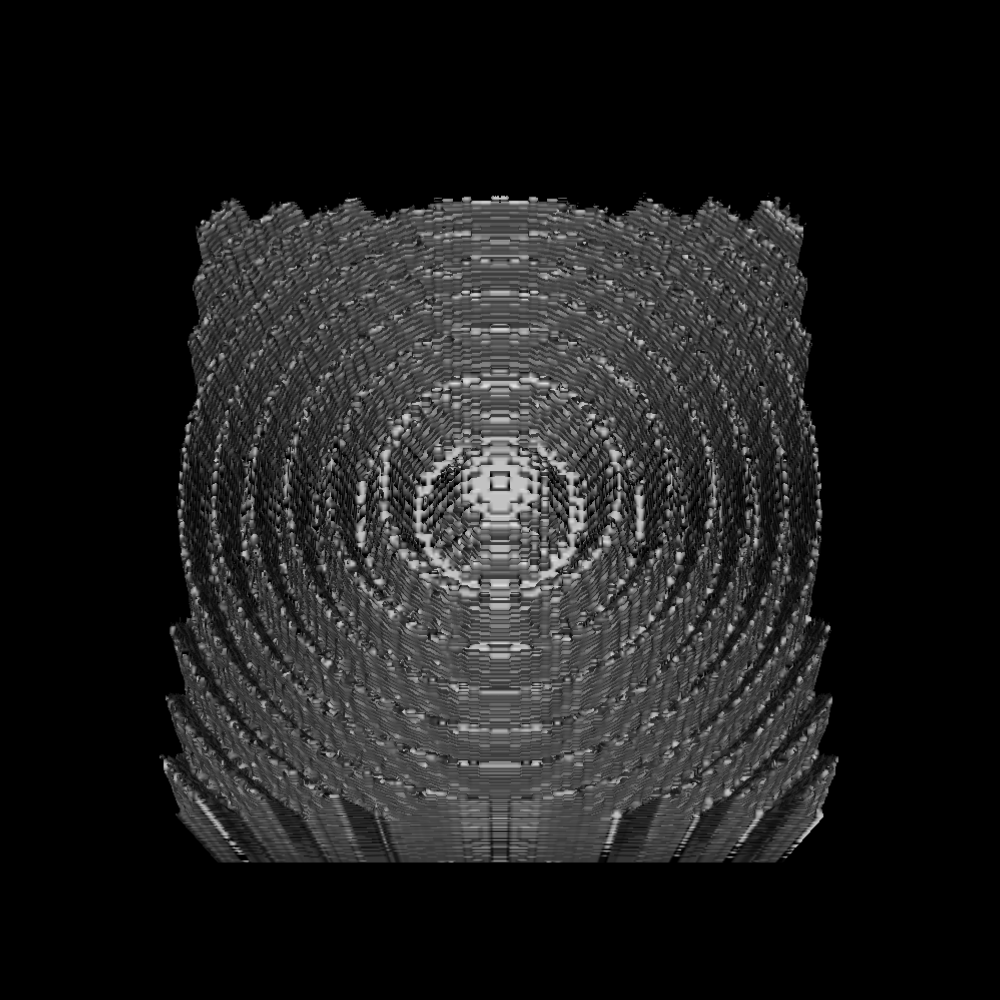}&
\includegraphics[width=\linewidth]{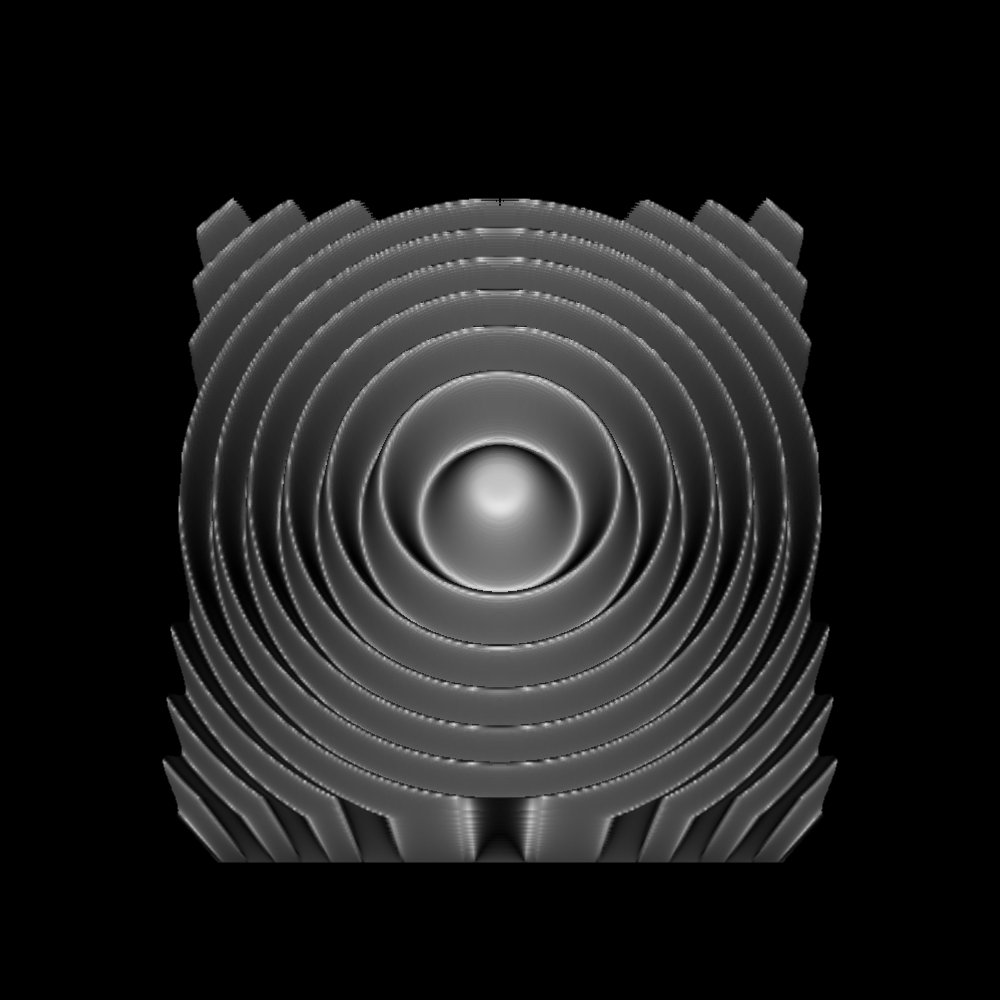}&
\includegraphics[width=\linewidth]{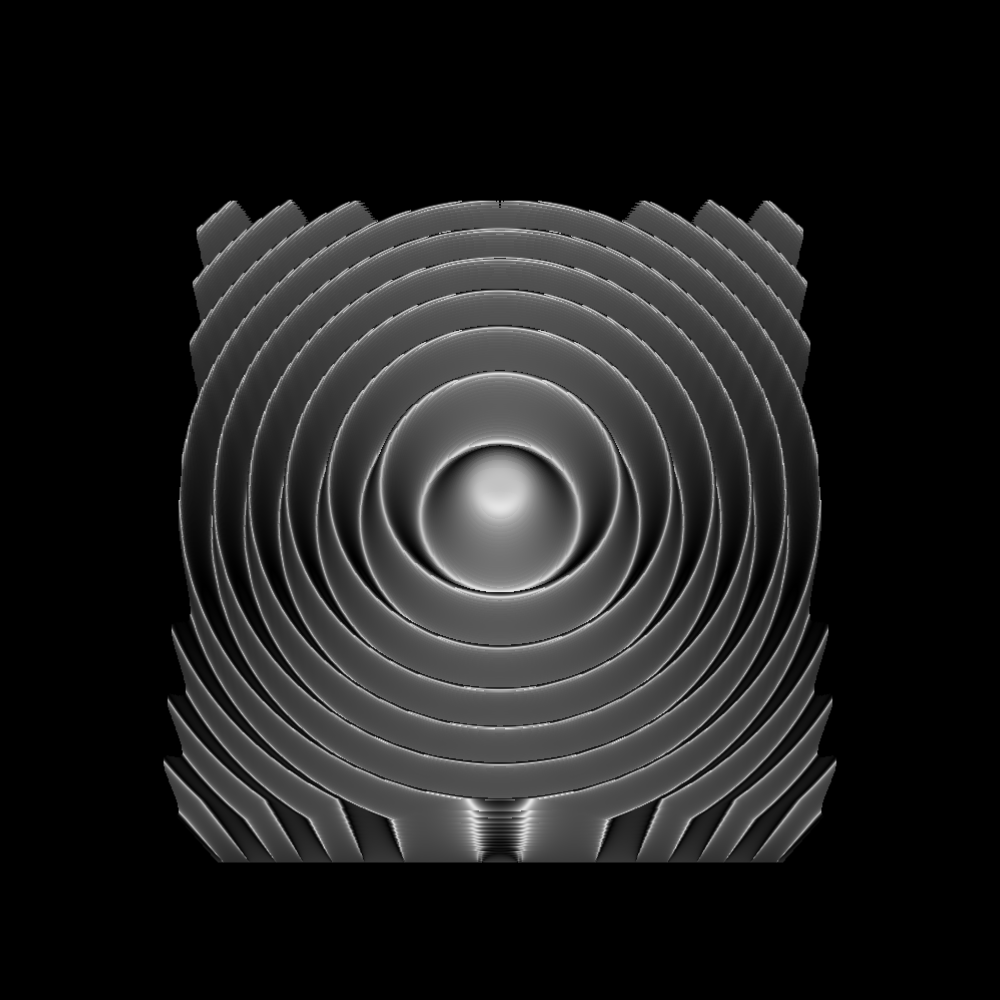}\\
 &PSNR: 20.66&PSNR: 24.397&PSNR: 20.674&PSNR: 24.406&PSNR: \textbf{24.580}\\
\rotatebox[origin=l]{90}{\thead{CR $\approx 134$}}&
\includegraphics[width=\linewidth]{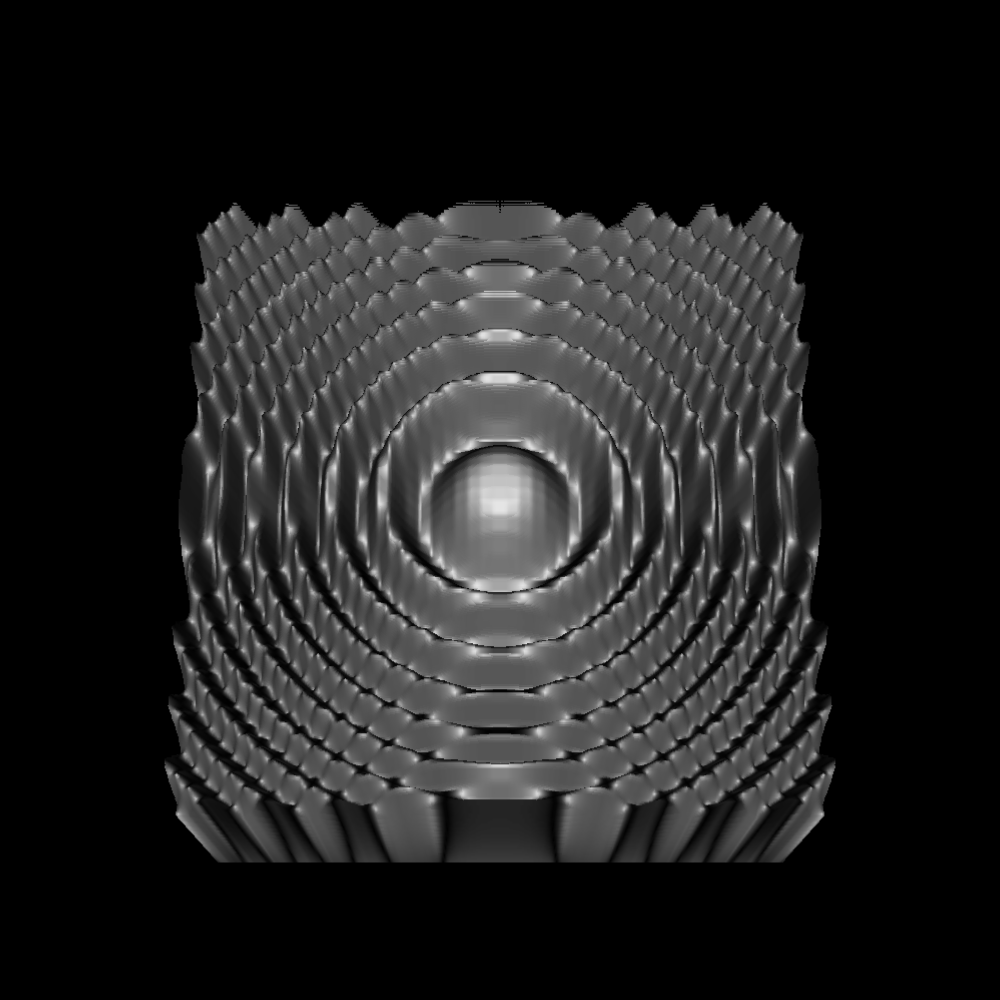}&
\includegraphics[width=\linewidth]{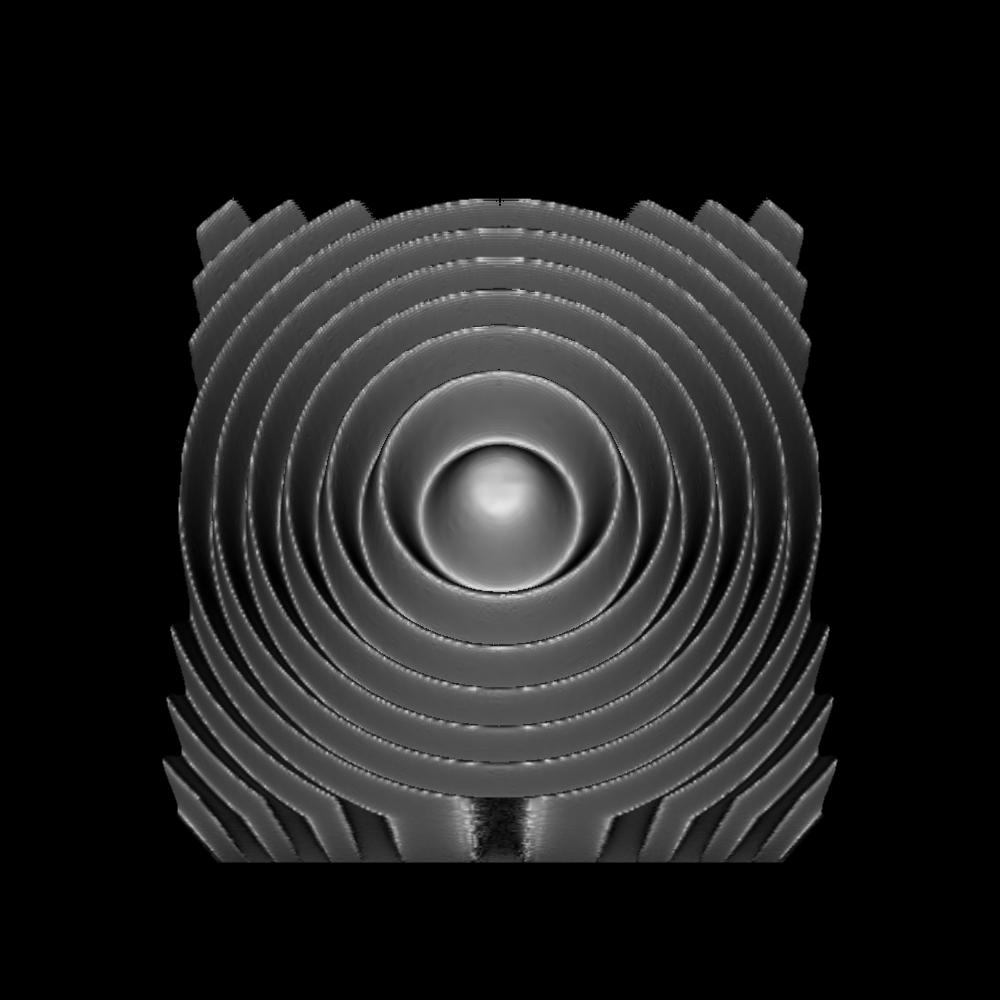}&
\includegraphics[width=\linewidth]{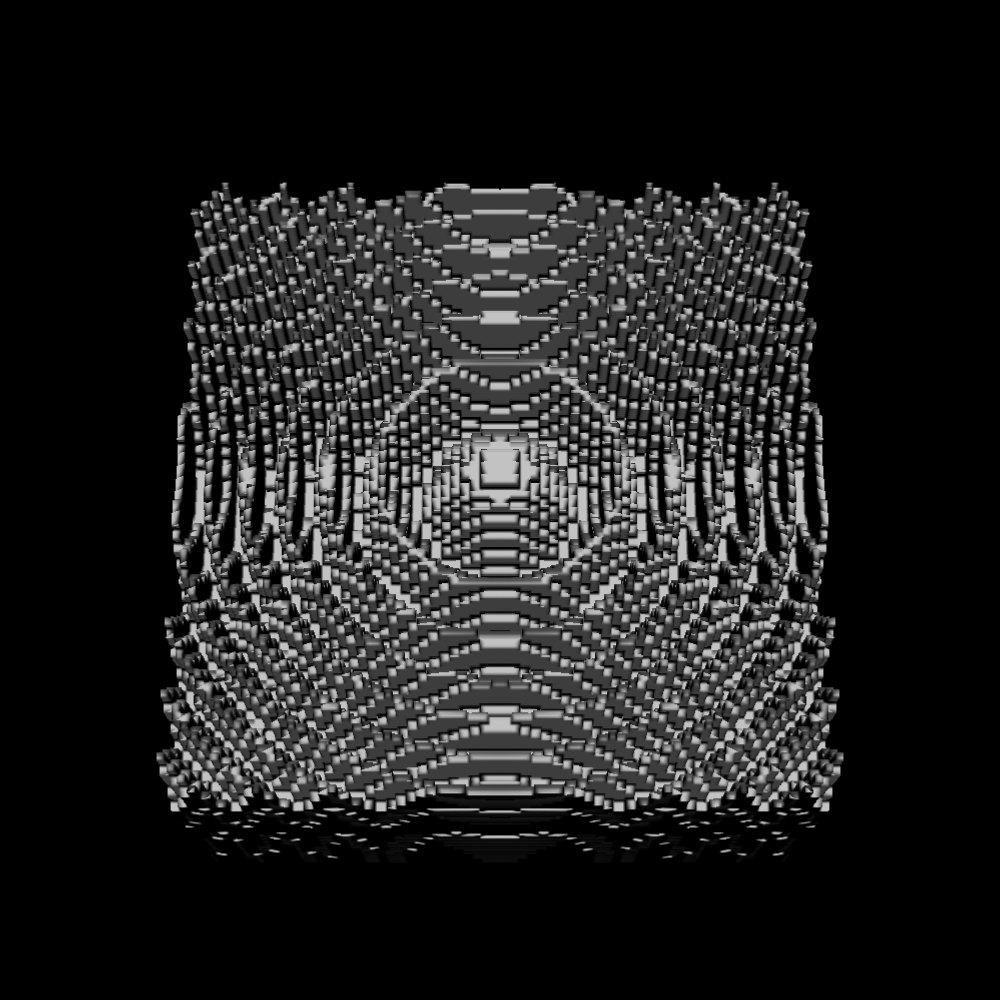}&
\includegraphics[width=\linewidth]{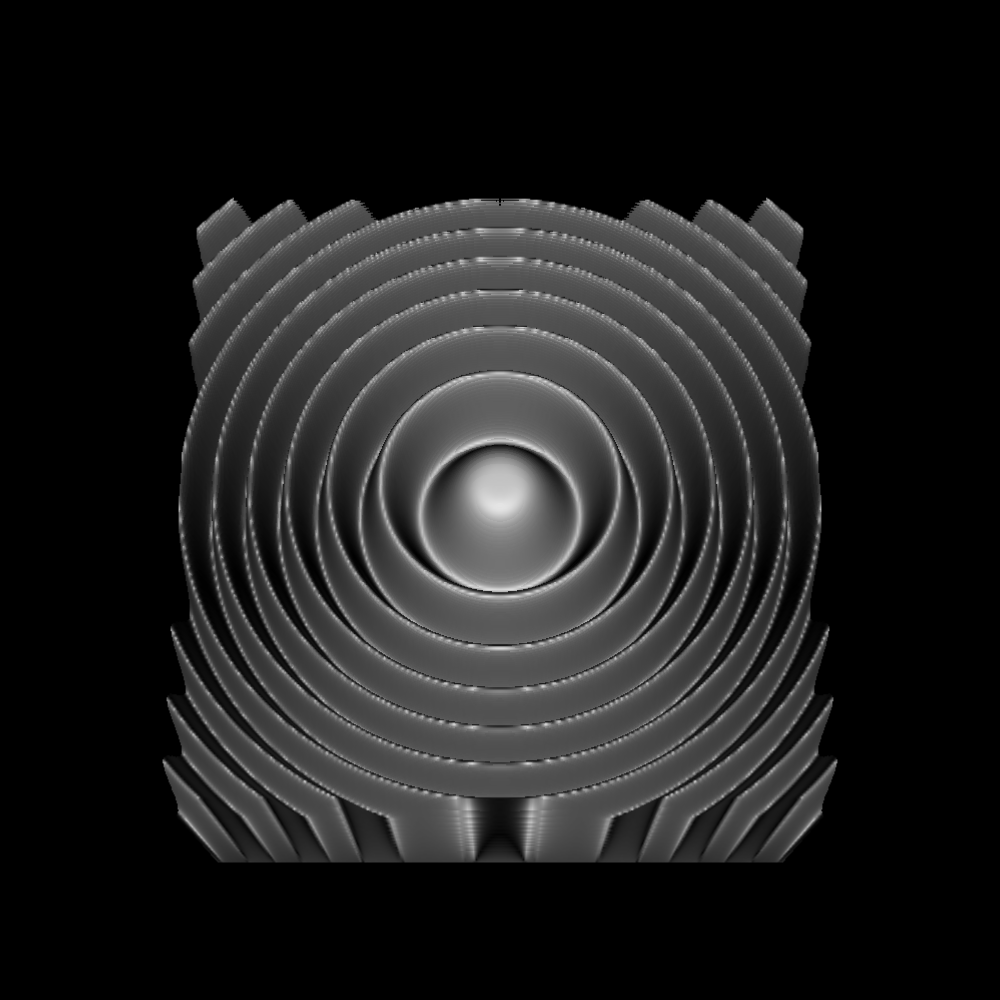}&
\includegraphics[width=\linewidth]{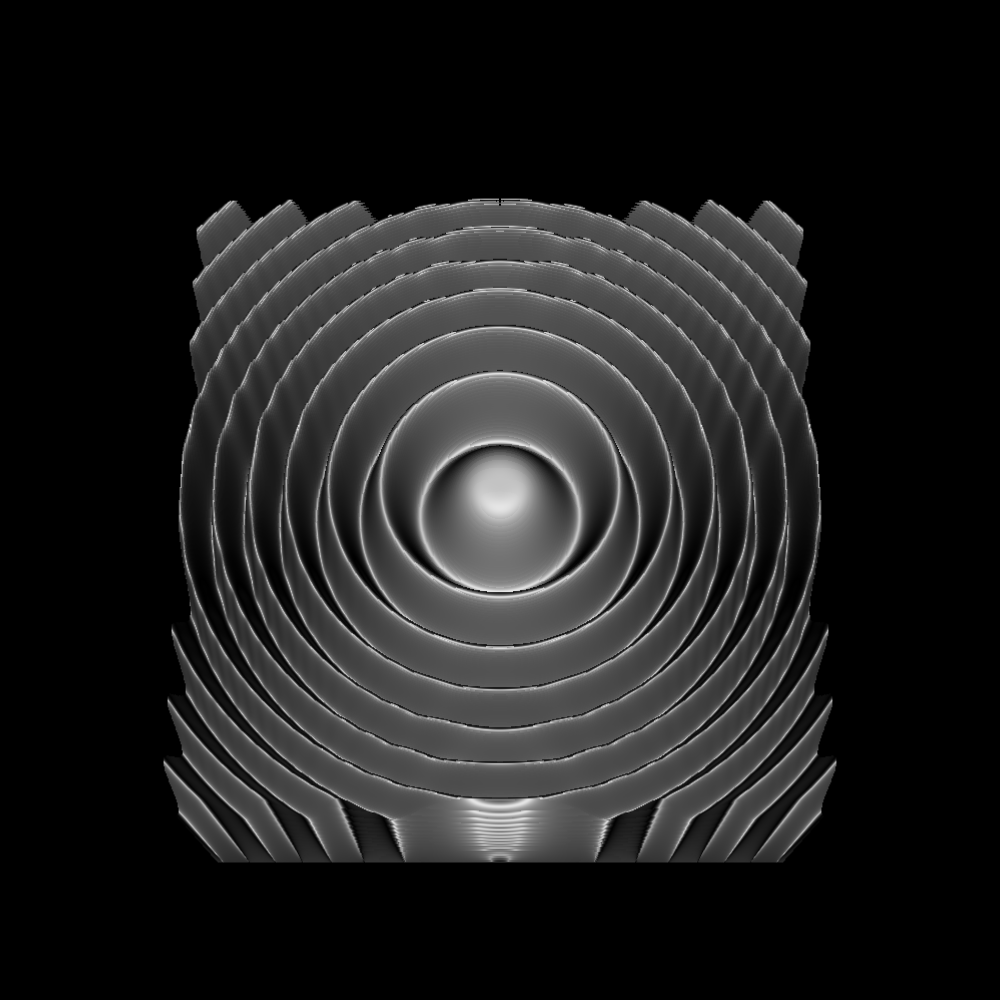}\\
 &PSNR: 19.800&PSNR: 24.319&PSNR: 15.116&PSNR: \textbf{24.406}&PSNR: 23.838\\
\rotatebox[origin=l]{90}{\thead{CR $\approx 212$}}&
\includegraphics[width=\linewidth]{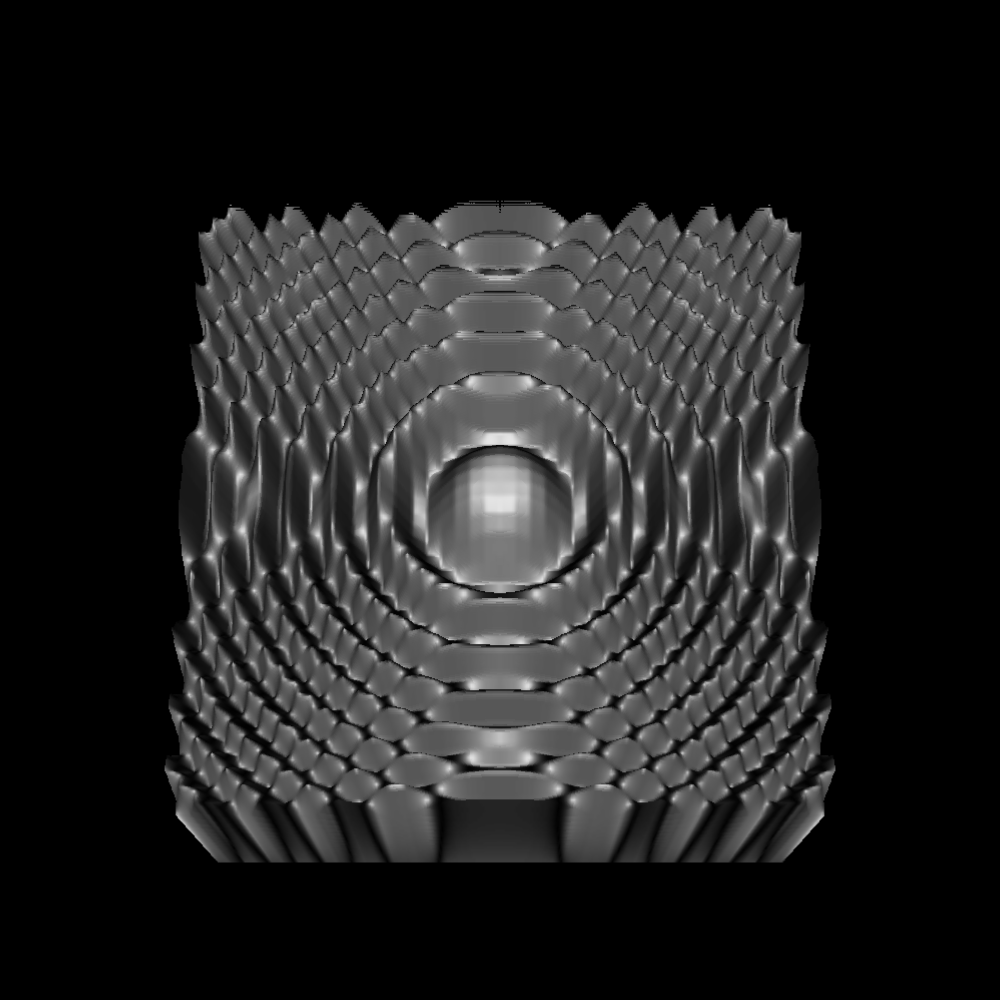}&
\includegraphics[width=\linewidth]{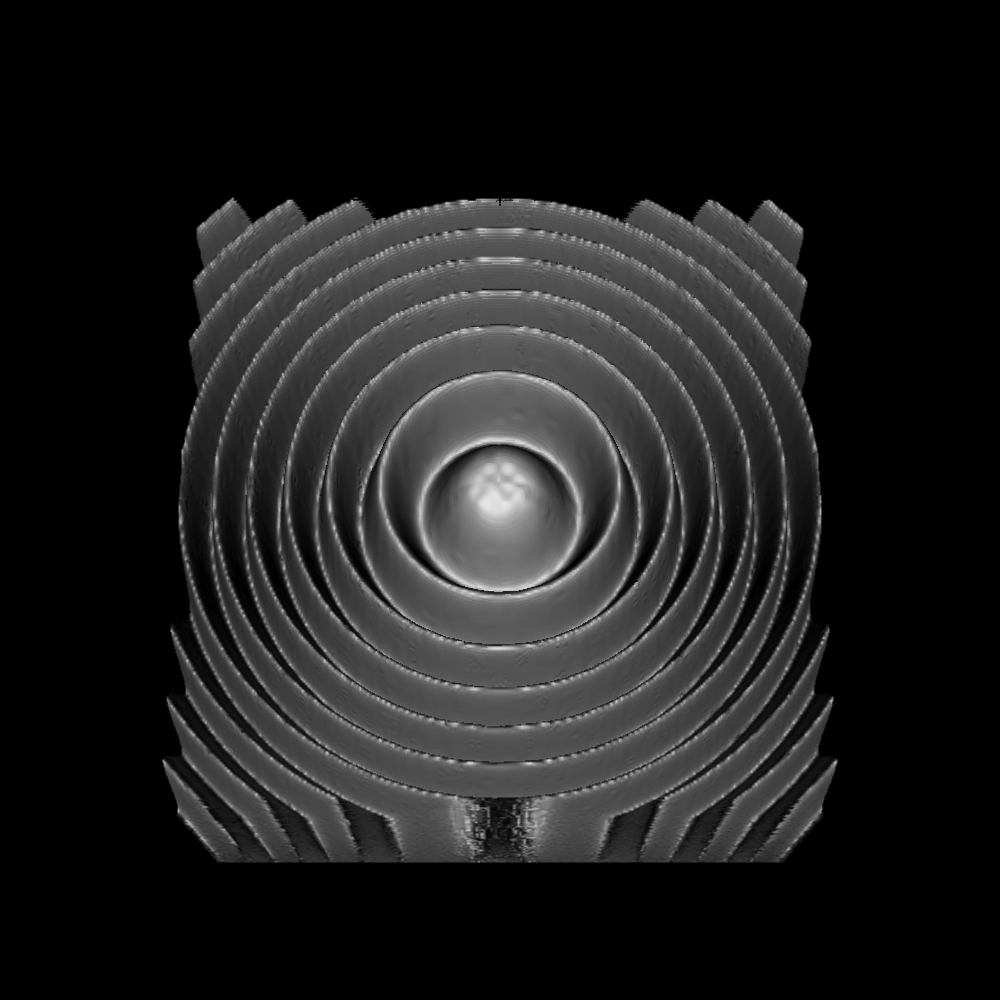}&
\includegraphics[width=\linewidth]{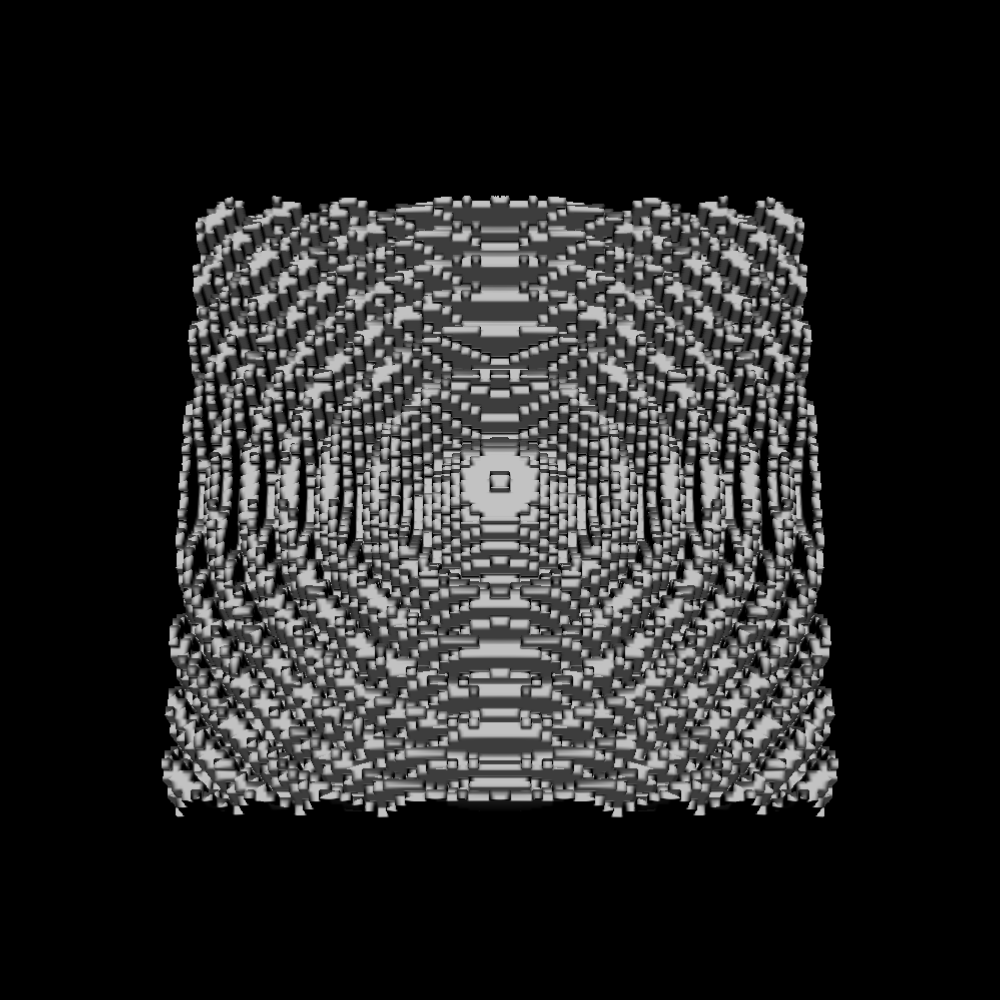}&
\includegraphics[width=\linewidth]{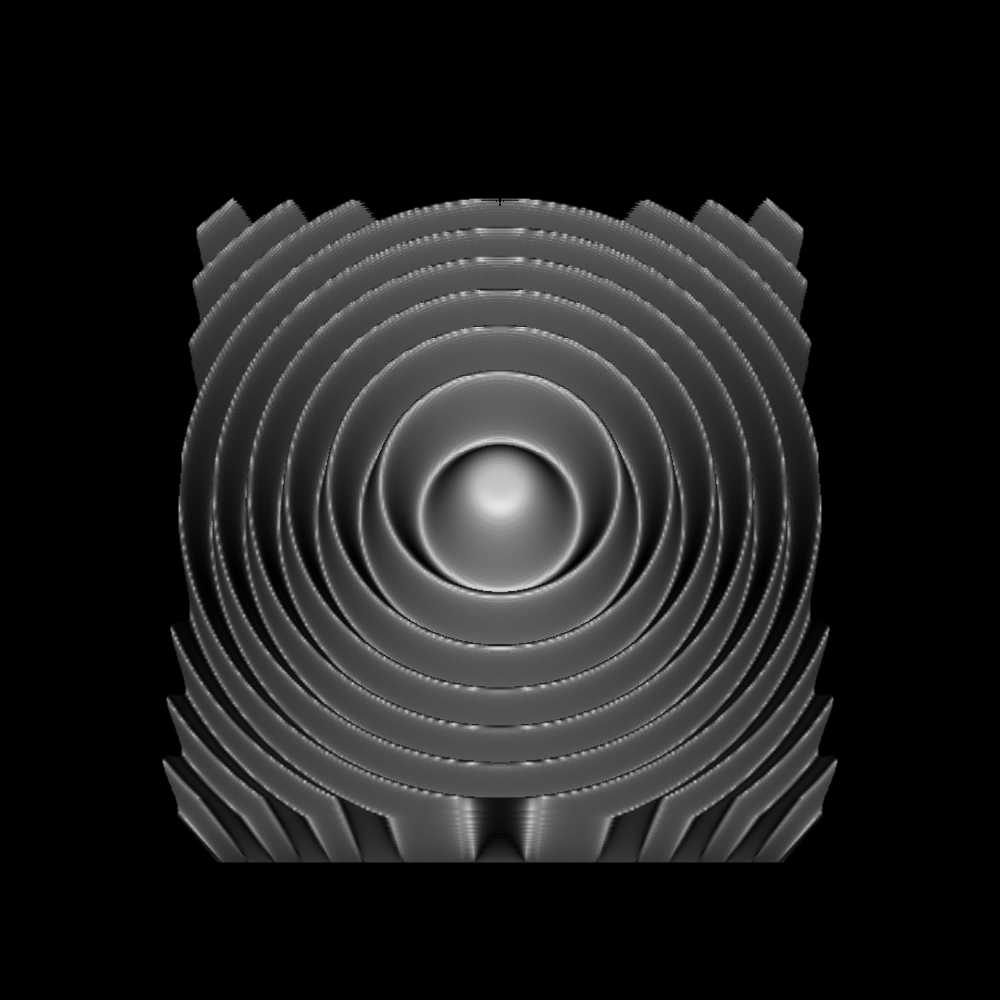}&
\includegraphics[width=\linewidth]{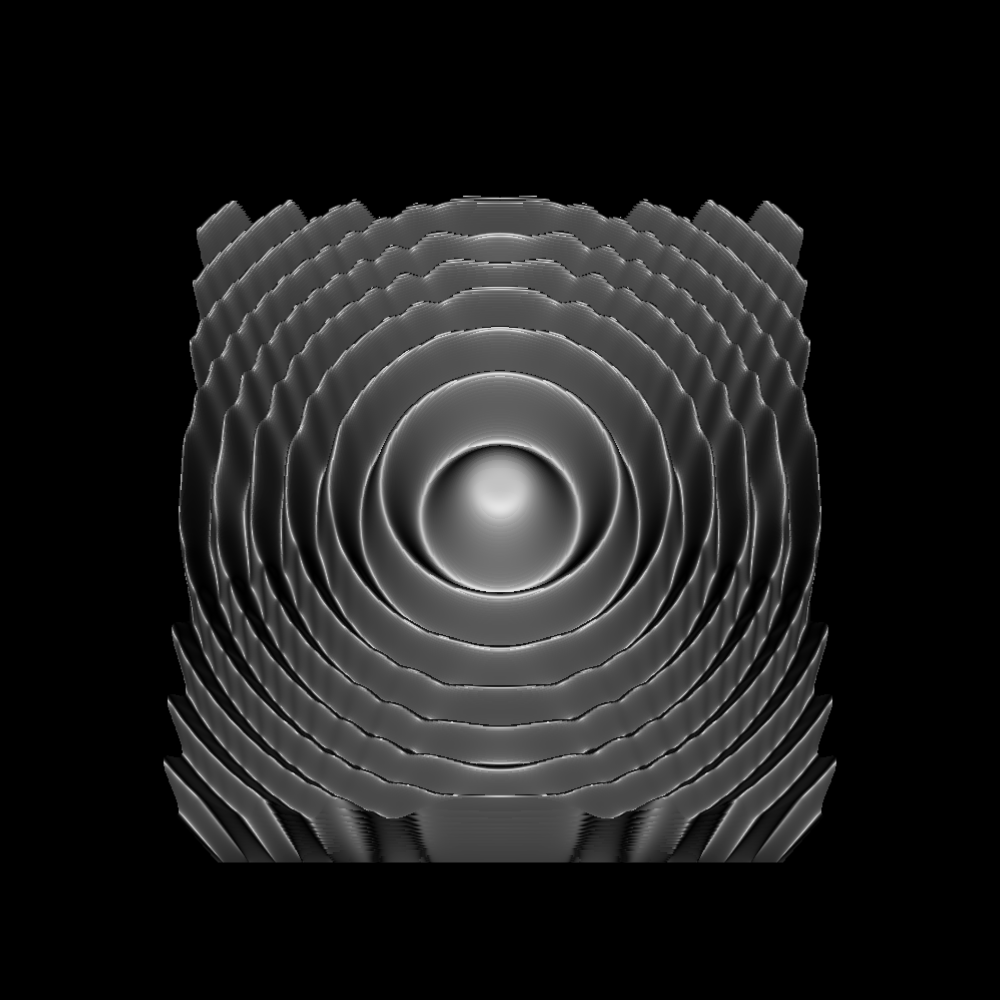}\\
 &PSNR: 19.063&PSNR: 24.270&PSNR: 14.287&PSNR: \textbf{24.406}&PSNR: 23.090
\end{tabular}
    \caption{Volume rendering results under different compression ratio (CR) using proposed MFA-DVV with compressed MFA model and other volume compression algorithms.}
    \label{fig:imageWithCompression} 
\end{figure}

\begin{figure*}[t]
  \centering
  \tiny
\begin{tabular}{M{0.01\textwidth}M{0.14\textwidth}@{}M{0.14\textwidth}@{}M{0.14\textwidth}@{}M{0.14\textwidth}@{}M{0.14\textwidth}@{}M{0.14\textwidth}@{}M{0.14\textwidth}}
\rotatebox[origin=c]{90}{Nucleon}&
\includegraphics[width=\linewidth]{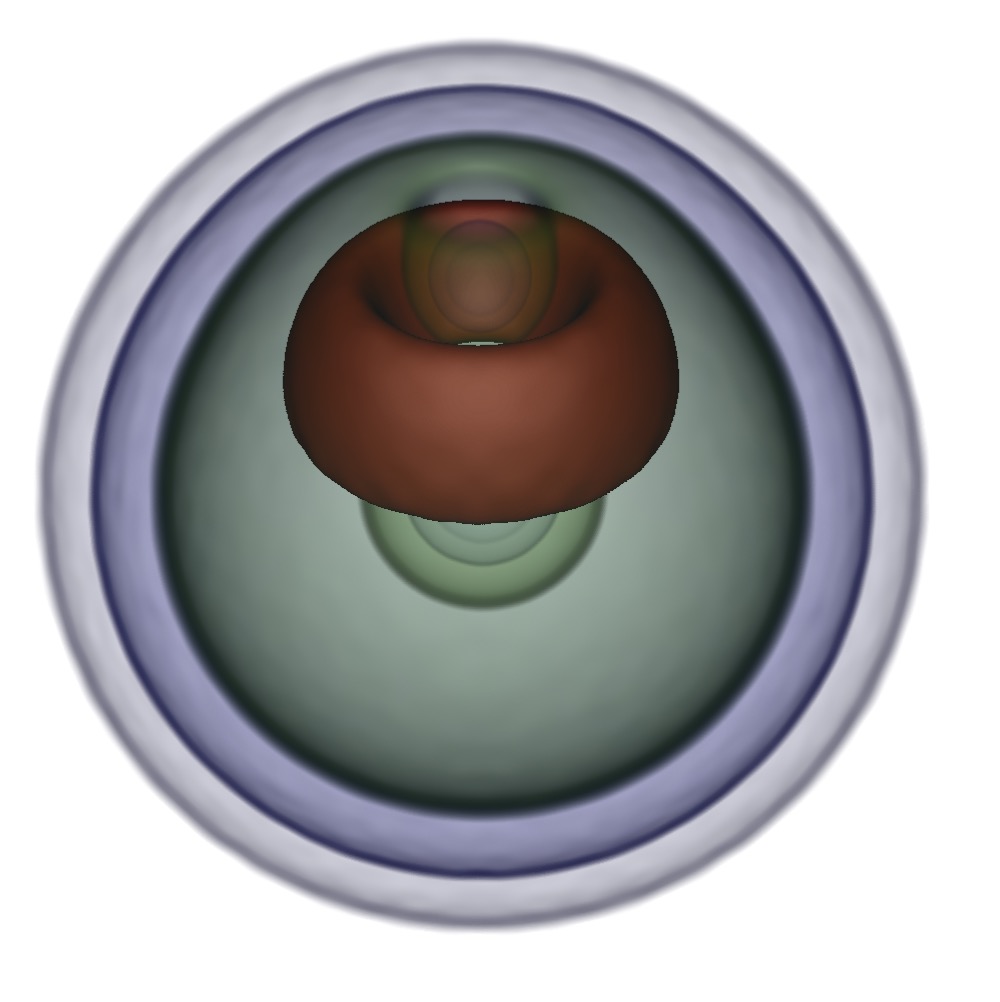}&
\includegraphics[width=\linewidth]{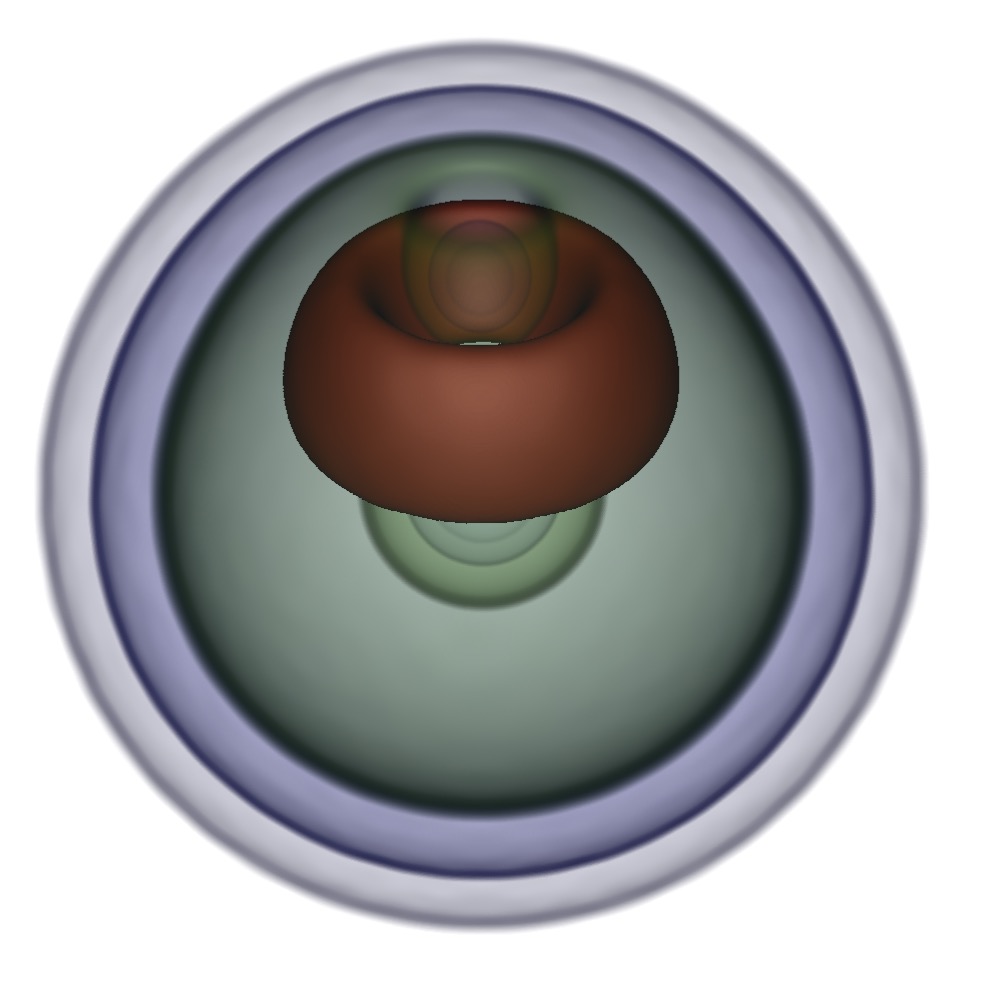}&
\includegraphics[width=\linewidth]{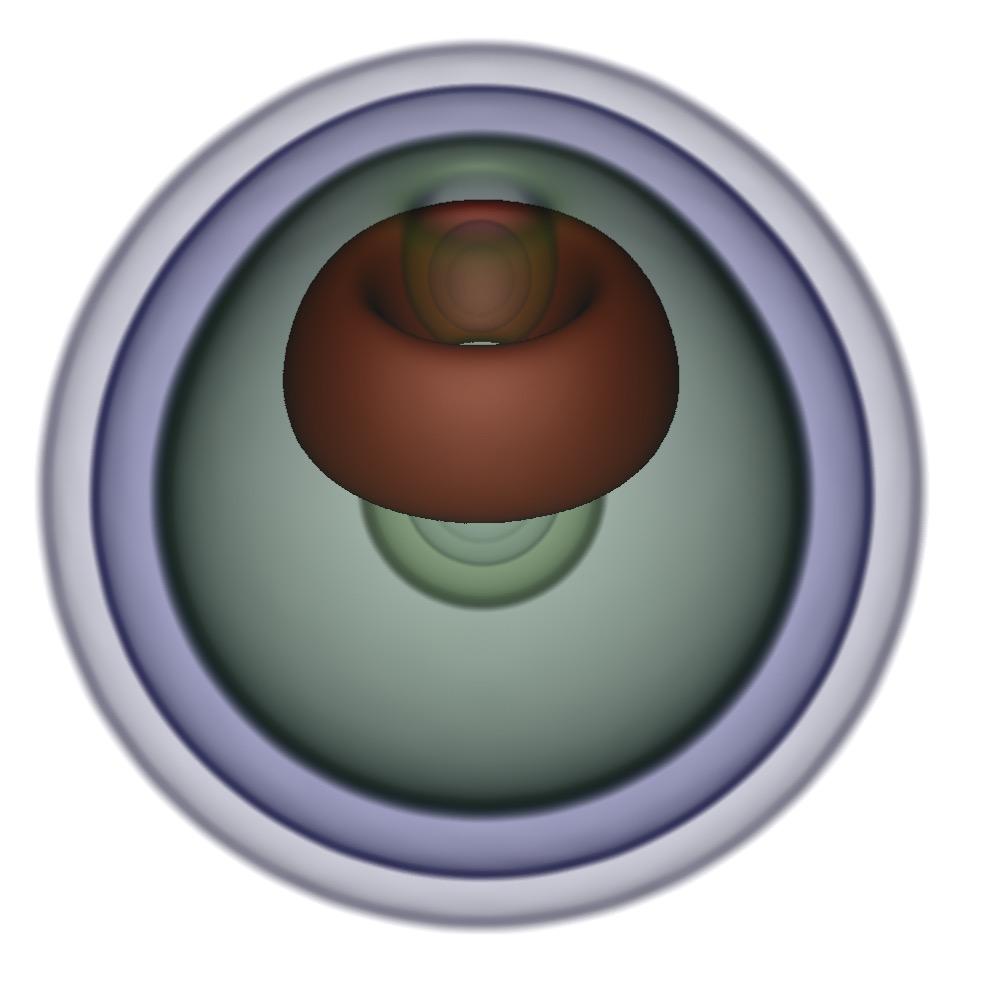}&
\includegraphics[width=\linewidth]{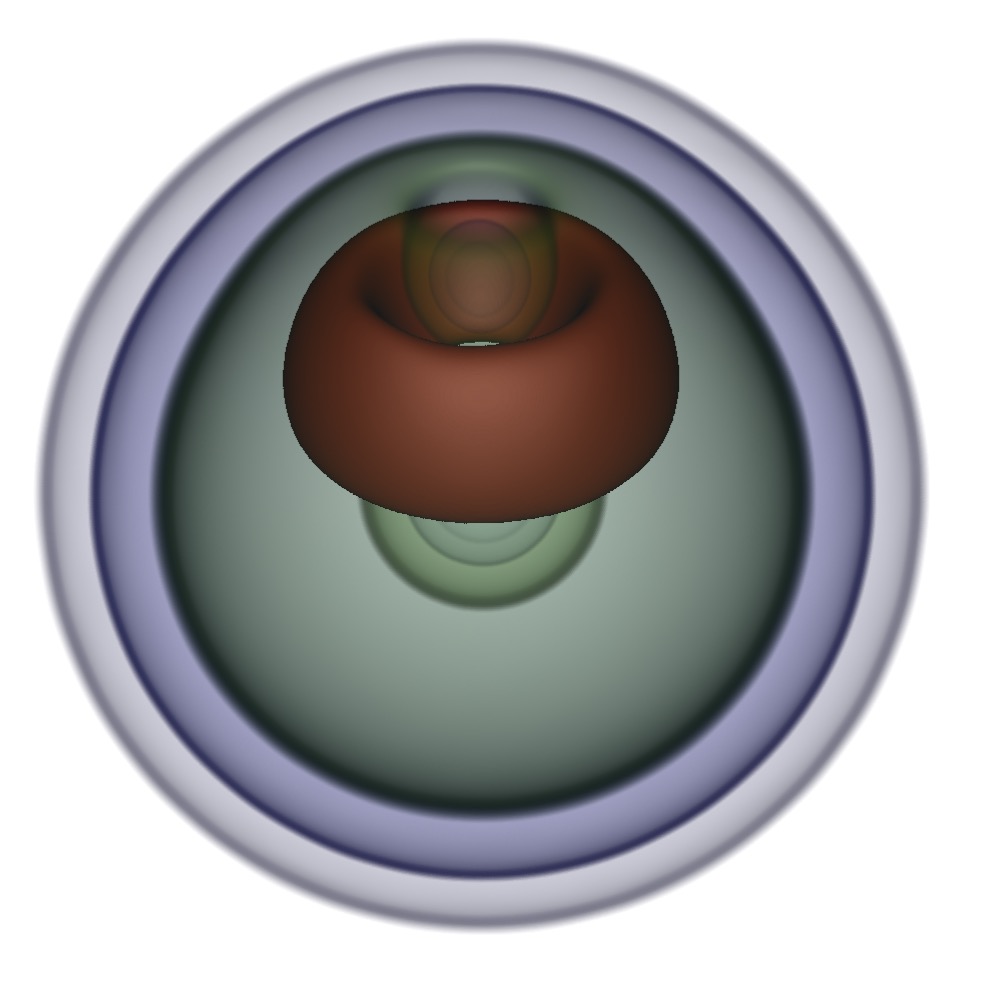}&
\includegraphics[width=\linewidth]{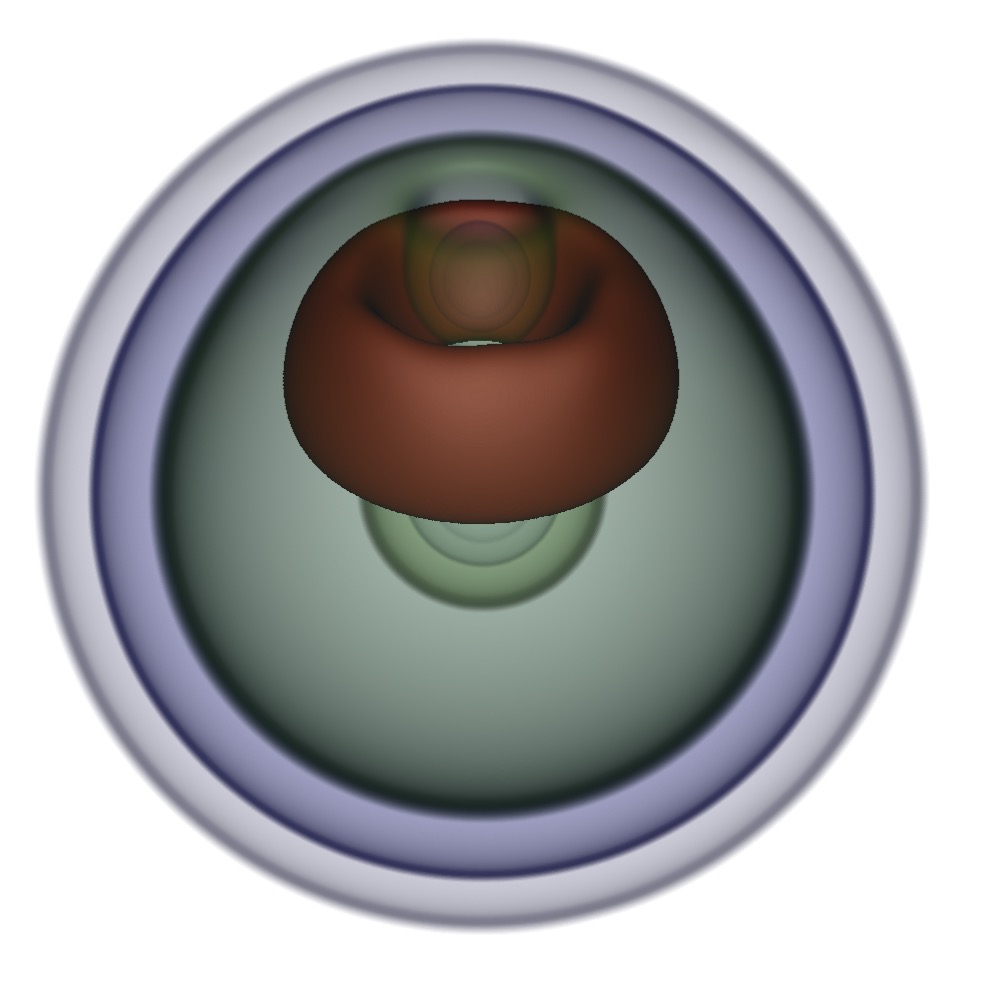}&
\includegraphics[width=\linewidth]{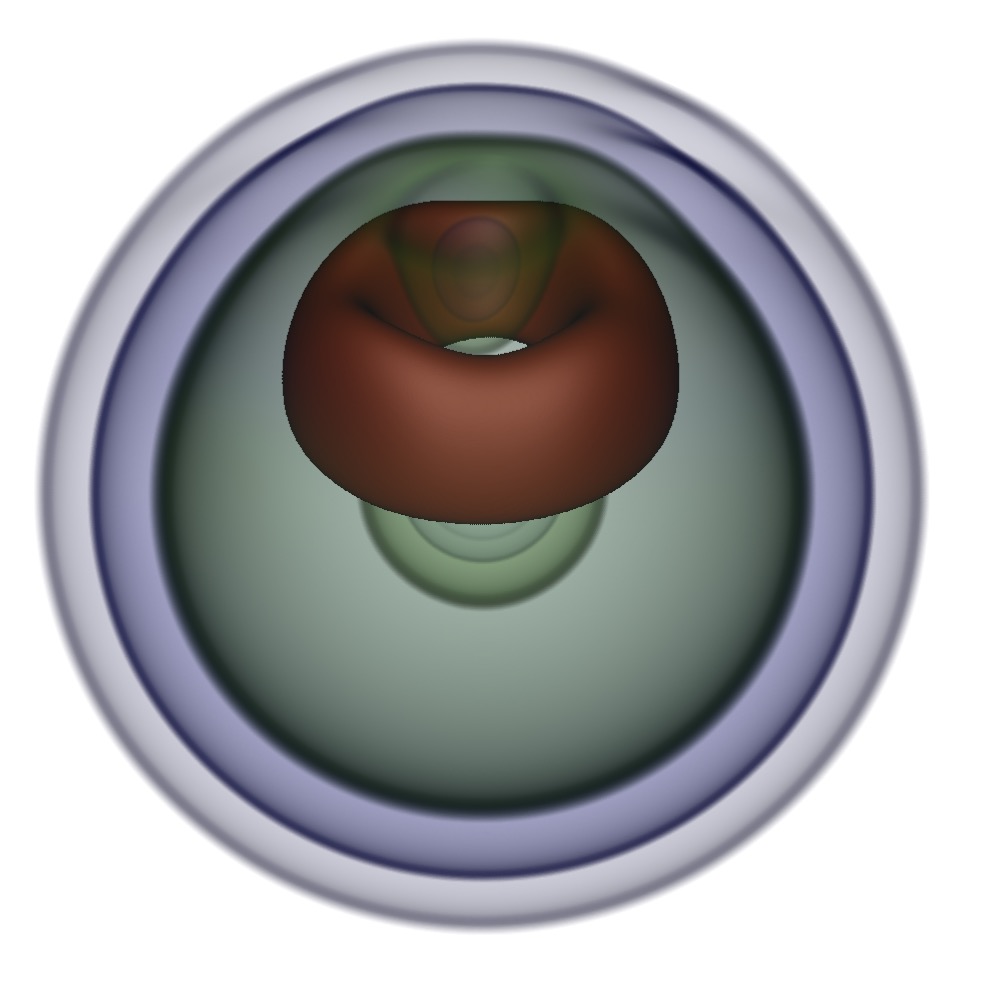}&
\includegraphics[width=\linewidth]{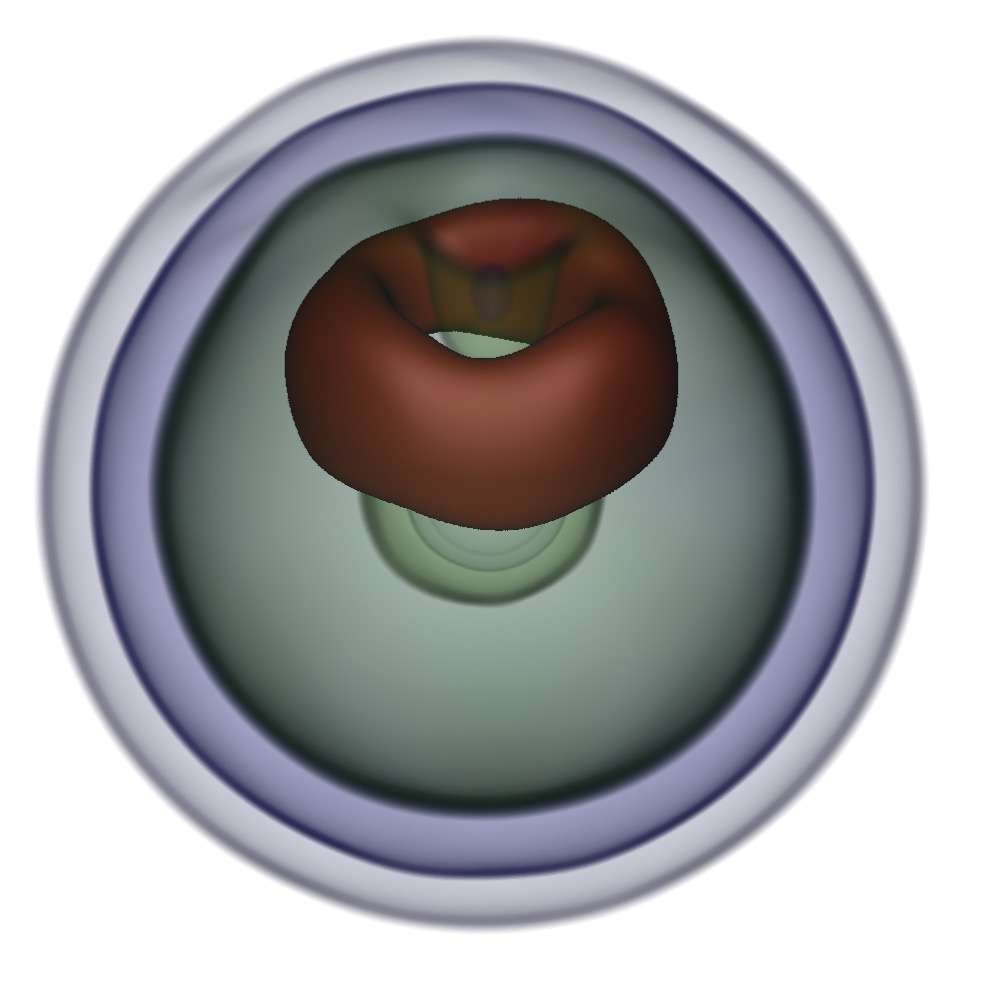}\\
 &CR = 1.0&CR = 1.477&CR = 2.313&CR = 3.921&CR = 7.442&CR = 16.826&CR = 51.781\\
\rotatebox[origin=c]{90}{Fuel}&
\includegraphics[width=\linewidth]{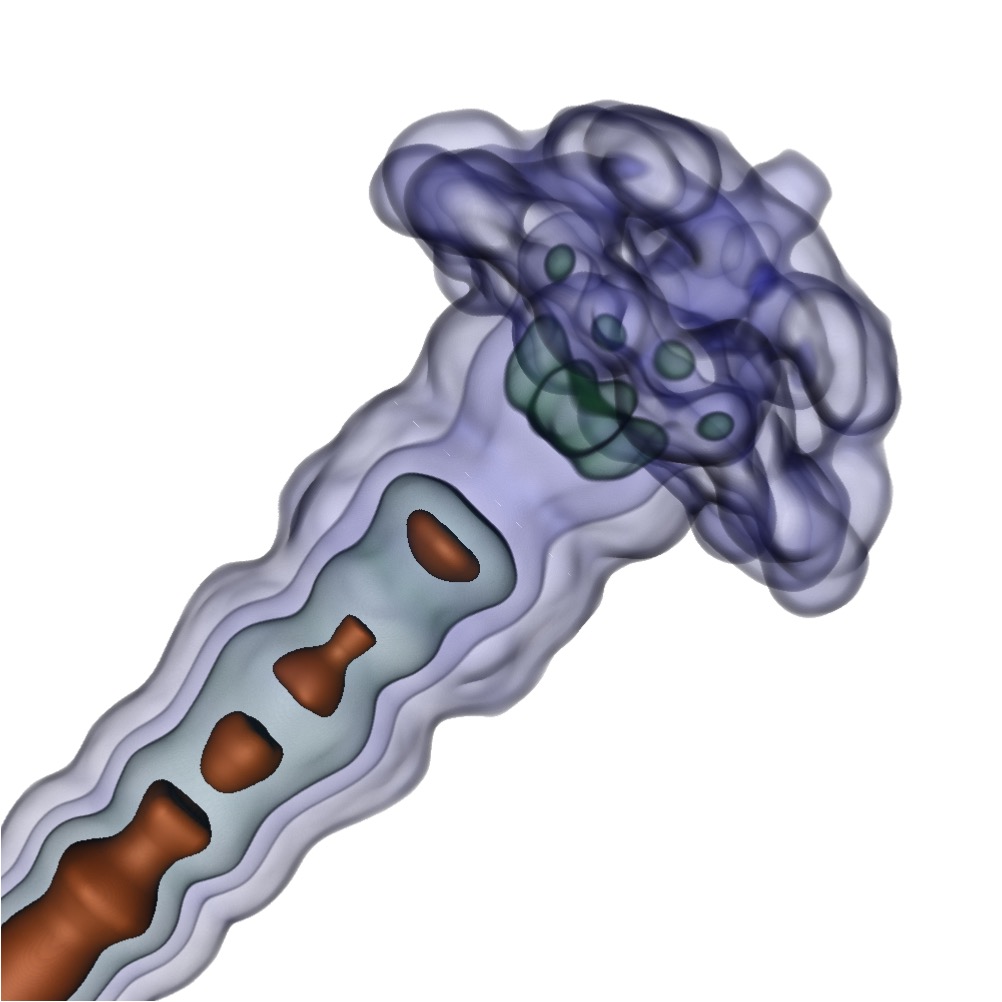}&
\includegraphics[width=\linewidth]{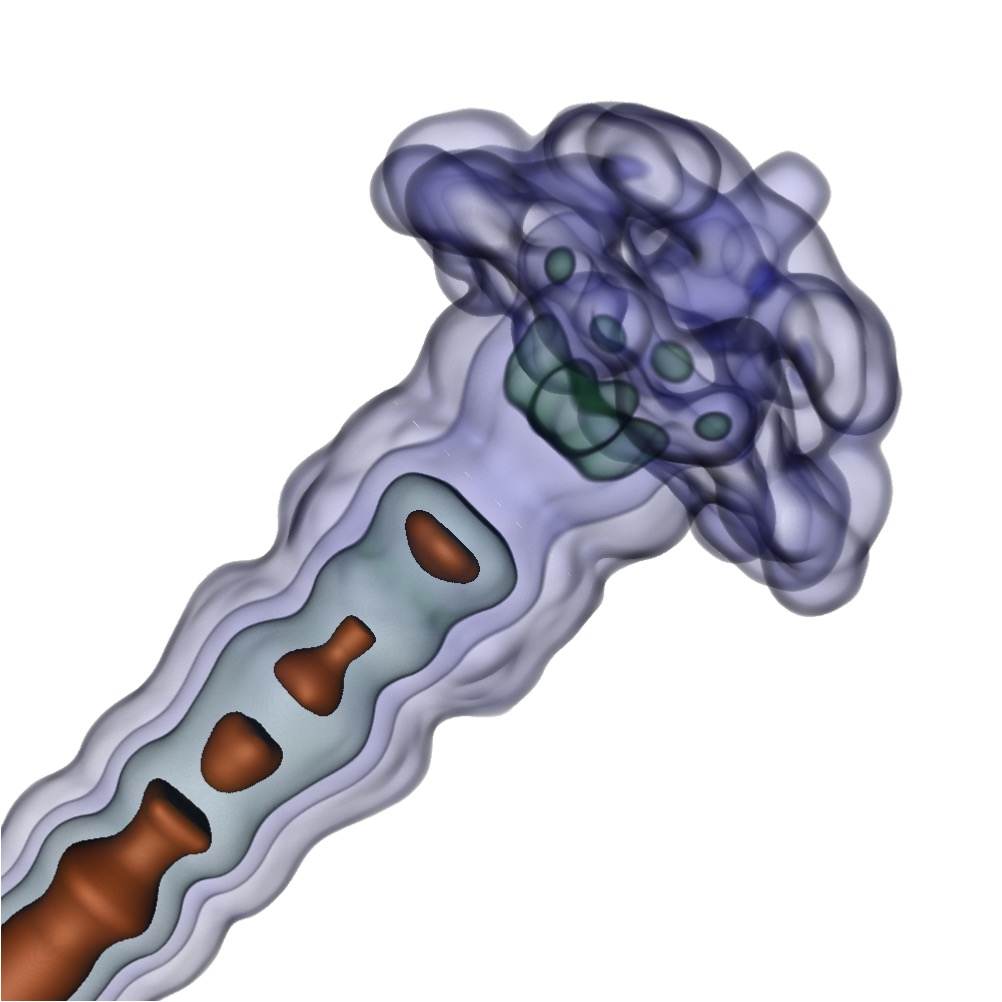}&
\includegraphics[width=\linewidth]{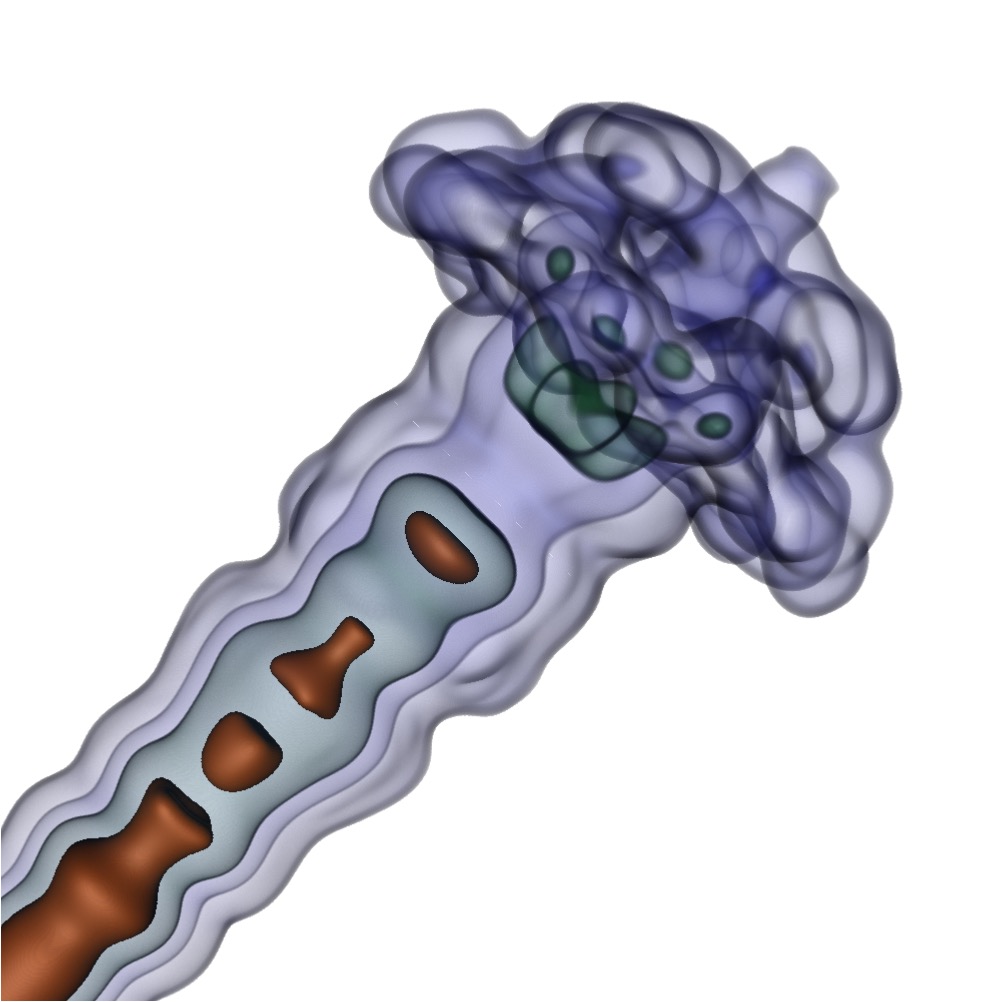}&
\includegraphics[width=\linewidth]{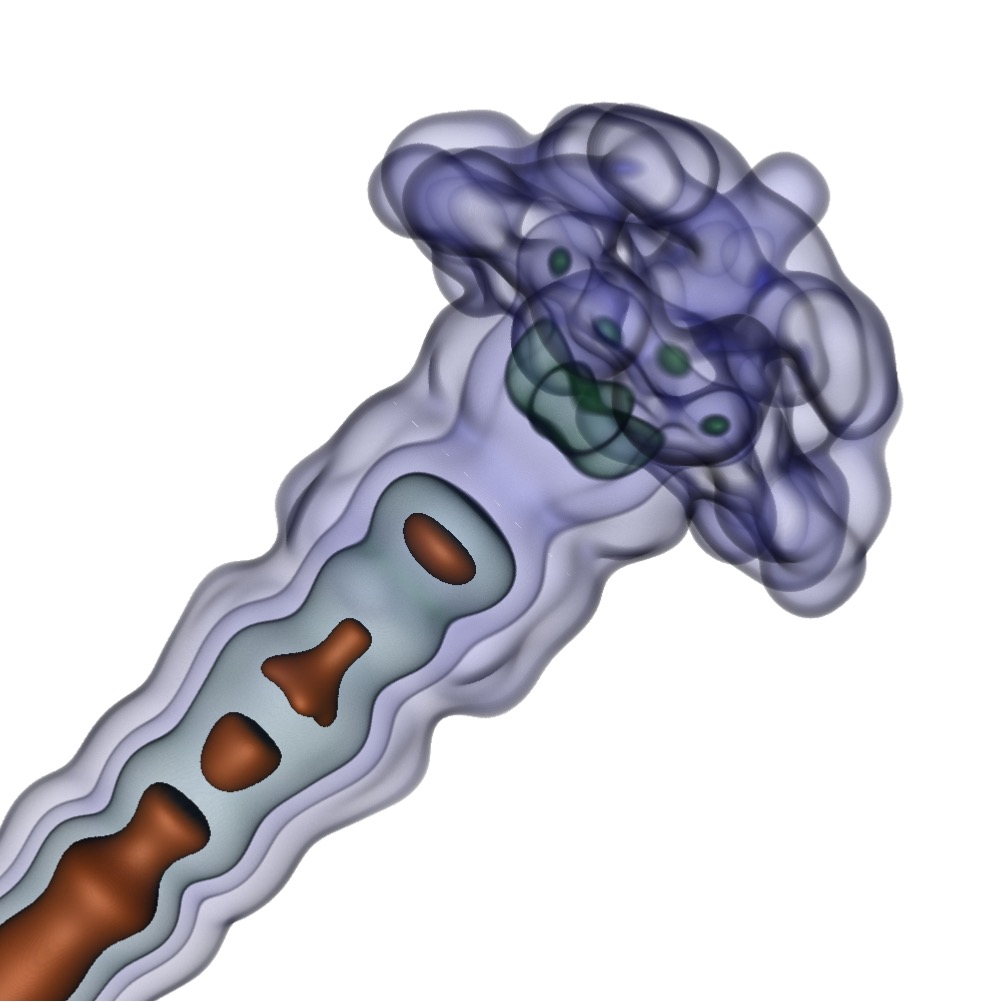}&
\includegraphics[width=\linewidth]{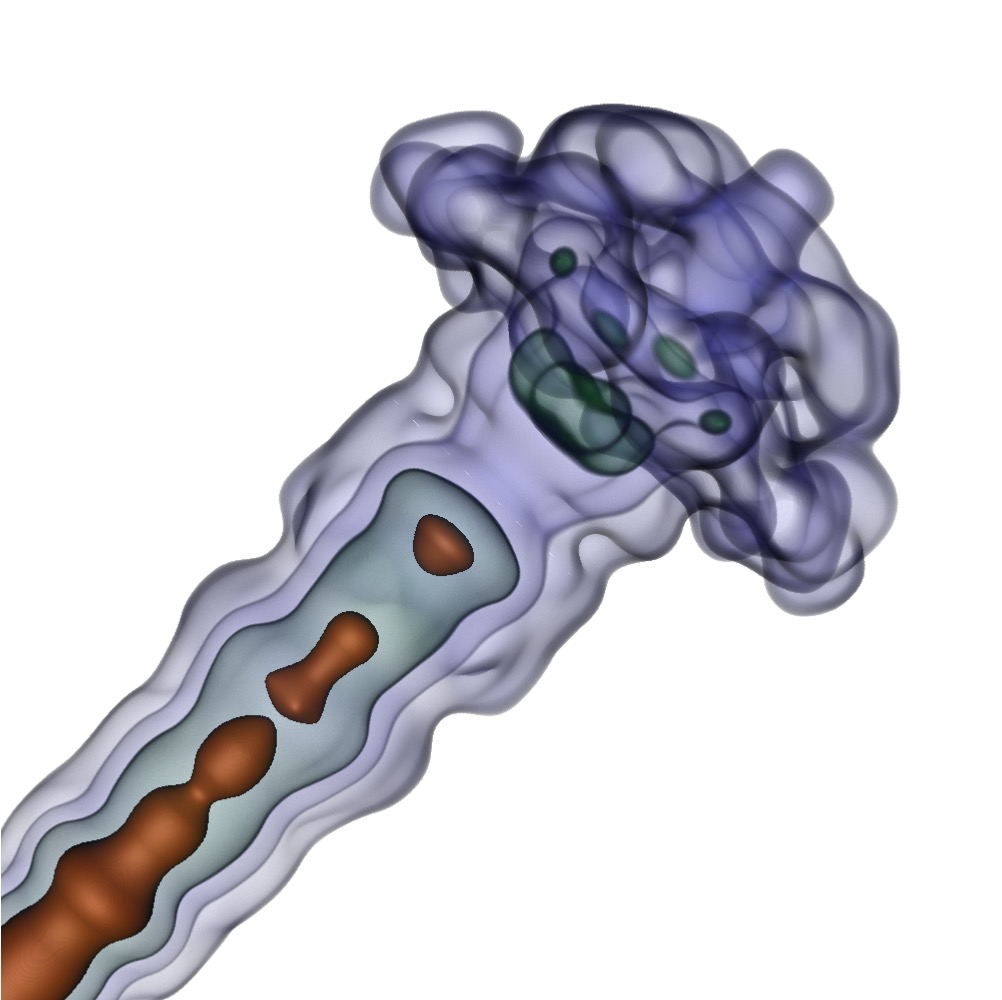}&
\includegraphics[width=\linewidth]{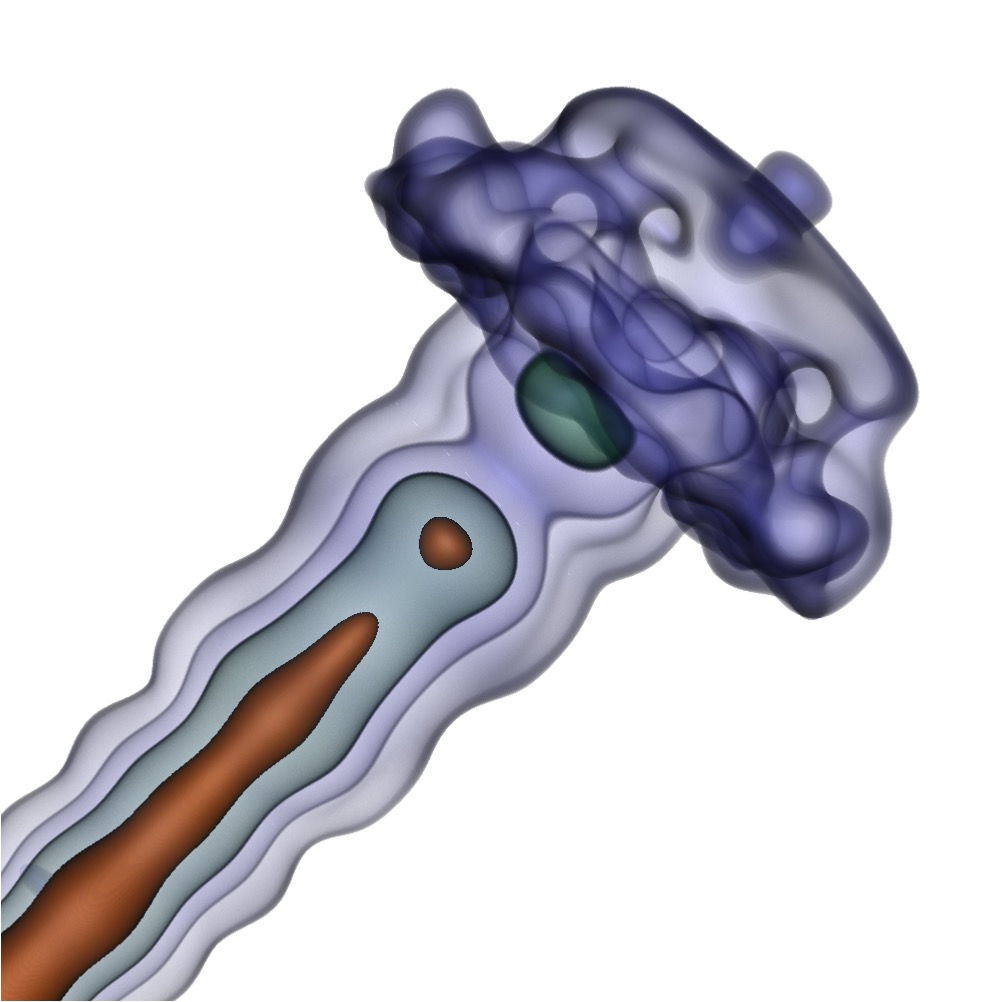}&
\includegraphics[width=\linewidth]{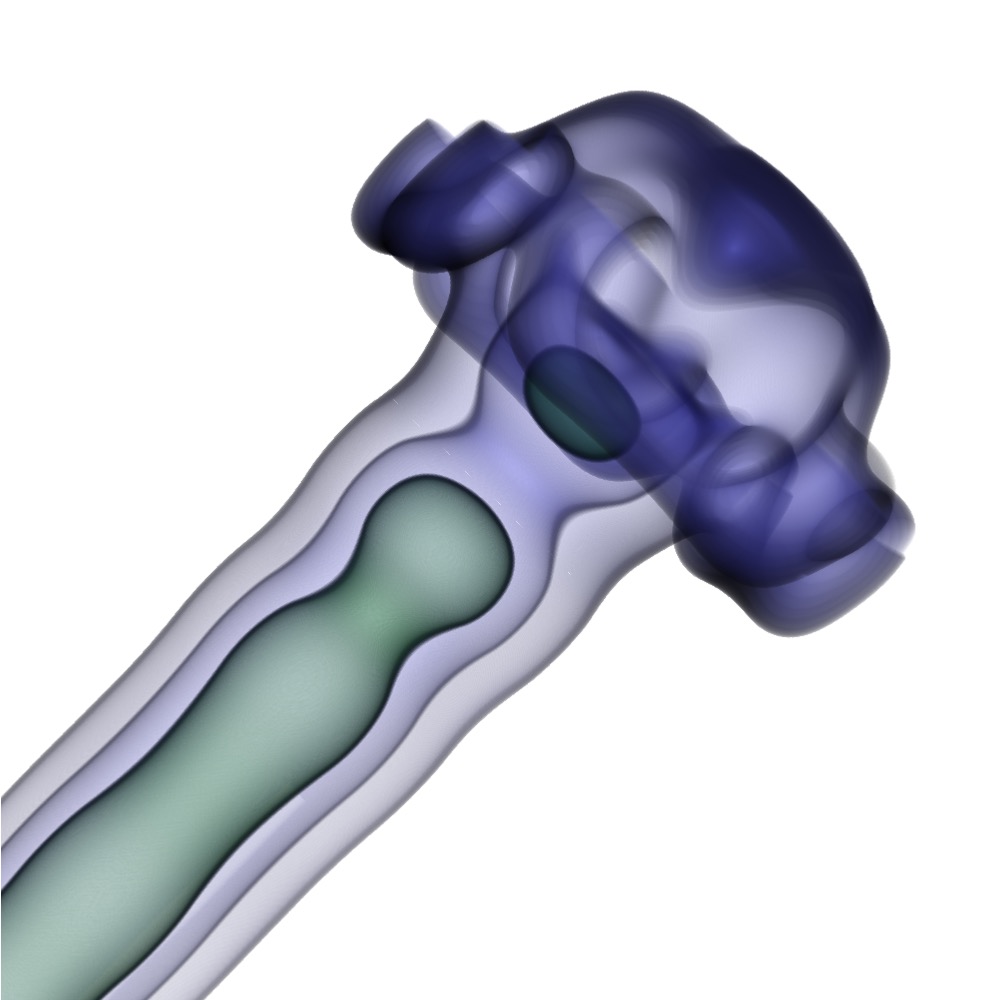}\\
 &CR = 1.0&CR = 1.492&CR = 2.370&CR = 4.096&CR = 8.0&CR = 18.963&CR = 64.0\\
\rotatebox[origin=c]{90}{Aneurism}&
\includegraphics[width=\linewidth]{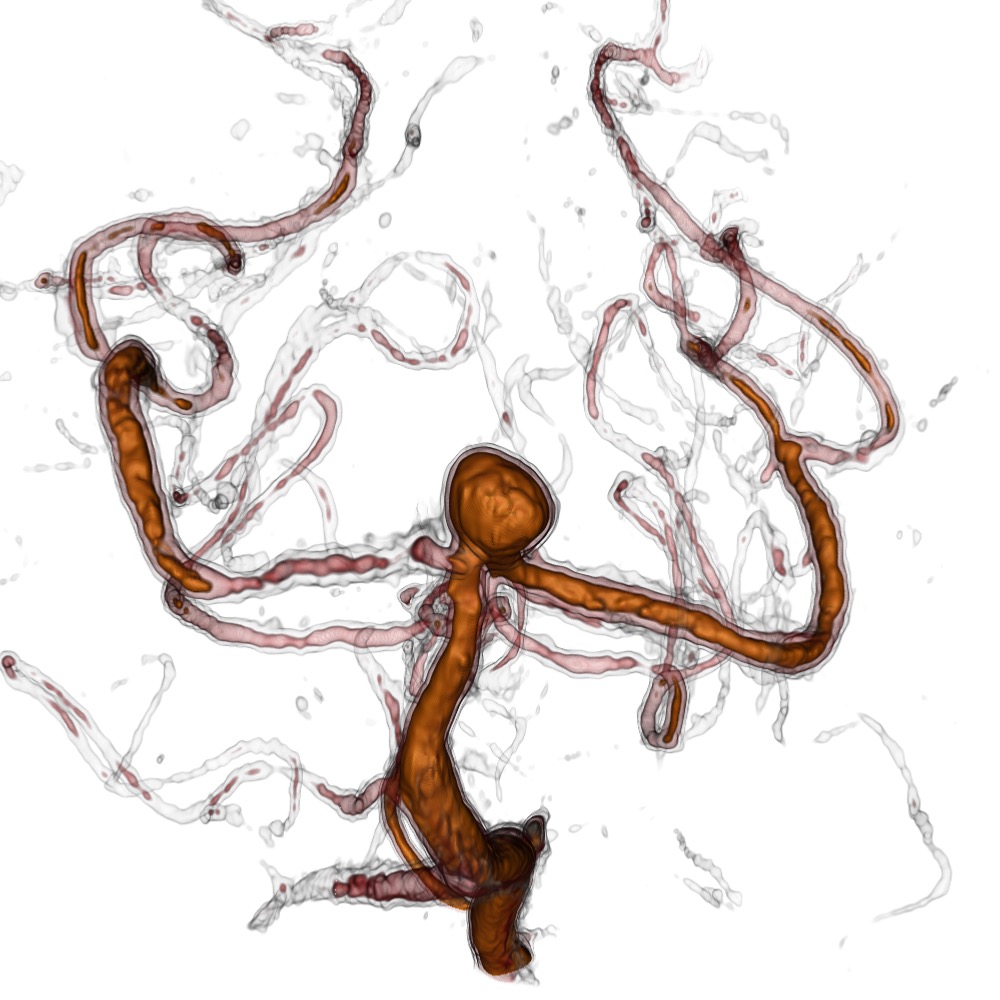}&
\includegraphics[width=\linewidth]{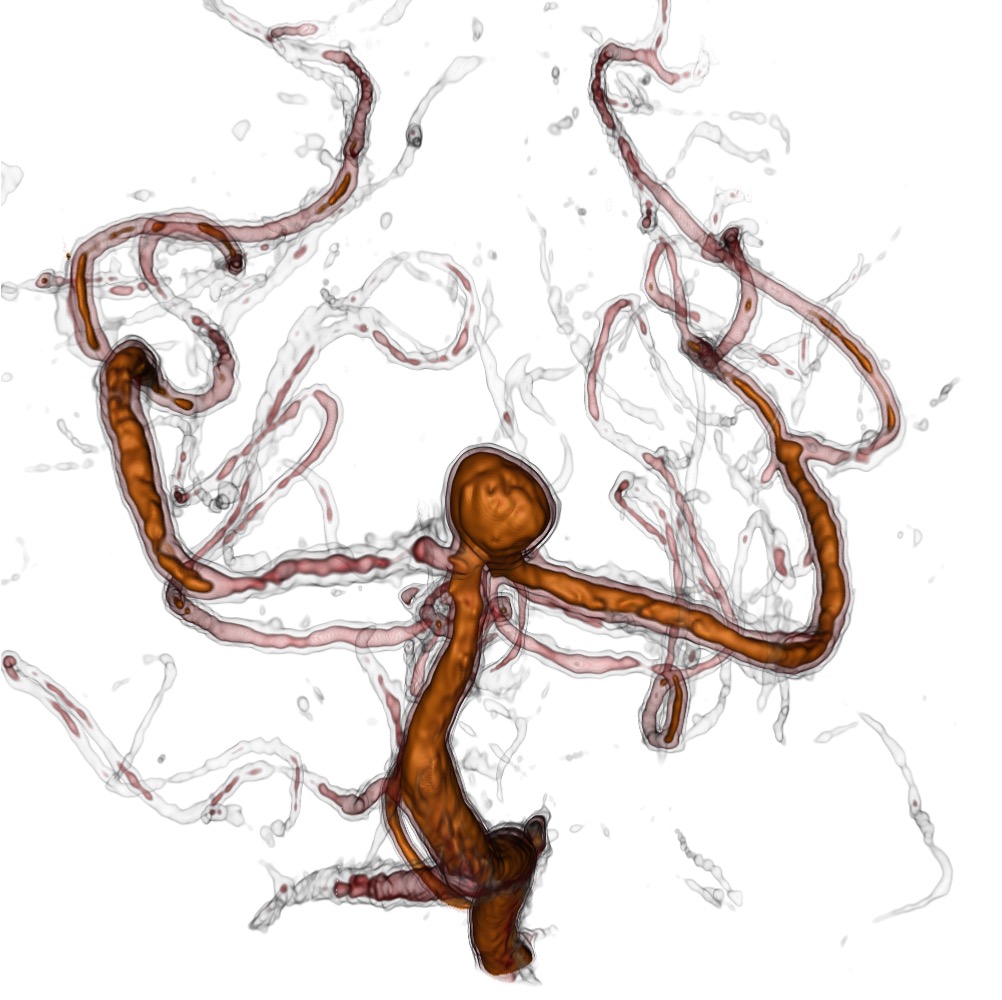}&
\includegraphics[width=\linewidth]{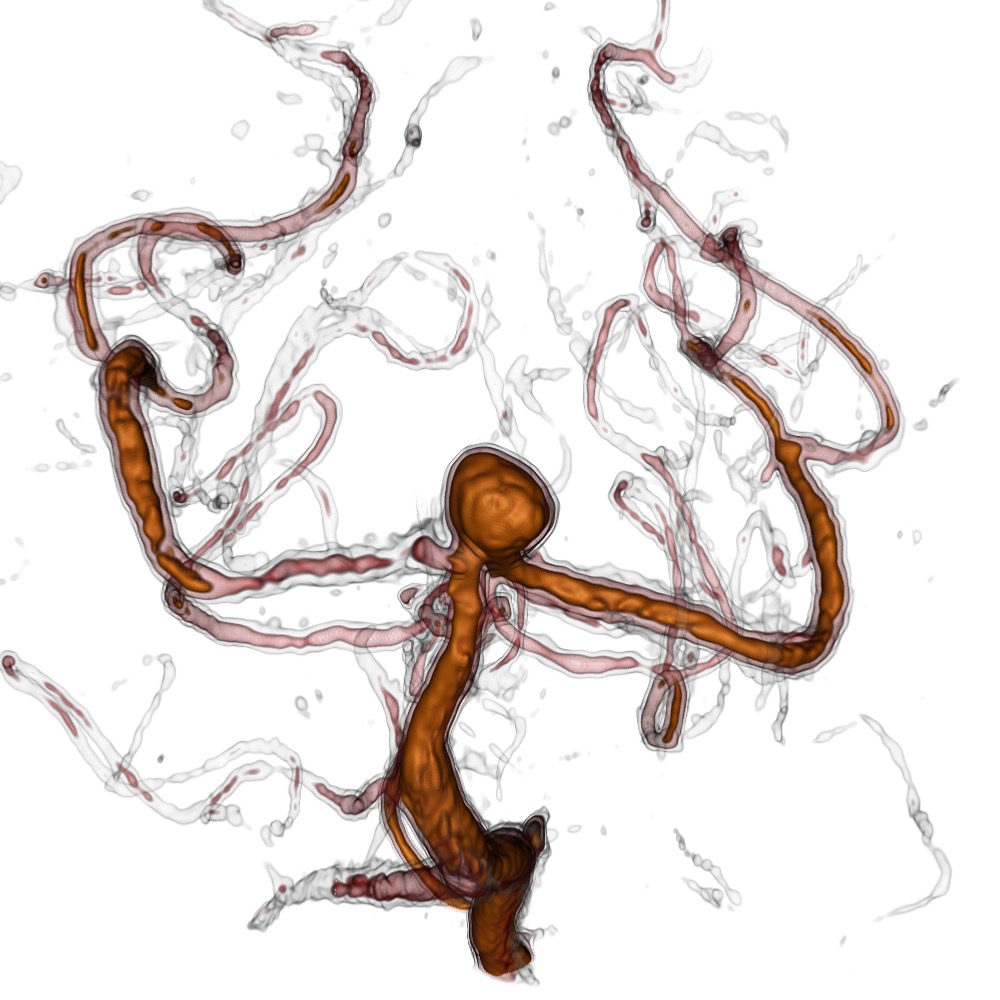}&
\includegraphics[width=\linewidth]{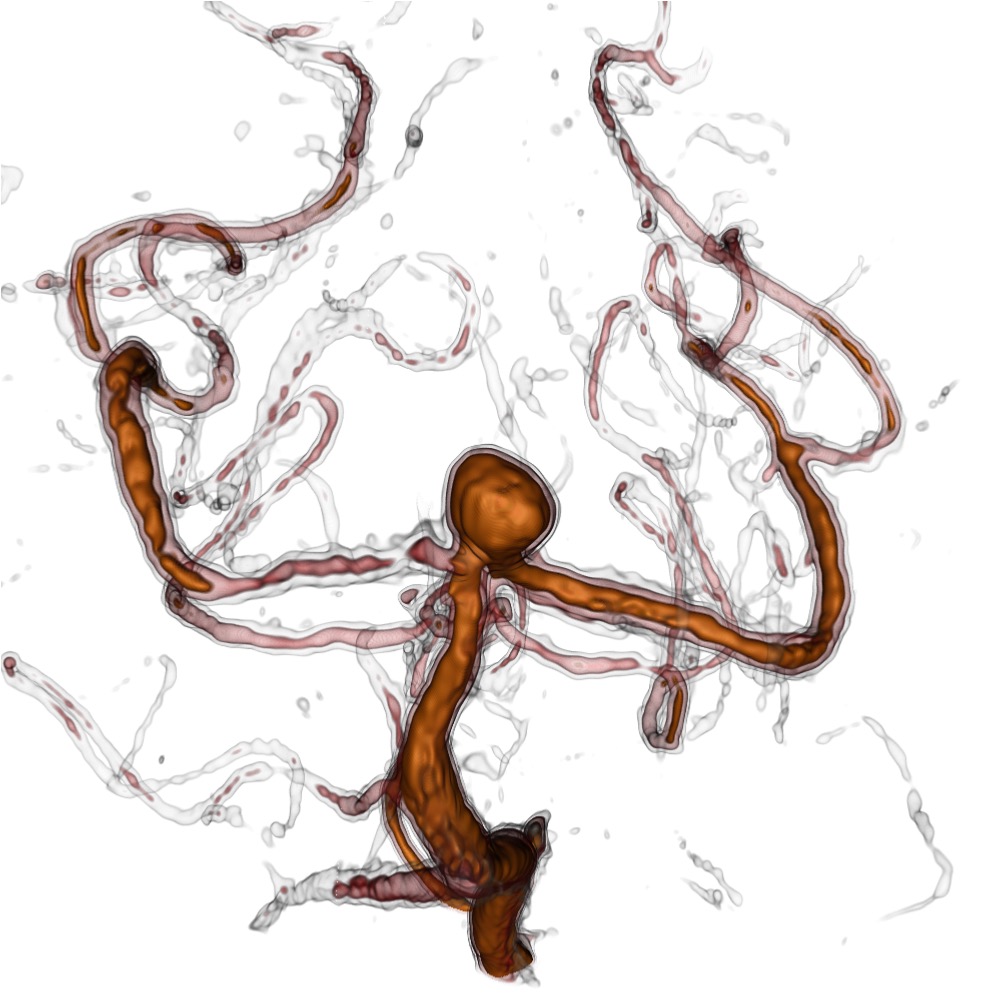}&
\includegraphics[width=\linewidth]{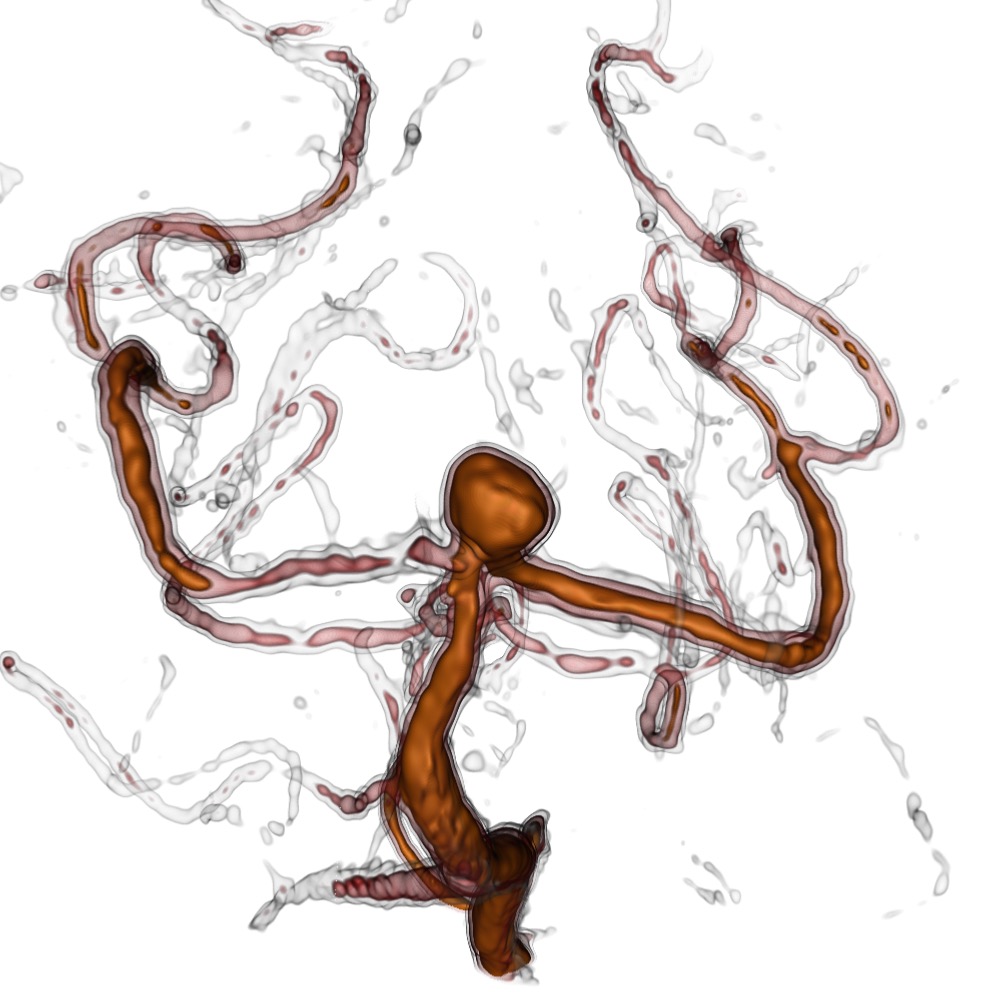}&
\includegraphics[width=\linewidth]{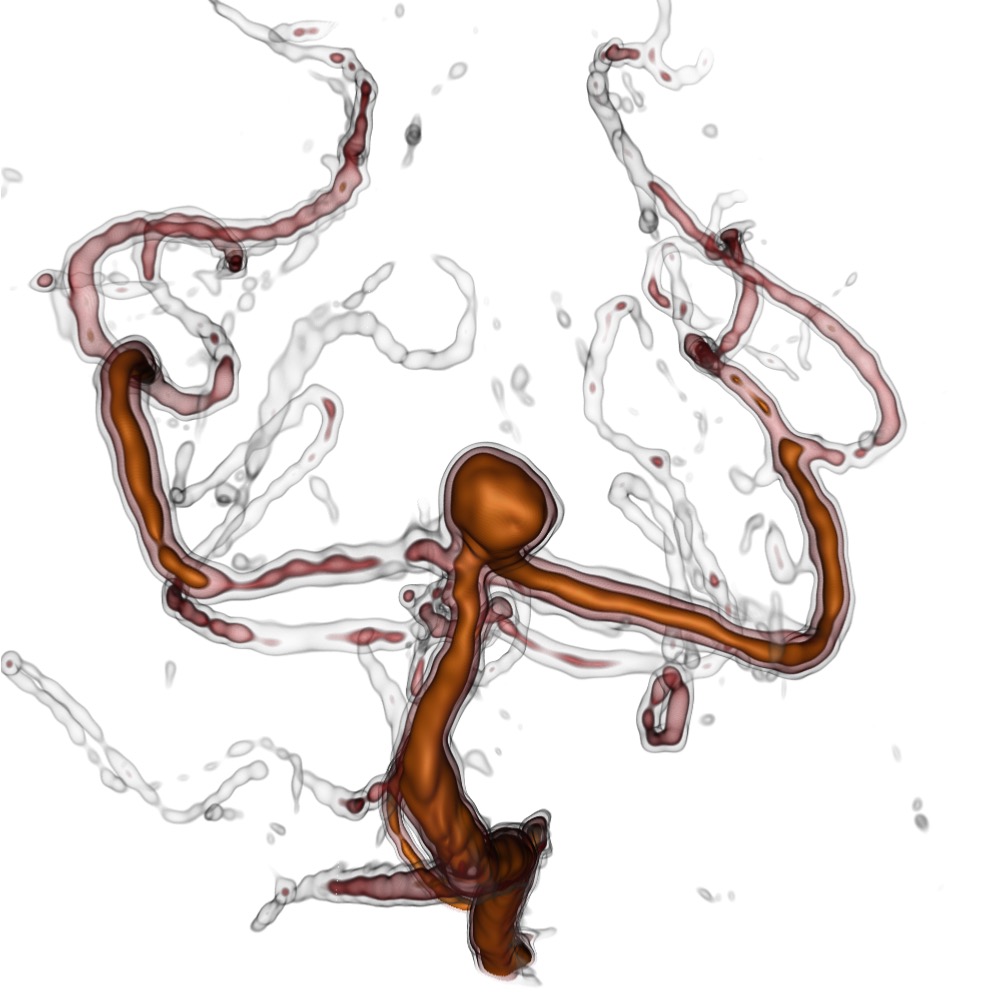}&
\includegraphics[width=\linewidth]{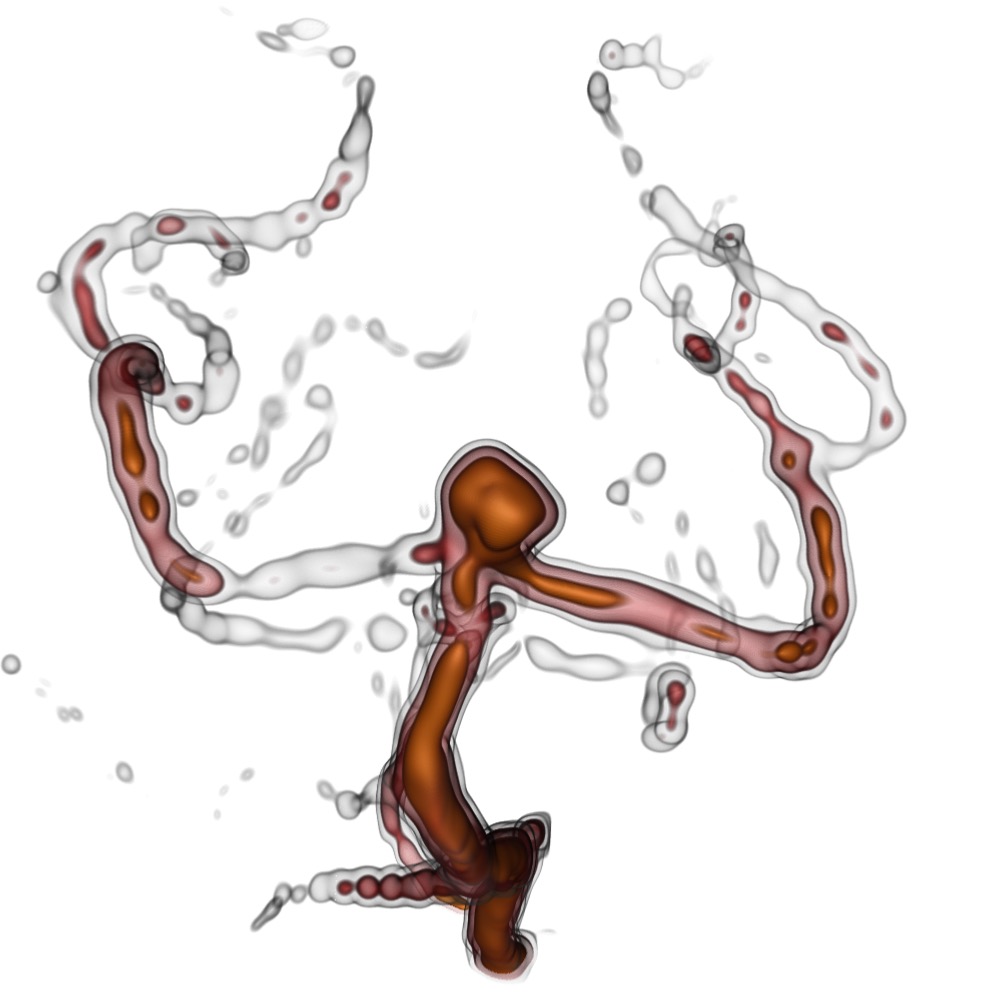}\\
 &CR = 1.0&CR = 1.492&CR = 2.370&CR = 4.096&CR = 8.0&CR = 18.963&CR = 64.0\\
\rotatebox[origin=c]{90}{Bonsai}&
\includegraphics[width=\linewidth]{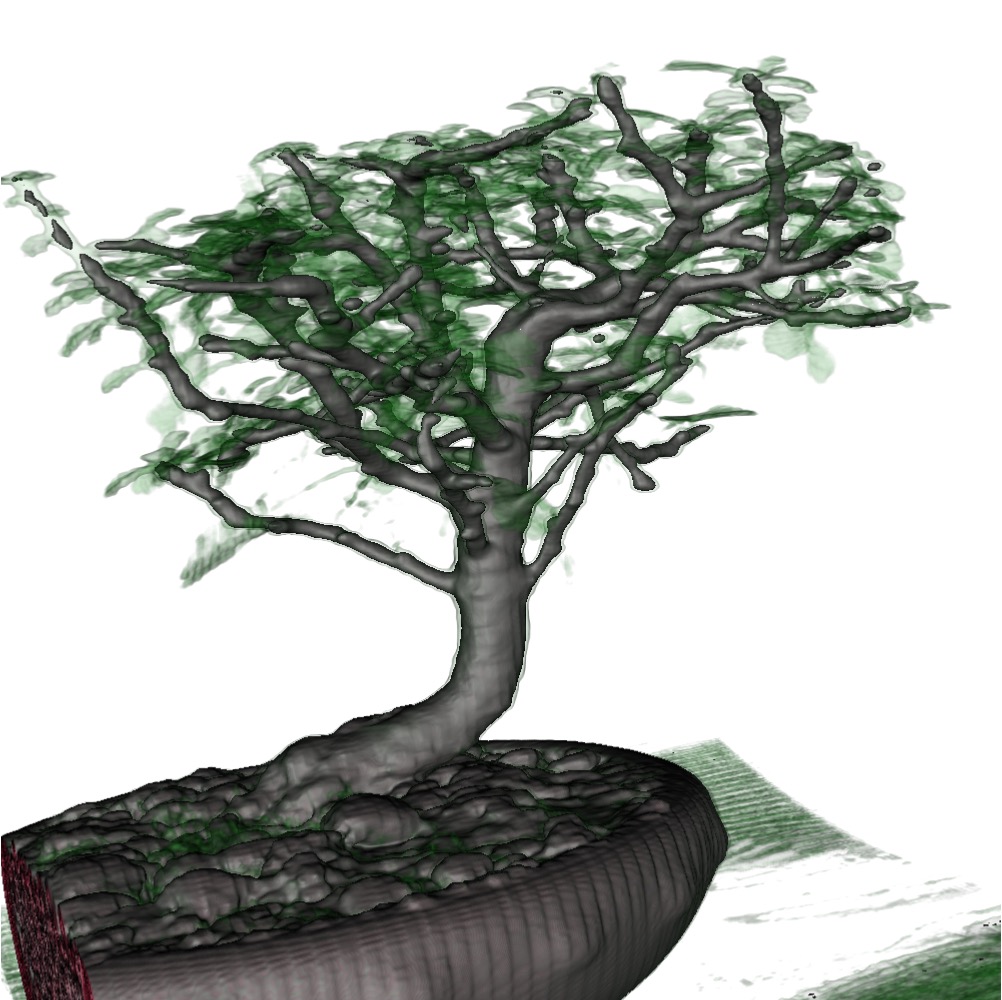}&
\includegraphics[width=\linewidth]{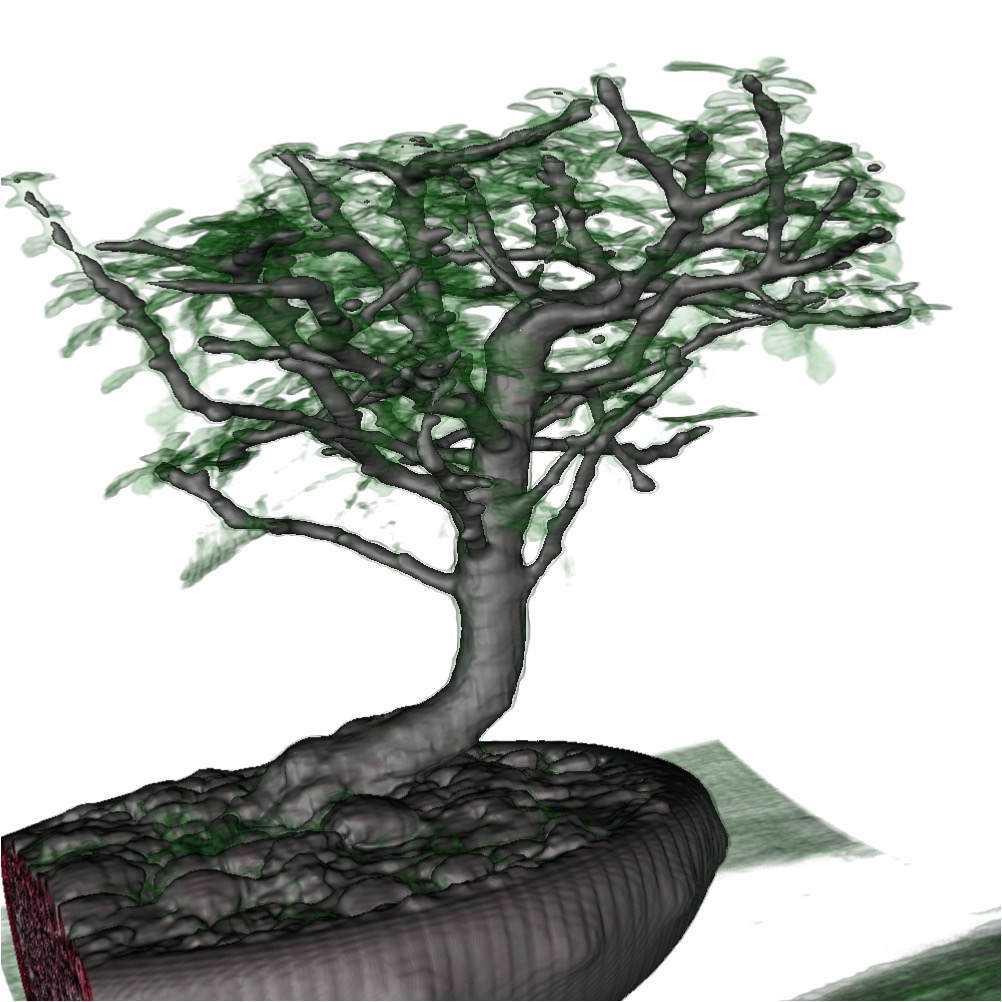}&
\includegraphics[width=\linewidth]{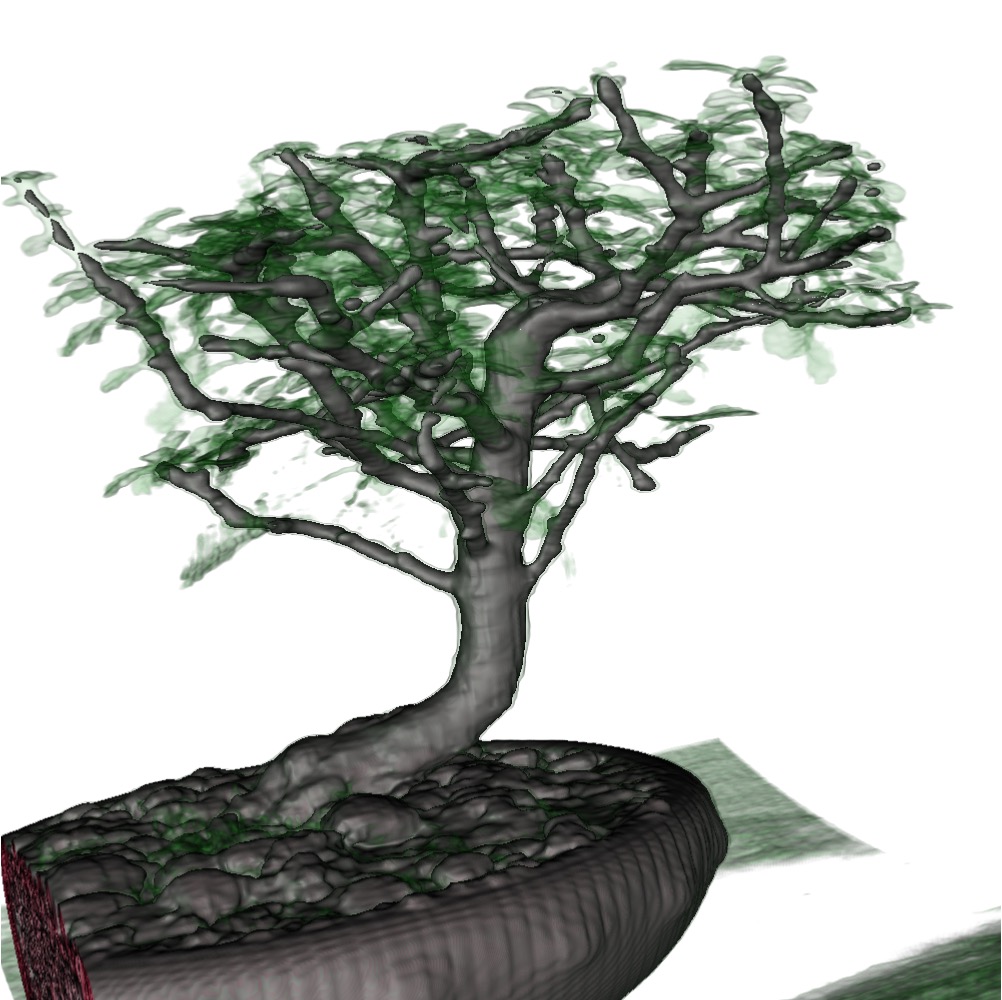}&
\includegraphics[width=\linewidth]{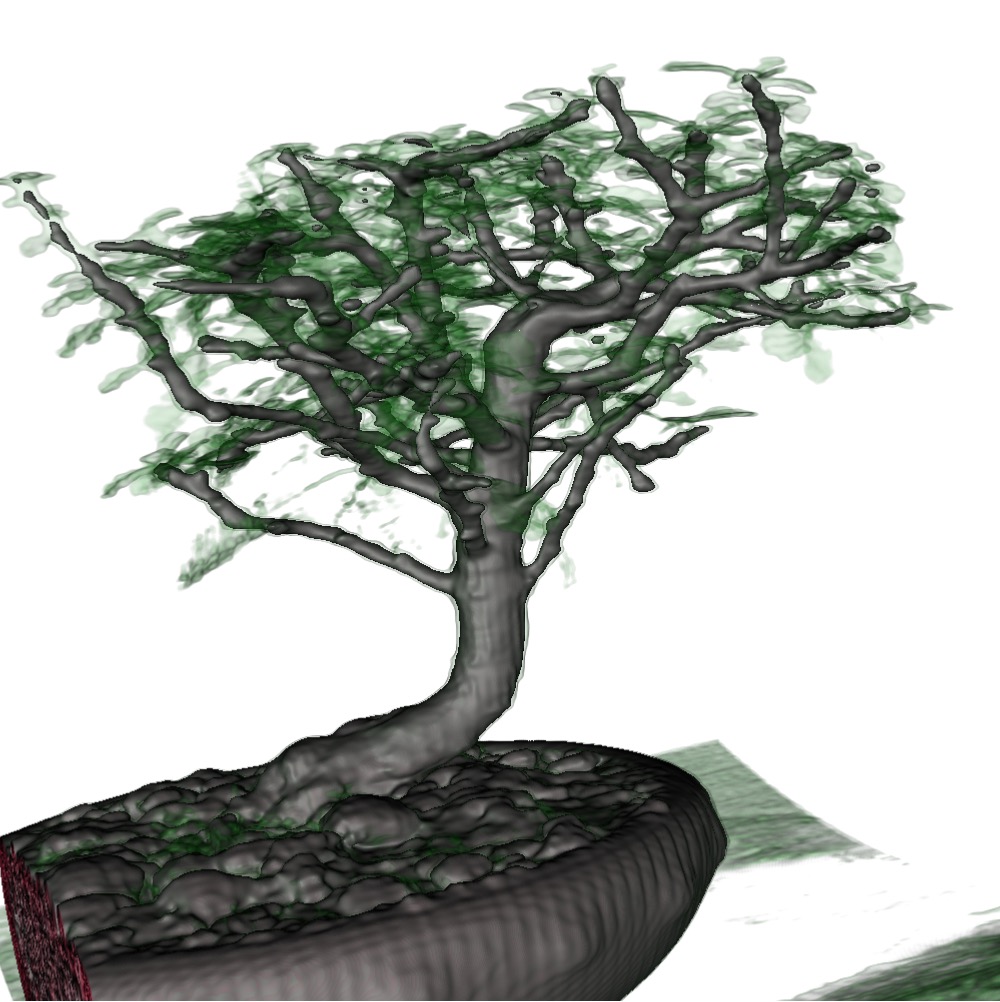}&
\includegraphics[width=\linewidth]{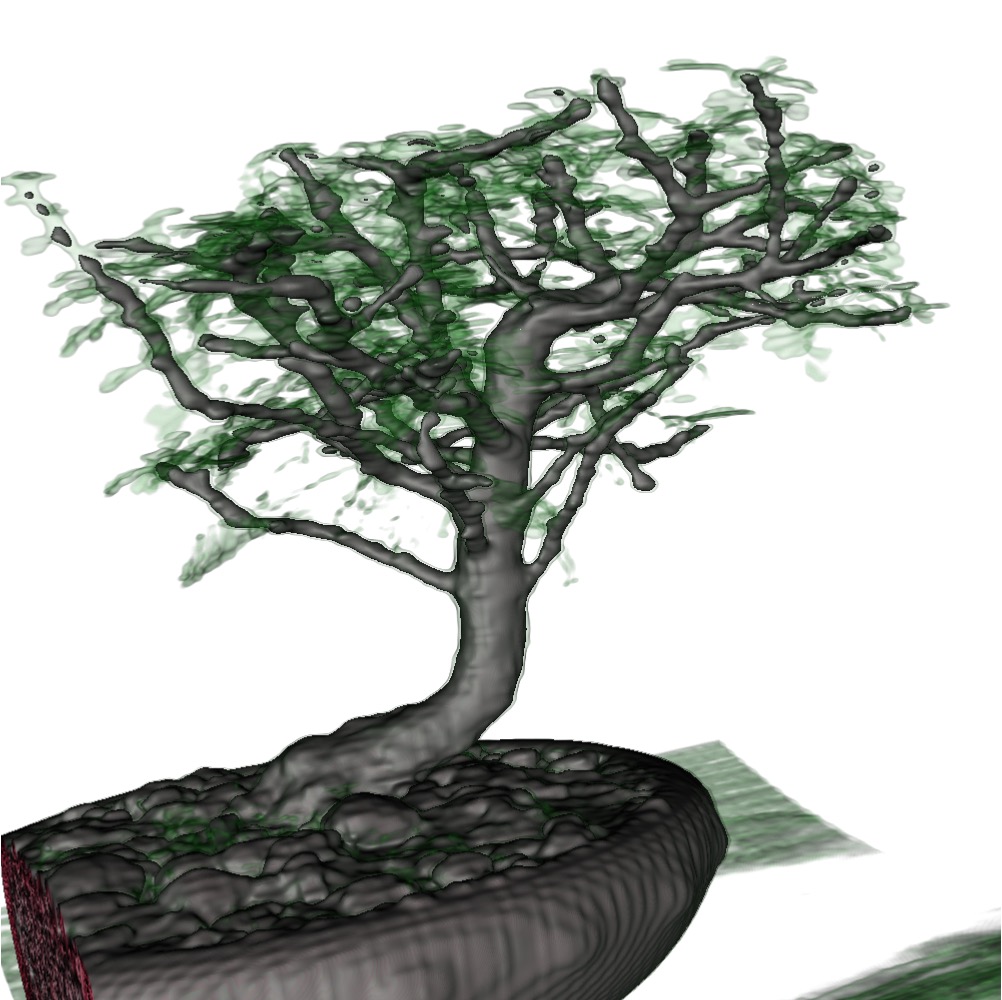}&
\includegraphics[width=\linewidth]{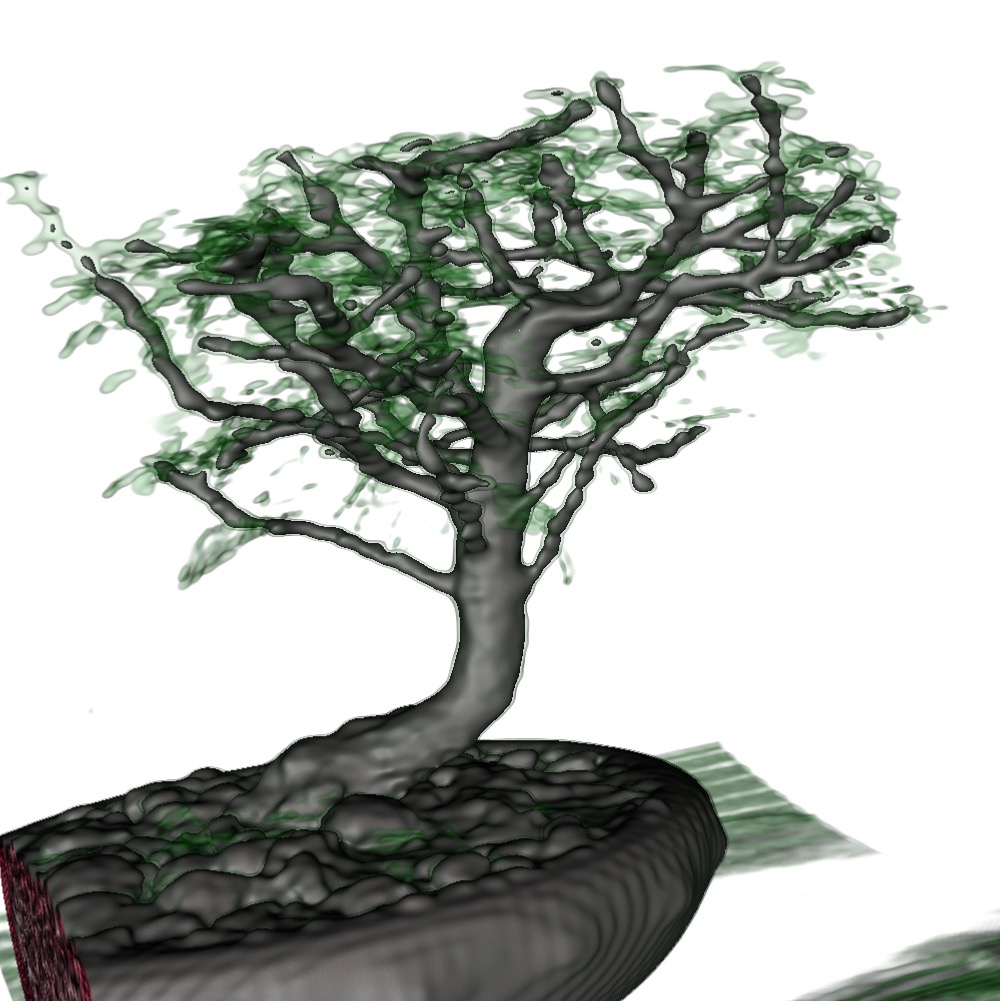}&
\includegraphics[width=\linewidth]{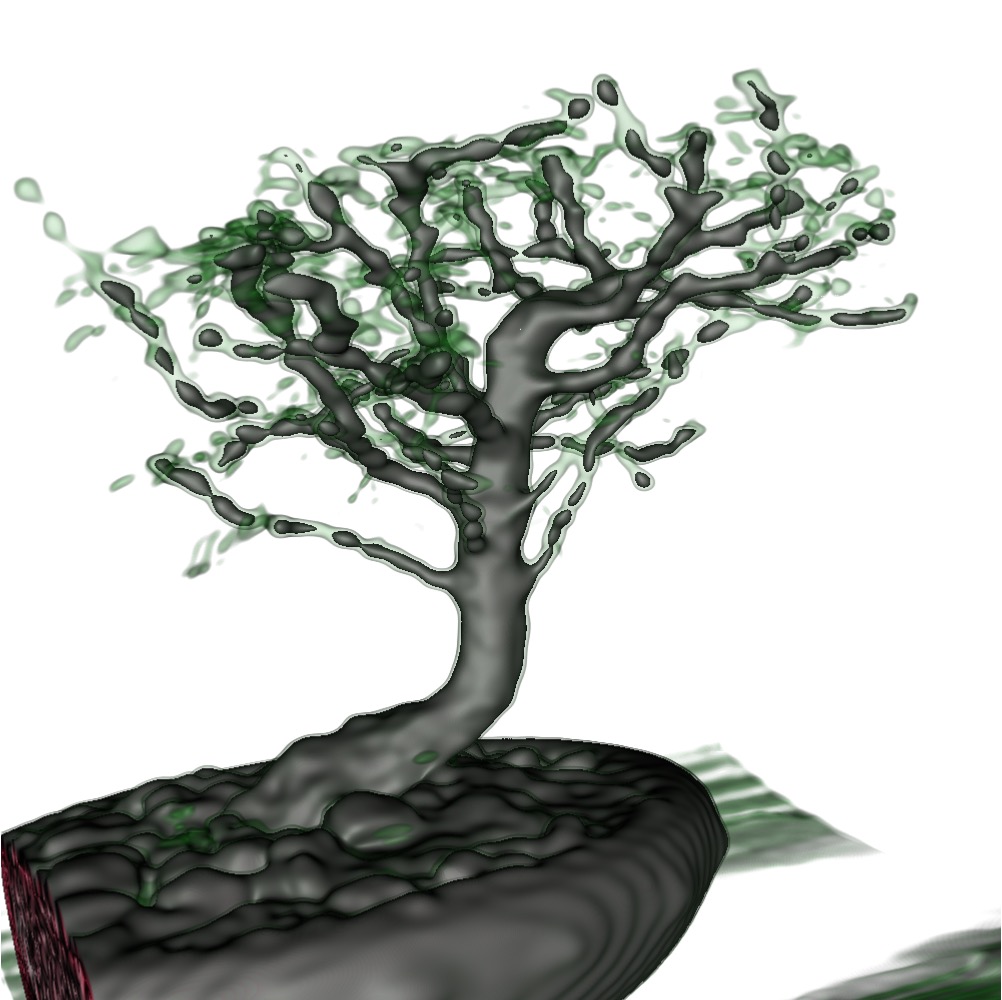}\\
 &CR = 1.0&CR = 1.492&CR = 2.370&CR = 4.096&CR = 8.0&CR = 18.963&CR = 64.0
\end{tabular}
    \caption{MFA-DVV volume rendering results of MFA model with different compression ratio (CR) encoded using various numbers of control points.}
    \label{fig:imageWithCompressionRealData} 
\end{figure*}

We quantitatively evaluate the rendering quality of MFA-DVV using compressed MFA models compared with other popular compression algorithms, including TTHRESH, ZFP, SZ3, and downsampling (DS), in both image and volume domains. MFA-DVV generates volume rendering results directly from the MFA model, while the employment of other compression algorithms needs to decompress data first and then render the decompressed discrete data. The Marschner-Lobb dataset of the size $256^3$ is used as the original data for compression. \autoref{fig:ratioVsPSNR} shows the image quality PSNR scores on volume-rendered images with respect to compression ratio. We observe that MFA-DVV always gives better rendering quality than ZFP and DS. Its rendering result is better than SZ3 and TTHRESH for a compression ratio of less than 70. For aggressive compression, SZ3 and TTHRESH give better results than MFA-DVV. The volume-rendered images with their PSNR scores under various levels of compression ratio are shown in \autoref{fig:imageWithCompression}. We can see that MFA-DVV does not have high-frequency artifacts on the rings of the ripple seen in other compression algorithms. This shows the effectiveness of high-order value and gradient estimation for off-grid locations, while other algorithms rely on linear interpolation. For compression evaluation in the volume domain, the reconstruction error is calculated by comparing values on original discrete sample locations. As shown in \autoref{fig:ratioVsPSNR_volume}, MFA-DVV always performs better than ZFP and better than DS for higher compression ranges. However, SZ3 and TTRESH are generally better than MFA-DVV. This means the MFA model is not the best data compressor targeting reconstruction of data values on-grid locations, which is what lossy floating-point compressors are designed to do. However, MFA's accurate query of off-grid locations improves its volume rendering quality in the image domain, as shown in \autoref{fig:ratioVsPSNR}, compensating for its accuracy in the volume domain.

\begin{figure}[t]
  \begin{subfigure}[b]{0.49\columnwidth}
    \includegraphics[trim=0 0 40 0,clip,width=\linewidth]{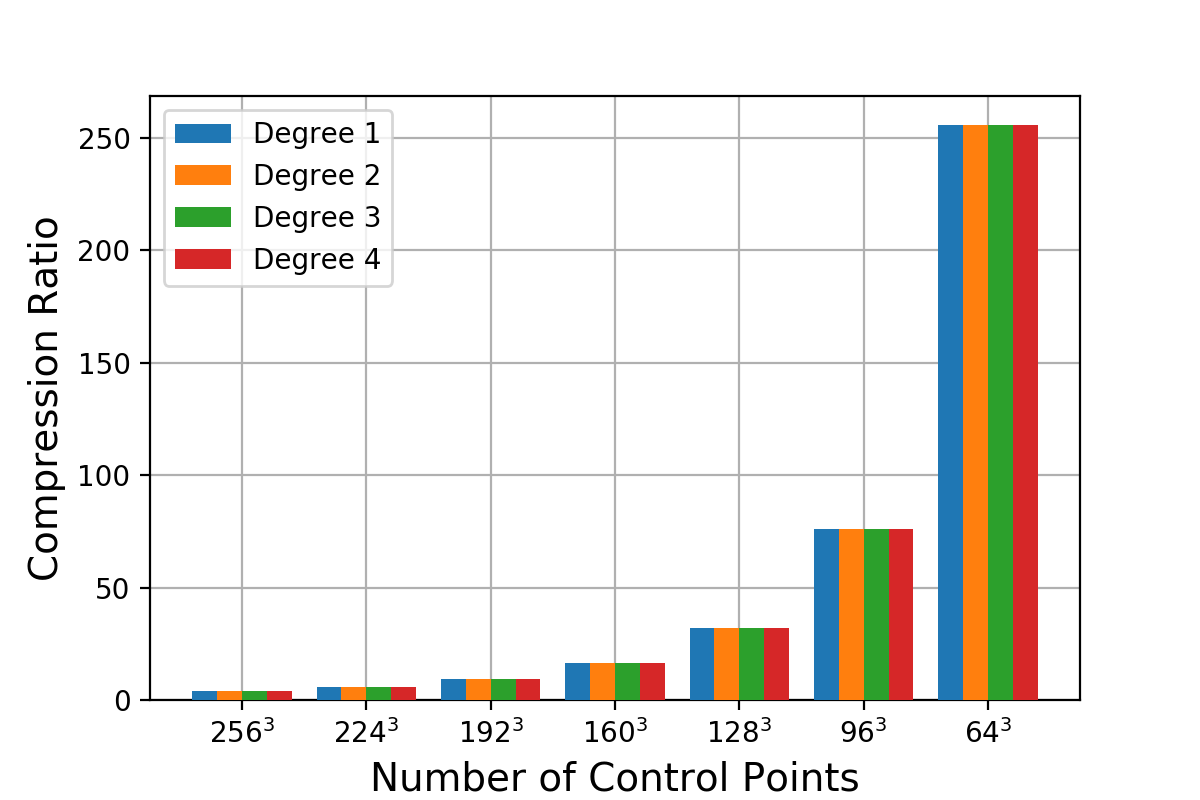}
    \caption{}
    \label{fig:parameterVsCompression}
  \end{subfigure}%
  \hspace*{\fill}   
  \begin{subfigure}[b]{0.49\columnwidth}
    \includegraphics[trim=0 0 40 0,clip,width=\linewidth]{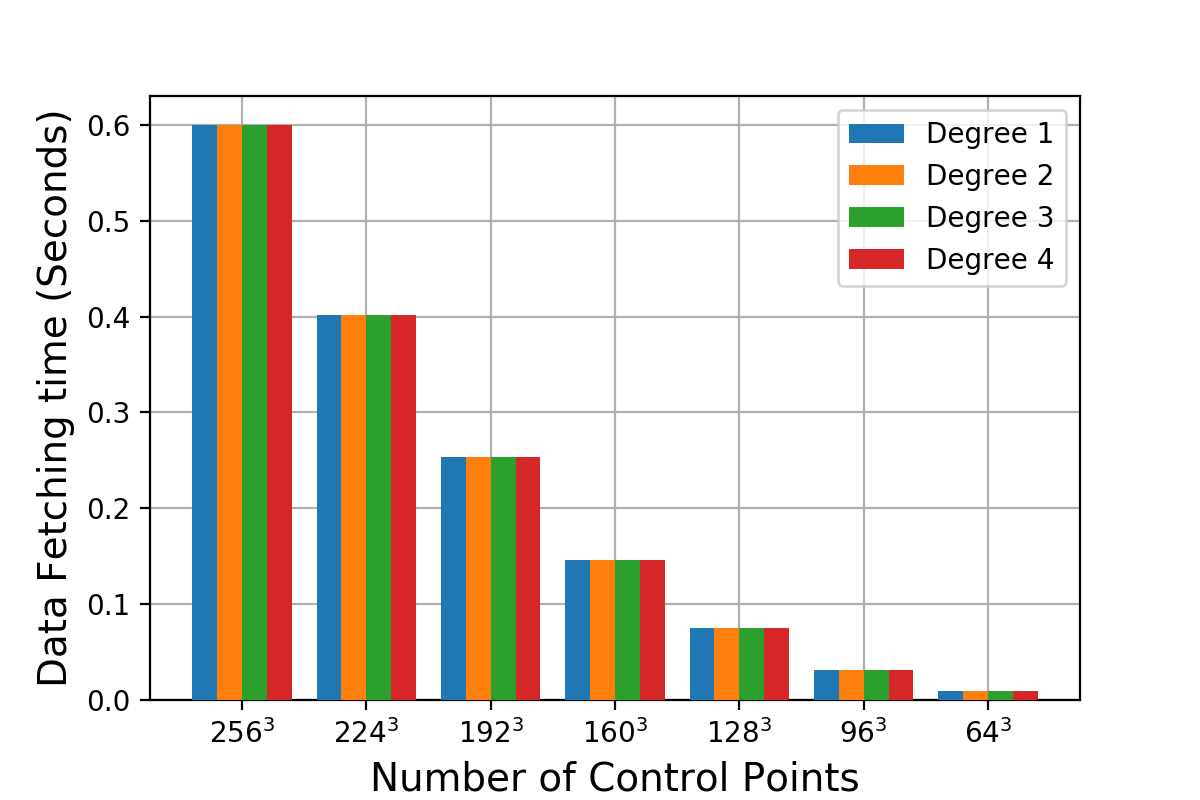}
    \caption{}
    \label{fig:parameterVsIO}
  \end{subfigure}%
\caption{MFA compression ratio and data fetching time using different numbers of control points and polynomial degree.}
\label{fig:parameterVsCompressionIO}
\end{figure}

\begin{figure*}[t]
  \begin{subfigure}[b]{0.32\textwidth}
    \includegraphics[trim=0 0 0 0,clip,width=\linewidth]{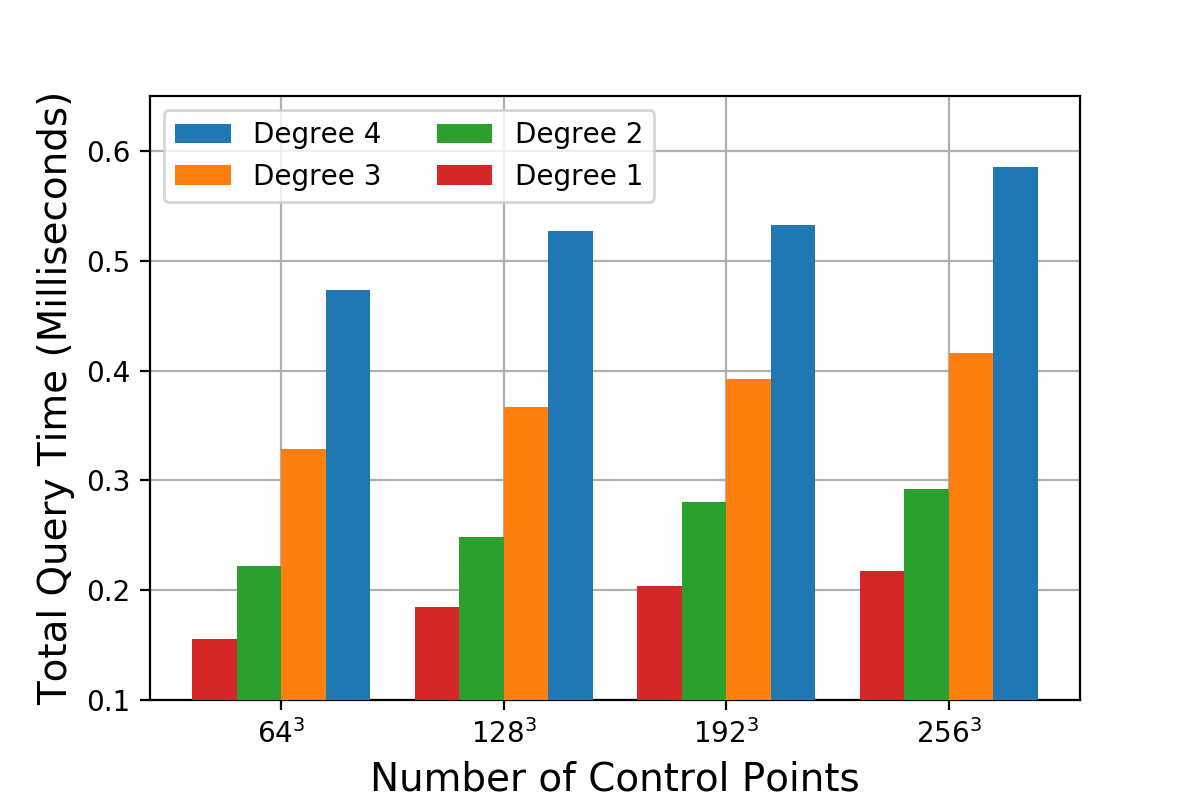}
    \caption{Value query time}
  \end{subfigure}%
  \hspace*{\fill}   
  \begin{subfigure}[b]{0.32\textwidth}
    \includegraphics[trim=0 0 0 0,clip,width=\linewidth]{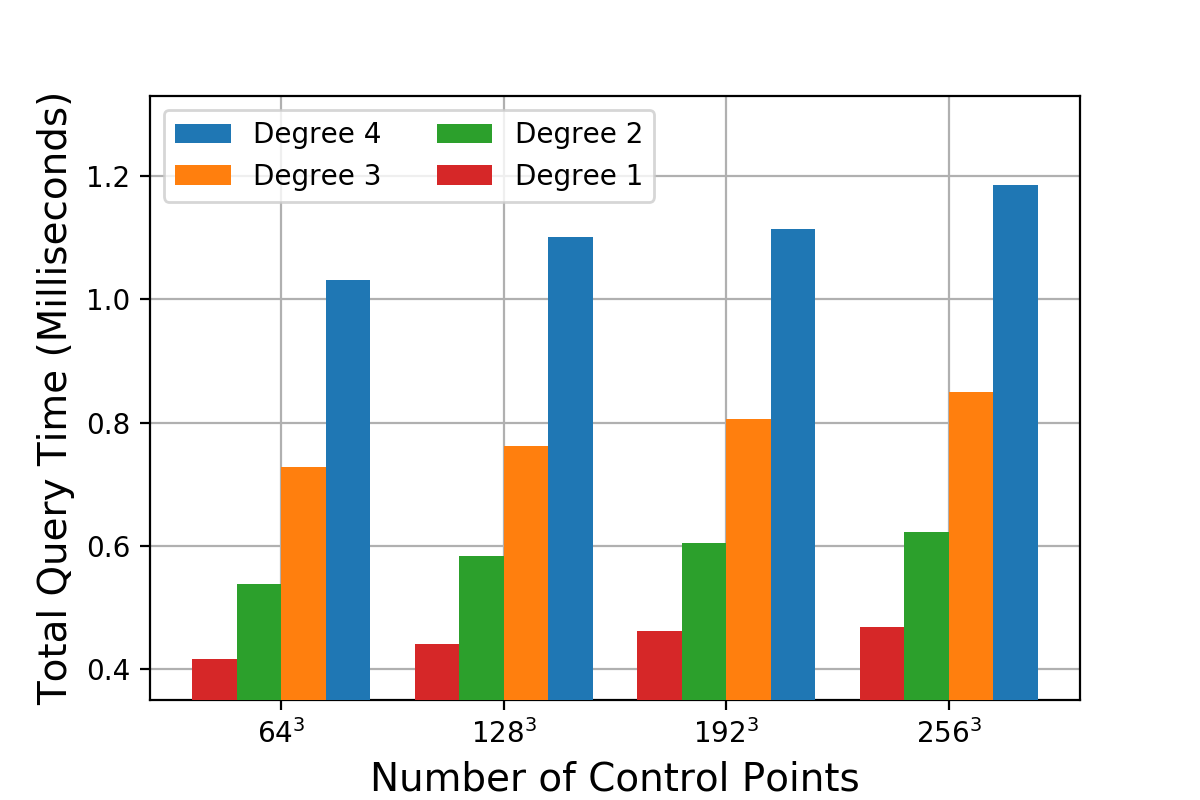}
    \caption{Gradient query time}
  \end{subfigure}%
  \hspace*{\fill}   
  \begin{subfigure}[b]{0.32\textwidth}
    \includegraphics[trim=0 0 0 0,clip,width=\linewidth]{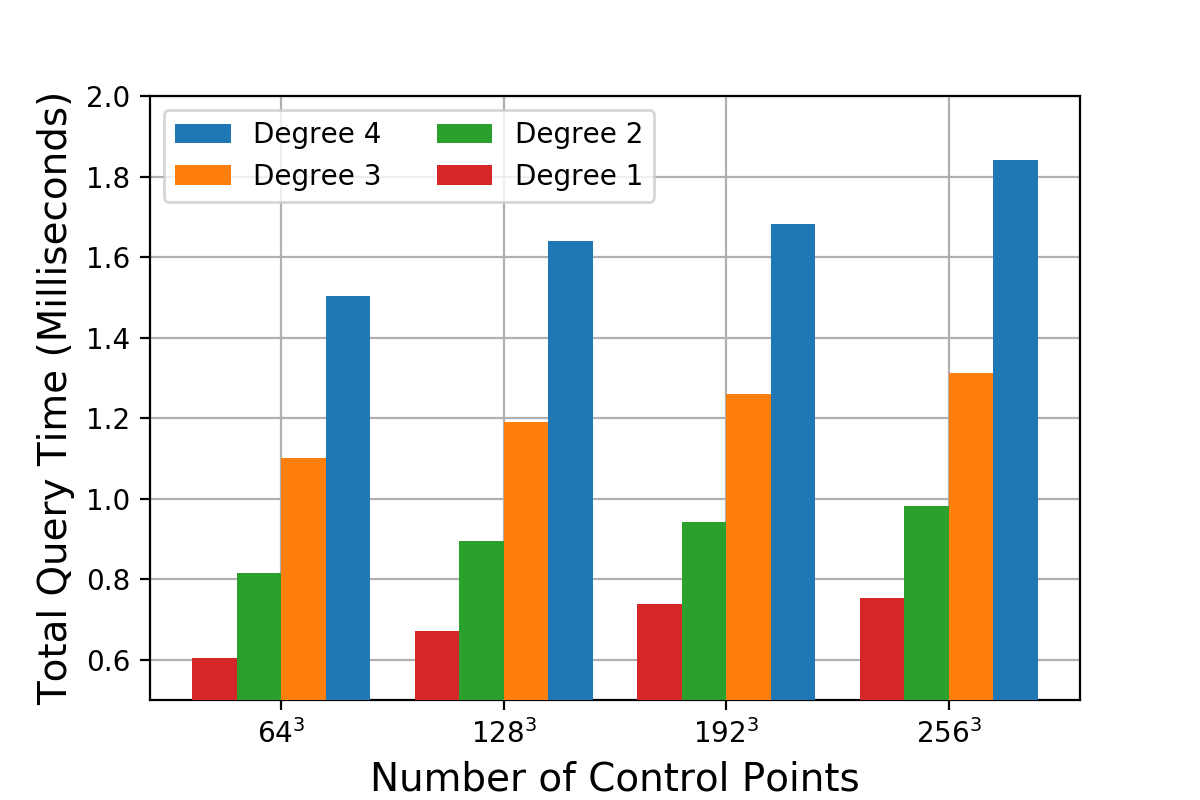}
    \caption{Overall query time}
  \end{subfigure}%
\caption{Query performance from MFA model for value and gradient using various number of control points. The overall query time is the sum of the query times for value and gradient.}
\label{fig:mfaQuery}
\end{figure*}

\begin{figure*}[h]
  \begin{subfigure}[b]{0.32\textwidth}
    \includegraphics[trim=0 0 0 0,clip,width=\linewidth]{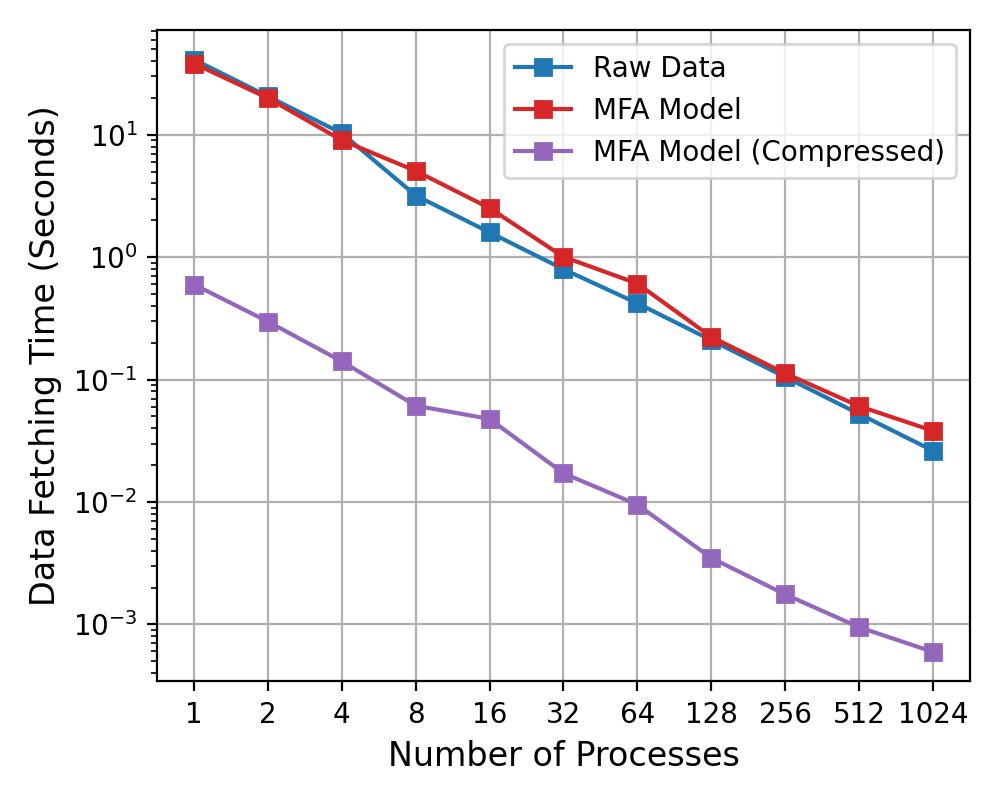}
    \caption{Input loading I/O time}
    \label{fig:compressIoTime}
  \end{subfigure}%
  \hspace*{\fill}   
  \begin{subfigure}[b]{0.32\textwidth}
    \includegraphics[trim=0 0 0 0,clip,width=\linewidth]{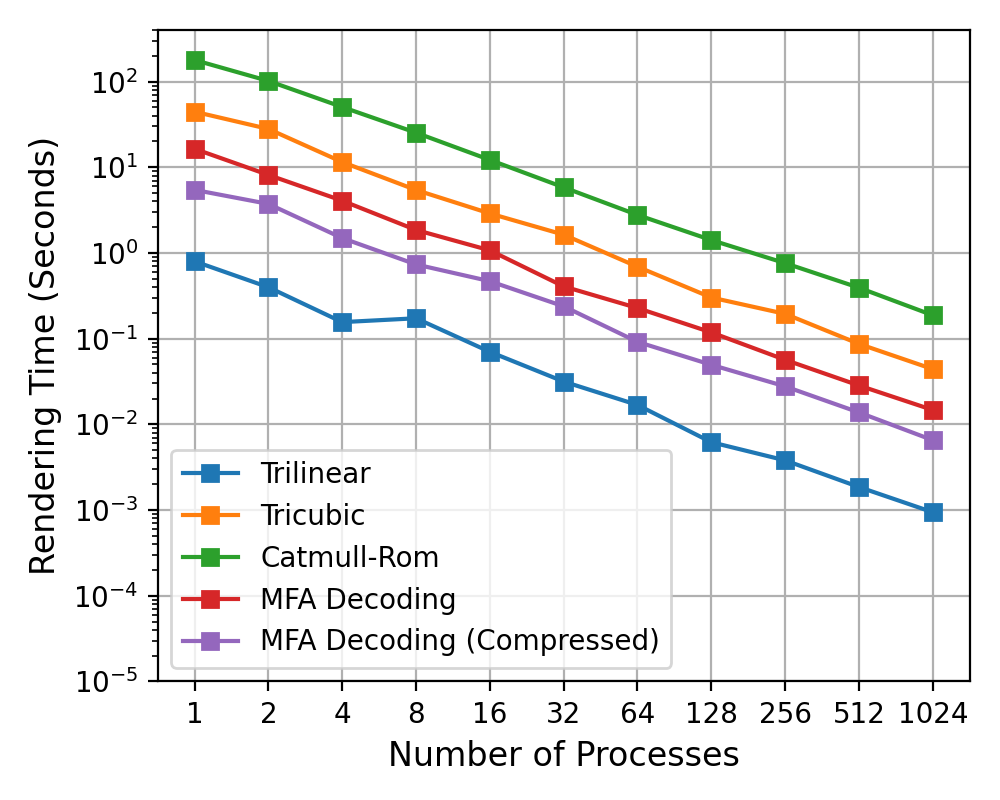}
    \caption{Rendering time}
    \label{fig:compressRendering}
  \end{subfigure}%
  \hspace*{\fill}   
  \begin{subfigure}[b]{0.32\textwidth}
    \includegraphics[trim=0 0 0 0,clip,width=\linewidth]{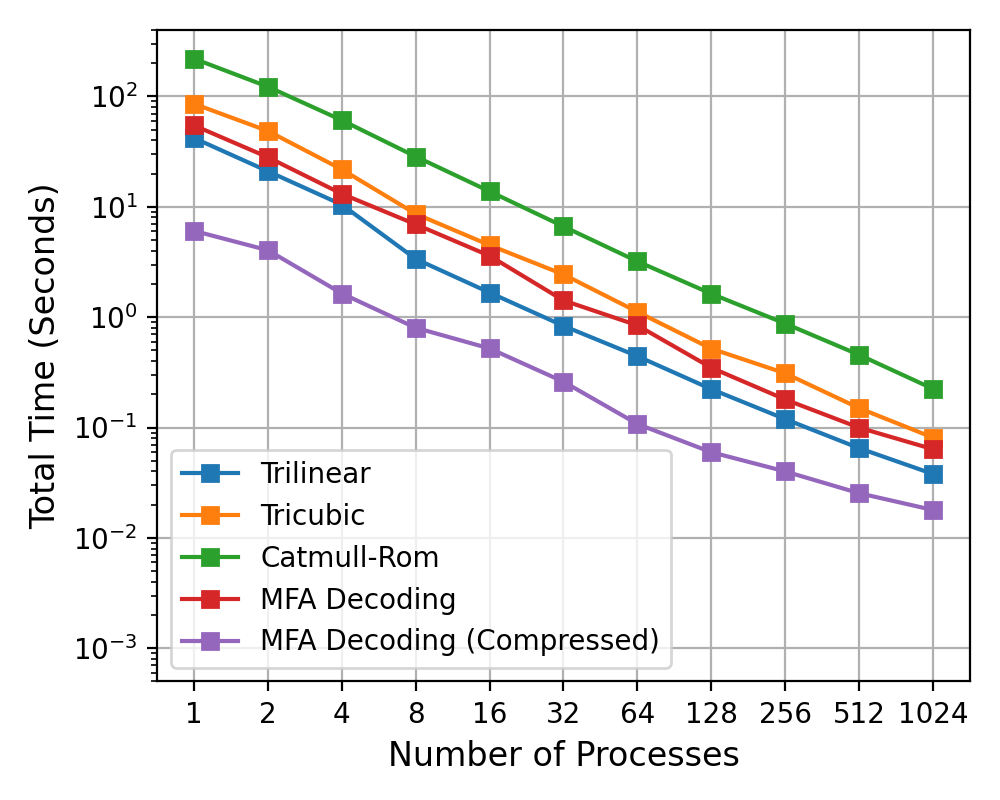}
    \caption{Total time}
    \label{fig:compressedTotal}
  \end{subfigure}%
\caption{Scalability of data fetching time, rendering time, and total time using various ways of information query.}
\label{fig:mfaParameterQuality}
\end{figure*}

We also qualitatively evaluate the volume rendering using real datasets under various levels of compression. The encoded MFA model of the input discrete raw dataset is compressed by adjusting the number of control points as shown in \autoref{fig:imageWithCompressionRealData}. The first column of the figure is the volume rendering result using MFA-DVV without compression, where the number of the control points is equal to the number of points of the input discrete dataset. As the compression ratio (CR) increases, the rendering quality decreases. For sparse datasets like Nucleon and Fuel, although MFA-DVV using moderate compression can capture the main structure of the data, the results have noticeable errors compared to using an uncompressed model. Because the Nucleon dataset is simpler than the Fuel dataset, its result with a similar level of compression gives more faithful results. However, due to the limited number of control points, an extremely compressed MFA model results in inaccurate volume rendering results, as shown in their right images. For datasets with high resolutions, like Aneurysm and Bonsai, the compression ratio can be more aggressive because there are more sample points to describe the dynamics of the data. As shown in the middle figures, the rendering error of using moderate compression is not very noticeable. For high-resolution data, the maximal compression ratio the MFA-DVV can achieve is determined by the complexity of the specific data. 

\subsubsection{MFA Parameter Study on Data Fetching Time}

The number of control points is the main MFA parameter that determines the compression ratio of the MFA model. Specifically, as shown in \autoref{fig:parameterVsCompression}, the size of the compressed MFA model is proportional to the number of control points used for encoding, while the polynomial degree almost has no influence on the size of the MFA model. As the MFA model decreases due to compression, its data fetching time also decreases, as shown in \autoref{fig:parameterVsIO}. This time-saving becomes more significant when handling large-scale datasets. 

\begin{figure}[t]
\centering
\subcaptionbox{Trilinear\label{fig:11}}{\includegraphics[width=0.33\columnwidth]{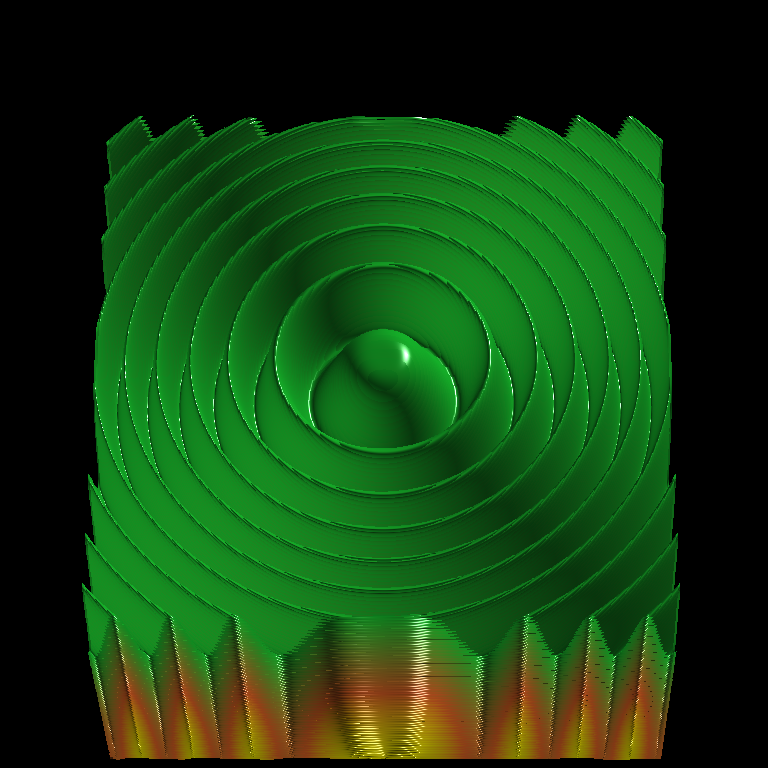}}\hfill 
\subcaptionbox{MFA-DVV (compressed MFA model)\label{fig:22}}{\includegraphics[width=0.33\columnwidth]{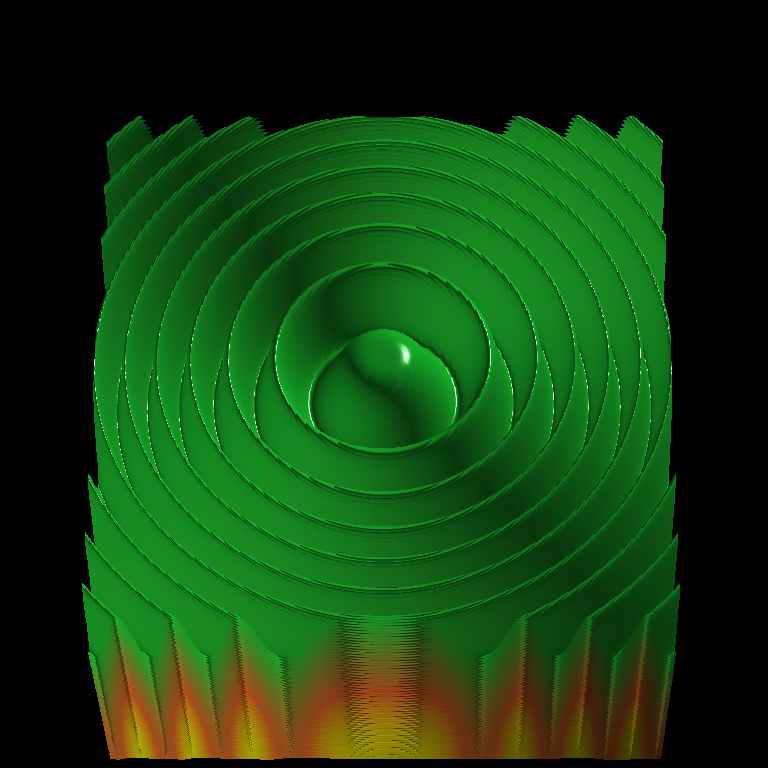}}\hfill
\subcaptionbox{Ground Truth\label{fig:33}}{\includegraphics[width=0.33\columnwidth]{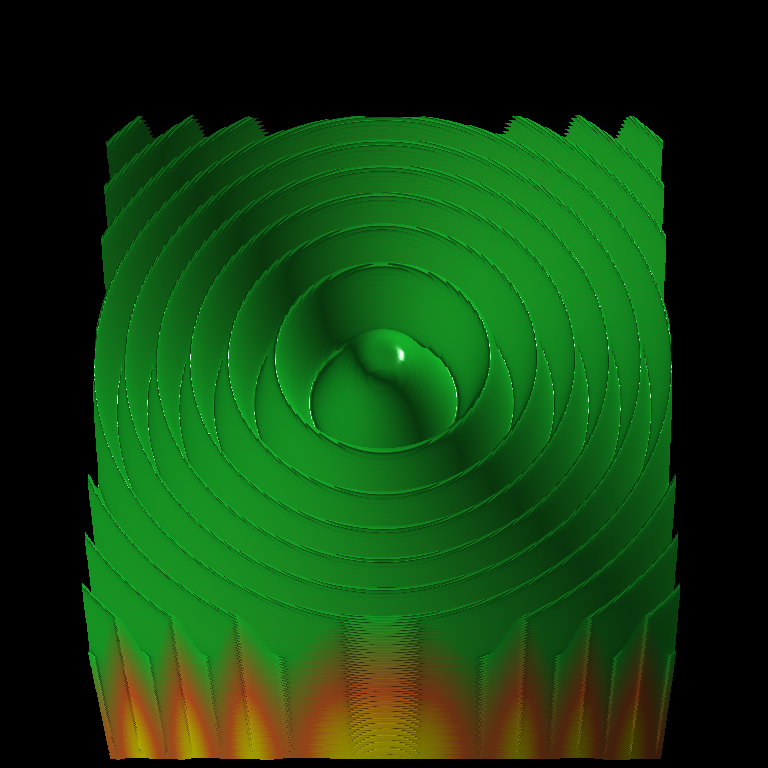}}
\caption{Rendering results using trilinear interpolation, MFA-DVV of compressed MFA model, and the ground truth.}
\label{fig:renderRenderingResults}
\end{figure}

\begin{table}[t]
  \caption{Quantitative comparison of rendering quality between trilinear and MFA-DVV using compressed MFA model.}
  \label{tab:renderCompare}
  \scriptsize%
	\centering%
      \begin{tabu}{ c | c c c }
      \toprule
      Methods & MSE & PSNR (dB) & SSIM\\
      \midrule
      Trilinear & 205.46 & 25.00 & 0.89\\
      \midrule
      MFA-DVV (Compressed) & \textbf{50.76} & \textbf{31.08} & \textbf{0.96}\\
      \bottomrule
      \end{tabu}
\end{table}

\subsubsection{MFA Parameter Study on Rendering Time}

MFA model encoded using fewer control points will give faster query times for both value and gradient because fewer items (i.e., basis functions and control points) are involved in the calculation. General volume visualization needs to query both value and gradient for better visualization with a shading effect. Thus, faster queries with a compressed MFA model encoded with fewer control points will reduce the rendering time. \autoref{fig:mfaQuery} shows how the number of control points affects the query time used to retrieve the value and gradient for a location of interest within the domain.

\begin{figure*}[t]
  \begin{subfigure}[b]{0.33\textwidth}
    \includegraphics[trim=0 0 0 0,clip,width=\linewidth]{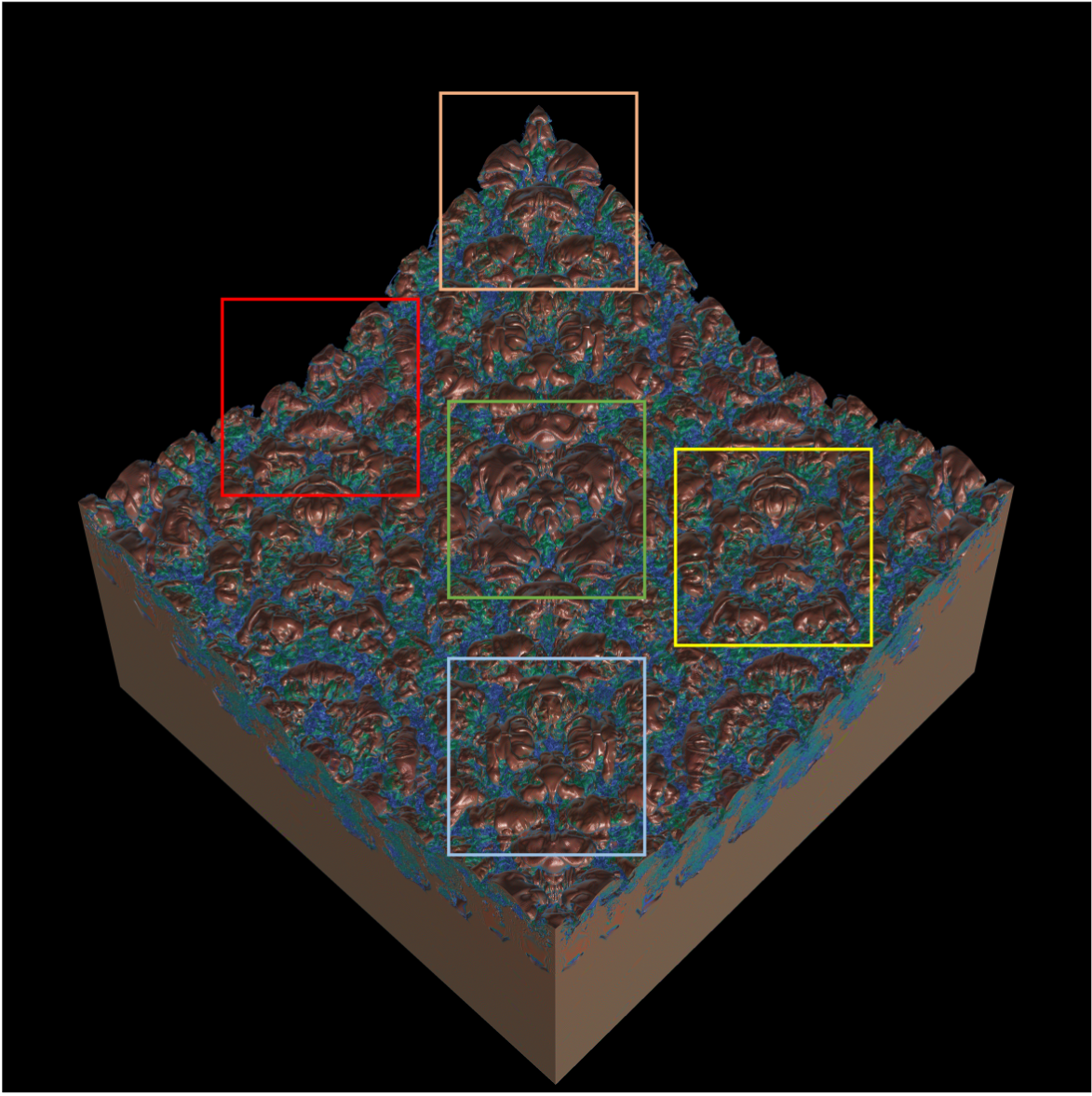}
    \caption{}
  \end{subfigure}%
  \hspace*{\fill}   
  \begin{subfigure}[b]{0.33\textwidth}
    \includegraphics[trim=0 0 0 0,clip,width=\linewidth]{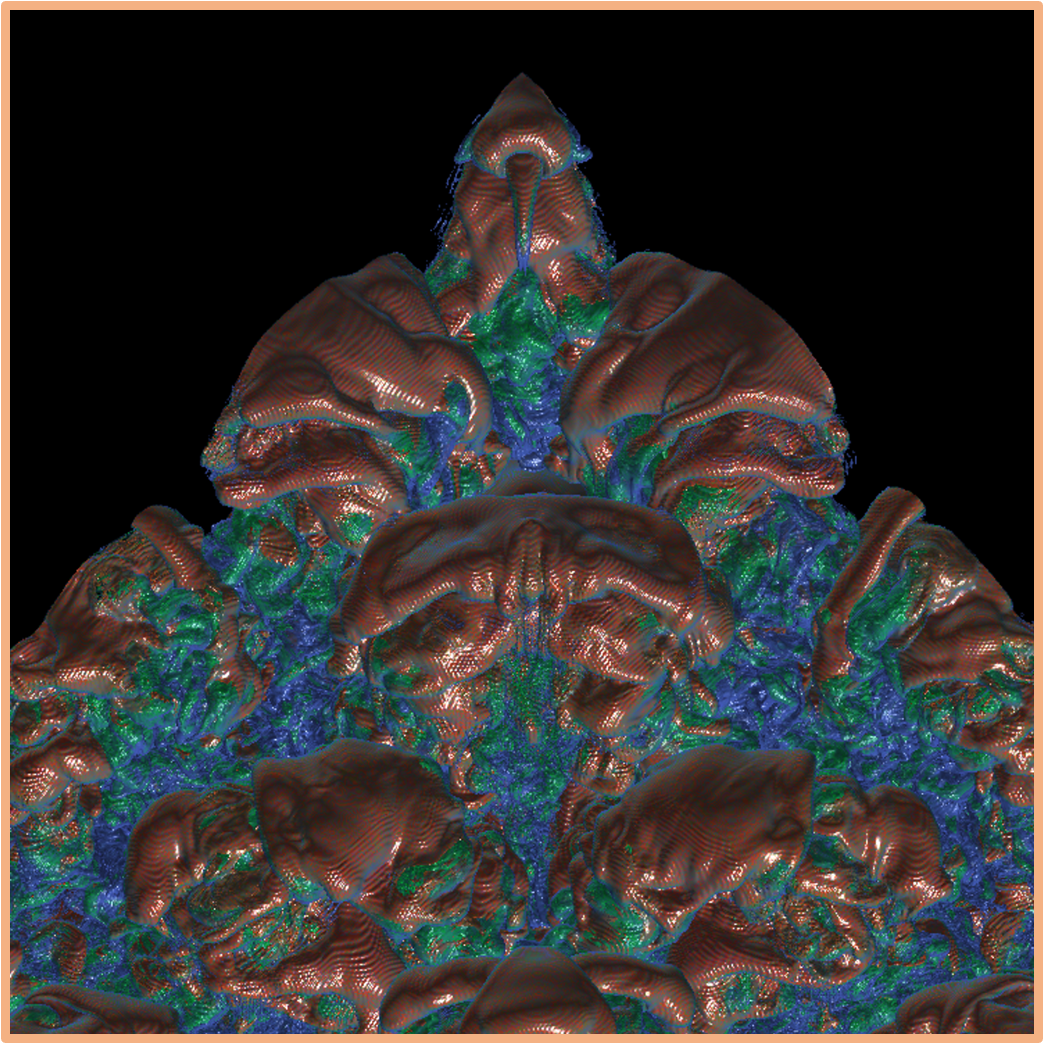}
    \caption{}
  \end{subfigure}%
    \hspace*{\fill}   
  \begin{subfigure}[b]{0.33\textwidth}
    \includegraphics[trim=0 0 0 0,clip,width=\linewidth]{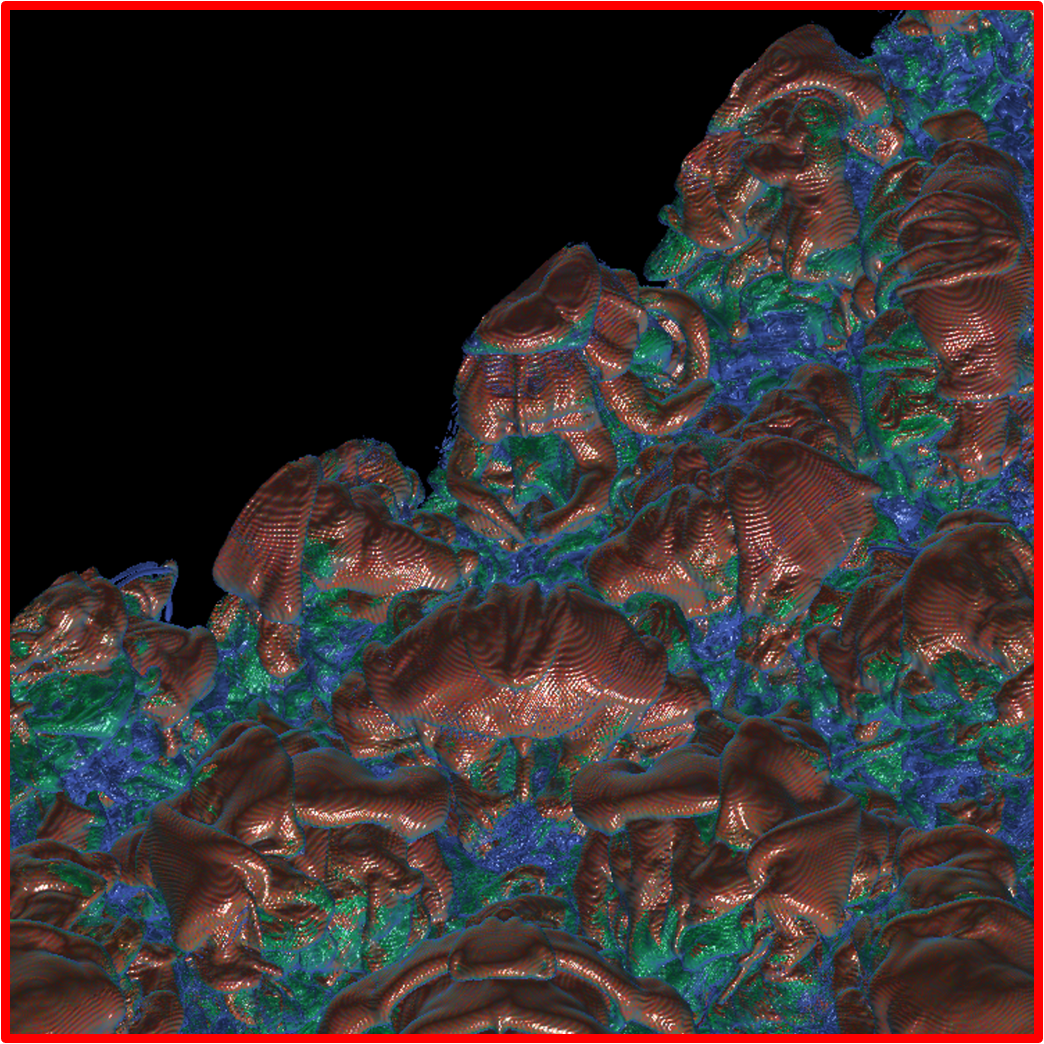}
    \caption{}
  \end{subfigure}%
  \\
    \begin{subfigure}[b]{0.33\textwidth}
    \includegraphics[trim=0 0 0 0,clip,width=\linewidth]{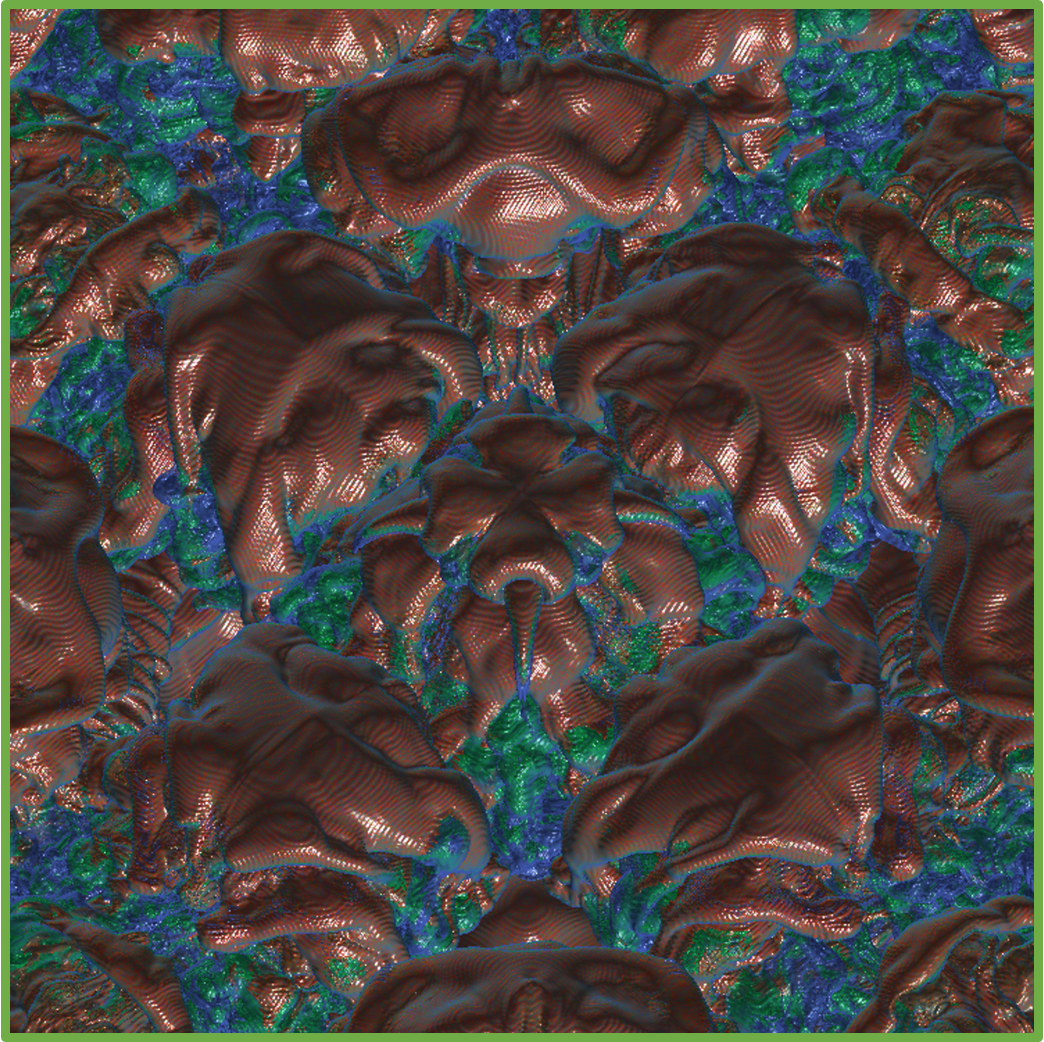}
    \caption{}
  \end{subfigure}%
  \hspace*{\fill}   
  \begin{subfigure}[b]{0.33\textwidth}
    \includegraphics[trim=0 0 0 0,clip,width=\linewidth]{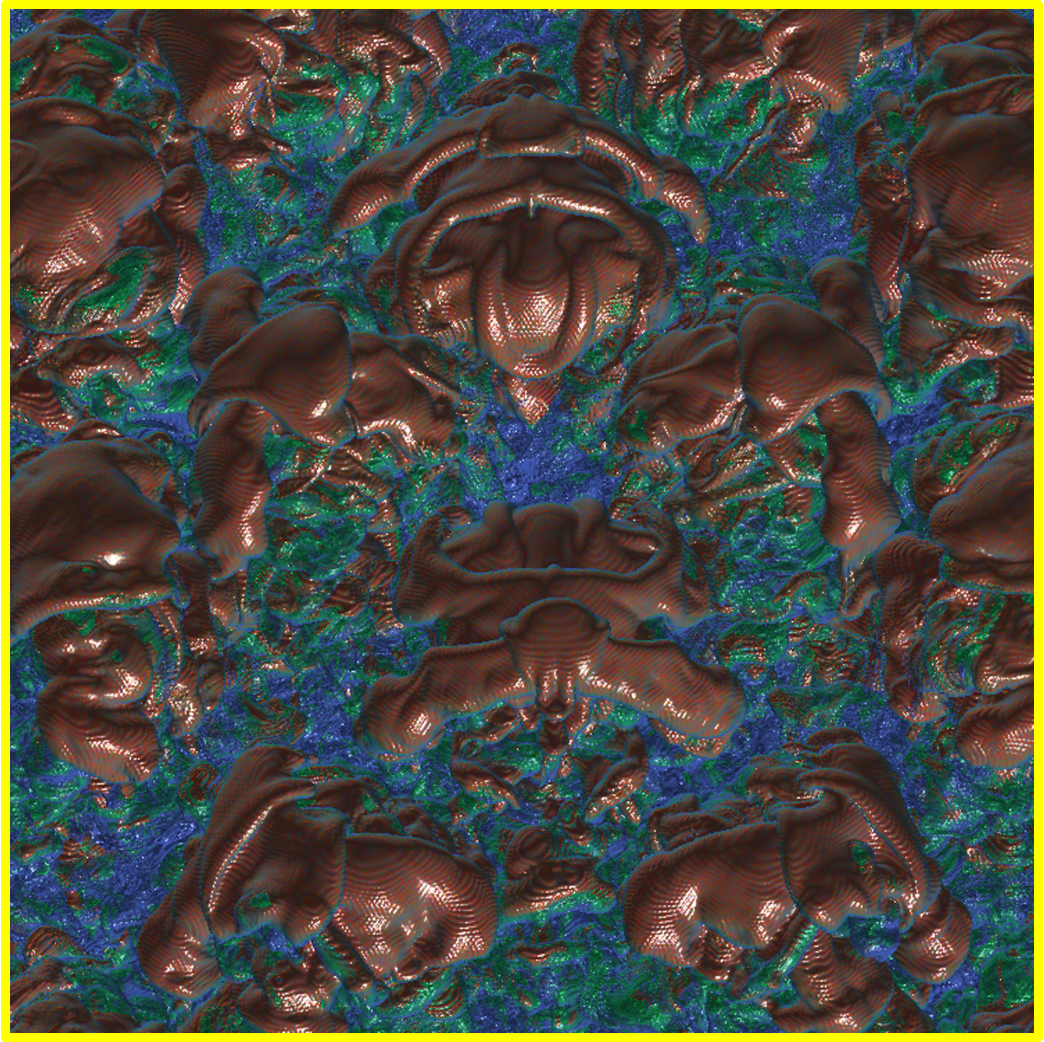}
    \caption{}
  \end{subfigure}%
    \hspace*{\fill}   
  \begin{subfigure}[b]{0.33\textwidth}
    \includegraphics[trim=0 0 0 0,clip,width=\linewidth]{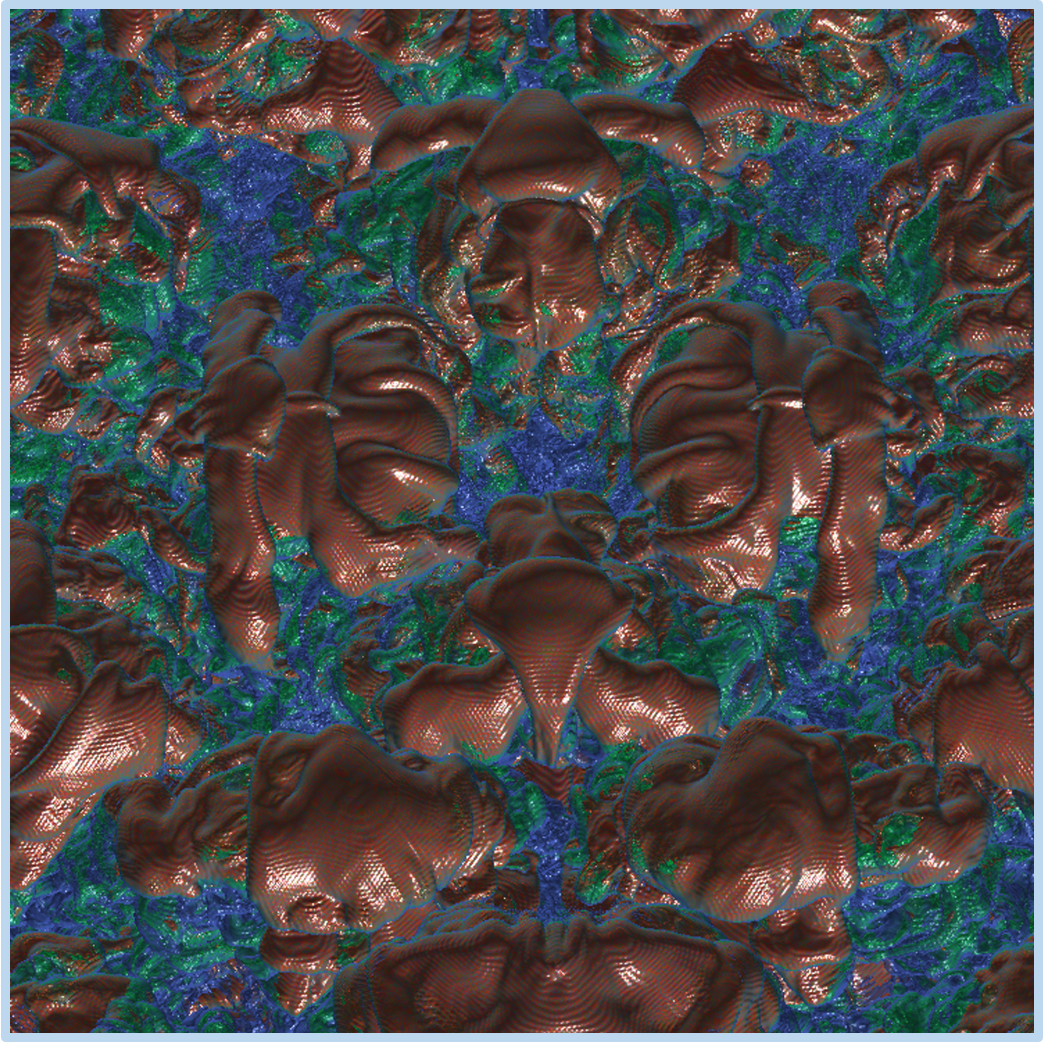}
    \caption{}
  \end{subfigure}%
\caption{Rendering of RMI large-scale dataset using MFA-DVV, where (a) is  an overview of the data with an image resolution of $4096\times4096$, and (b) to (f) show 5 zoom-in view with a $512\times512$ image resolution. Detail of the dataset can be clearly observed, enhancing the comprehension of the large dataset.}
\label{fig:example}
\end{figure*}

\subsubsection{Performance Evaluation}

We now evaluate the performance of MFA-DVV using a compressed MFA model encoded from a large-scale dataset. The large-scale dataset is derived from the Marschner-Lobb function with a spatial resolution of $1024\times1024\times1024$. We generate two MFA models, with and without compression, respectively. The non-compressed MFA model uses $1024\times1024\times1024$ control points while the compressed MFA model uses $256\times256\times256$ control points. Thus, the size of the compressed MFA model is $1/64$ of the original dataset. We used various query methods for performance comparison on rendering time. We consider the most commonly used trilinear interpolation and two high-order local filters which are tricubic and Catmull-Rom. \autoref{fig:compressIoTime} shows the data fetching scalability that depends on the size of the input file, raw data, MFA model, and the compressed MFA model. We can see that the compressed MFA model saves considerable time on data fetching. \autoref{fig:compressRendering} shows the rendering scalabity. All high-order approximations, two MFA models, tricubic and Catmull-Rom, have longer rendering time than trilinear interpolation. The MFA model is generally faster than the other two high-order filters. Due to the computation reduction, the compressed MFA model performs better than the non-compressed version. As shown in \autoref{fig:compressedTotal}, with the help of improved data fetching latency and rendering latency, MFA-DVV gives the best overall performance when rendering a compressed MFA model. \autoref{fig:renderRenderingResults} and \autoref{tab:renderCompare} clearly show that MFA-DVV not only surpasses the already fast trilinear interpolation but also achieves more accurate rendering results.



\subsection{Visualization Example}

In \autoref{fig:example}, we showcase the visualization result of the MFA model encoded from a large-scale real dataset, RMI, with an overall view and several zoom-in views. 
This dataset boasts a high image resolution of $4096 \times 4096$. The results highlight the effectiveness of our high-resolution, high-precision distributed solution of MFA-DVV, empowering scientists to discern intricate details from vast amounts of data.

\section{Conclusions}
\label{sec:Conclusions}
We present MFA-DVV, a distributed volume visualization framework to render large-scale scientific datasets modeled by functional approximation. The proposed framework leverages Multivariate Functional Approximation (MFA) to improve rendering accuracy and achieve low input latency through a distributed architecture. Compared to existing compression methods, our MFA-DVV can balance better between image space quality and volume space quality. We examine the impact of essential MFA encoding parameters on both data fetching latency and rendering latency. Experimental results demonstrate that MFA-DVV showcases strong scalability, and its performance can be significantly enhanced by utilizing a compressed MFA model while still maintaining a high-quality rendering result for scientific datasets. Our MFA-DVV has provided scientists with high-quality and high-performance visualization to gain more detailed observations from their big scientific datasets. In the future, we would like to integrate new data models like implicit neural representations into our pipeline. Additionally, we are also interested in exploring alternative volume visualization techniques, such as inverse rendering, to offer diverse approaches for effectively visualizing large-scale volumes in parallel.
\section{Acknowledgement}
\label{sec:Acknowledgement}
This work is supported by Advanced Scientific Computing Research, Office of Science, U.S. Department of Energy, under Contracts DE-AC02-06CH11357, program manager Margaret Lentz. This work was completed utilizing the Holland Computing Center of the University of Nebraska, which receives support from the UNL Office of Research and Economic Development, and the Nebraska Research Initiative.

\bibliographystyle{abbrv}
\bibliography{IEEEexample}


\end{document}